\documentclass[12pt]{article}
\usepackage{graphicx}
\usepackage[margin=1.25in]{geometry}
\usepackage[usenames,dvipsnames]{xcolor}
\usepackage{url}
\usepackage{multirow}
\usepackage{adjustbox}
\usepackage{arydshln}
\usepackage[perpage]{footmisc}

\usepackage{enumitem}
\usepackage{caption,subcaption, cite}
\usepackage{scalerel}
\usepackage{makecell}
\usepackage[colorlinks = true,
            linkcolor = blue,
            urlcolor  = blue,
            citecolor = blue,
            anchorcolor = blue]{hyperref}

\usepackage{amsmath,amssymb,slashed,xcolor,cancel}           
\usepackage{rotating}
\usepackage{array,booktabs,colortbl,multirow,tablefootnote}
\usepackage{colortbl,colordvi,xcolor}
            
\def\figureautorefname~#1\null{Fig.\,#1\null}

\newcommand{\eqal}[1]{\begin{align}#1\end{align}}
\newcommand{\eqalsplit}[1]{\begin{align}\begin{split}#1\end{split}\end{align}}
\def\equationautorefname~#1\null{Eq.\,(#1)\null}

\newcommand\snowmass{\begin{center}\rule[-0.2in]{\hsize}{0.01in}\\\rule{\hsize}{0.01in}\\
\vskip 0.1in Submitted to the  Proceedings of the US Community Study\\ 
on the Future of Particle Physics (Snowmass 2021)\\ 
\rule{\hsize}{0.01in}\\\rule[+0.2in]{\hsize}{0.01in} \end{center}}

\def\iab{\mbox{ab$^{-1}$}}%  Inverse attobarns.
\def\ifb{\mbox{fb$^{-1}$}}%  Inverse femtobarns.

\def\SLAC{SLAC,
    Stanford University, Menlo Park, CA 94025, USA}
\def\Tokyo{ICEPP, The University of Tokyo, Hongo 7-3-1, Tokyo
  113-0033, JAPAN}
\def\DESY{DESY,
    Notkestrasse 85,  22607 Hamburg, GERMANY}

\def\Ugr{CAFPE and Departamento de F\'isica Te\'orica y del Cosmos, Universidad de Granada,
Campus de Fuentenueva, E–18071 Granada, Spain}
\def\Fudan{Department of Physics, Center for Field Theory and Particle Physics, Key Laboratory of Nuclear Physics and Ion-beam Application (MOE), Fudan University, Shanghai 200438, China}
\def\Rome{INFN, Sezione di Roma, Piazzale A. Moro 2, I-00185 Roma, Italy}
\def\ITPCAS{CAS Key Laboratory of Theoretical Physics, Institute of Theoretical Physics, Chinese Academy of Sciences, Beijing 100190, China}
\def\Valencia{IFIC, Universitat de Val\`encia and CSIC, c./ Catedr\'atico Jos\'e Beltr\'an 2, E-46980 Paterna, Spain}
\def\Manchester{University of Manchester, Oxford Road, Manchester M13 9PL, United Kingdom}
\def\CERN{CERN, Theoretical Physics Department, Geneva, Switzerland}

\def\Title#1{\begin{center} {\Large #1 } \end{center}}
\def\Author#1{\begin{center}{ \sc #1} \end{center}}

\newenvironment{Abstract}{\begin{quotation} \begin{center}
                       ABSTRACT
     \end{center}\bigskip  }{\end{quotation}}

\def\Acknowledgements{\bigskip  \bigskip \begin{center} \begin{large}
             \bf ACKNOWLEDGEMENTS \end{large}\end{center}}
\def\beq{\begin{equation}}
\def\eeq#1{\label{#1}\end{equation}}
\def\eeqn{\end{equation}}

\newenvironment{Eqnarray}%
   {\arraycolsep 0.14em\begin{eqnarray}}{\end{eqnarray}}
\def\beqa{\begin{Eqnarray}}
\def\eeqa#1{\label{#1}\end{Eqnarray}}
\def\eeqan{\end{Eqnarray}}

\def\leqn#1{(\ref{#1})}

\let\bar=\overbar

\def\lsim{\mathrel{\raise.3ex\hbox{$<$\kern-.75em\lower1ex\hbox{$\sim$}}}}
\def\gsim{\mathrel{\raise.3ex\hbox{$>$\kern-.75em\lower1ex\hbox{$\sim$}}}}

\def\hc{{\mbox{\rm h.c.}}}

\def\del{\partial}
\def\Dslash{\not{\hbox{\kern-4pt $D$}}}
\def\dslash{\not{\hbox{\kern-2pt $\del$}}}

\def\Dlr{\mathrel{\raise1.5ex\hbox{$\leftrightarrow$\kern-1em\lower1.5ex\hbox{$D$}}}}

\def\mz{m_Z}
\def\gz{\Gamma_Z}
\def\mw{m_W}

\def\mh{m_H}

\def\msb{{\bar{\scriptsize M \kern -1pt S}}}

\def\drb{{\bar{\scriptsize D \kern -1pt R}}}

\makeatletter
\def\section{\@startsection{section}{0}{\z@}{5.5ex plus .5ex minus
 1.5ex}{2.3ex plus .2ex}{\large\bf}}
\def\subsection{\@startsection{subsection}{1}{\z@}{3.5ex plus .5ex minus
 1.5ex}{1.3ex plus .2ex}{\normalsize\bf}}
\def\subsubsection{\@startsection{subsubsection}{2}{\z@}{-3.5ex plus
-1ex minus  -.2ex}{2.3ex plus .2ex}{\normalsize\sl}}

\renewcommand{\@makecaption}[2]{%
   \vskip 10pt
   \setbox\@tempboxa\hbox{\small #1: #2}
   \ifdim \wd\@tempboxa >\hsize     % IF longer than one line:
       \small #1: #2\par          %   THEN set as ordinary paragraph.
     \else                        %   ELSE  center.
       \hbox to\hsize{\hfil\box\@tempboxa\hfil}
   \fi}

\makeatother

\def\figureautorefname~#1\null{Fig.\,#1\null}

\def\equationautorefname~#1\null{Eq.\,(#1)\null}

\newcommand{\ttbar}{{\ensuremath{t\bar{t}}}}

\newcommand{\eeww}{e^+e^- \to W^+W^-}

\newcommand{\mtxt}[1]{\mathrm{#1}}

\newcommand{\lrD}{~\!\overset{\leftrightarrow}{\hspace{-0.1cm}D}\!}
\newcommand{\lrDa}{~\!\overset{\leftrightarrow}{\hspace{-0.1cm}D}\!^{\!~a}}
\newcommand{\SM}{\mathrm{SM}}

\newcommand{\SMEFT}{\mathrm{SMEFT}}

\begin{document}
\begin{titlepage}

\snowmass

%\begin{picture}(0,145)
%\put(-50,230){\rotatebox{30}{\textcolor{blue}{\LARGE \bf PRELIMINARY}}}
%\end{picture}
%\vspace{-6.5cm}

\begin{center}
[\today]
\end{center}

\vspace{-5cm}

\vfill
%\Title{Global SMEFT Fits at Future Colliders \\ \bigskip \it A Snowmass 2021 Whitepaper}
\Title{Global SMEFT Fits at Future Colliders}
\bigskip

\bigskip 

\Author{Jorge de Blas$^{a,b}$, Yong Du$^c$, Christophe Grojean$^d$, Jiayin Gu$^e$, \\
Víctor Miralles$^{f}$, Michael E. Peskin$^{g}$, Junping Tian$^h$, Marcel Vos$^i$, \\
and Eleni Vryonidou$^{^j}$}

\bigskip

\begin{center} { \it
$^a$ \Ugr \\
$^b$ \CERN \\
$^c$ \ITPCAS \\
$^d$ \DESY \\ 
$^e$ \Fudan \\
$^f$ \Rome \\
$^g$ \SLAC \\ 
$^h$ \Tokyo \\
$^i$ \Valencia \\
$^j$ \Manchester} \\
\end{center}

\vspace{2cm}

\newpage

\begin{Abstract}
Based on the framework of Standard Model Effective Field Theory,
we performed a few global fits, each containing a subset of dimension-6 operators,
for the measurements that are expected at future colliders.
The fit for the Higgs and electroweak sector improves what
has been done for the European Strategy Update in 2020 on
both EFT treatments and experimental inputs.
A new comprehensive fit is performed focusing on 4-fermion interactions at 
future colliders. Top-quark sector is studied in a dedicated fit
which restricts the operators and measurements to be directly related
to top-quark. A small subset of CP-violating operators involving
bosonic fields alone are also investigated. 
Various running scenarios for future $e^+e^-$ and Muon Colliders
that are suggested in the Snowmass 2021 discussion are considered
in the global fits.
The outcomes from each fit are expressed in terms of either direct constraint
on Wilson Coefficients or precision on Higgs and electroweak
effective couplings.
\end{Abstract}

\newpage
\tableofcontents
\end{titlepage}

\def\thefootnote{\fnsymbol{footnote}}
\setcounter{footnote}{0}

%%%%%%%%%%%%%%%%%%%%%%%%%%%%%%%%%%
\section{Introduction}
%%%%%%%%%%%%%%%%%%%%%%%%%%%%%%%%%%
%[editor: Michael]

In particle physics today, we have a Standard Model that, arguably,
accounts for all experimental measurements. At the same time, we are
convinced that this model is not a complete description of nature.
First, we know from astrophysical observations that the universe contains
elements such as dark matter and dark energy that this model does not
include.  But also, this Standard Model contains a large number of
free parameters that control many of its most important properties,
such as the mass scale of $W$ and $Z$ bosons, the mass spectrum of
fermions, and the appearance of CP violation.   It is not
straightforward to complete the Standard Model (SM) to repair these
difficulties.  Essentially, in all of these directions, the SM is
powerless, so that  a new model with additional fields and interactions is
needed to make progress.

There are many possibilities for what this new model should be. But,
none of these seem to be particularly favored, either from experiment anomalies or
on theoretical grounds.    All
approaches, including extensions of the SM particle content and
composite models of  the SM particles, are highly constrained by the data
from electroweak interactions and accelerator searches.  In
particular, the idea of TeV-scale supersymmetry, which held pride of
place among SM extensions in the 2000's, has been brought down in
stature by the absence of supersymmetric particles in the energy range
of the LHC.

In this situation, we would like to have a language for exploring
physics beyond the SM in a ``model agnostic'' way.   We would like to
have a theoretical framework that allows us to evaluate evidence for
the widest possible variety of new physics models, bringing together
data from the broadest set of experimental measurements.

Such a framework is actually at hand, under a particular
hypothesis---that
the mass scale of the new particles in the SM extension is much
larger than the energies used in our experimental probes.  This
hypothesis is suggested by the absence of new  particle discoveries at
the LHC.   If beyond-SM physics is manifested in very weakly coupled
light particles or in states that, because of details of their
production and decay, are difficult to observe at the LHC, this
hypothesis will not strictly apply and a more general analysis would be needed.
Still, the hypothesis leads to a tight conceptual structure that can organize our
exploration for physics beyond the SM.

Under the hypothesis that new physics has a high mass scale above the
reach of our current experiments,  the new fields of  any particular model can be integrated
out, producing an effective Lagrangian containing only SM fields
that can equally well describe the physics that we observe.  If the number of
parameters in this Lagrangian is restricted by gauge invariance and
other observed symmetries, it might be possible to determine the
parameters of the  effective Lagrangian from experiment without any
further model assumptions.   Then we can use these determinations as a
guide to formulate models of the new, underlying, theory.

The general effective field theory built from SM fields is called
Standard Model Effective Field Theory (SMEFT).    The Lagrangian of
SMEFT is organized by the dimension of the possible operators that can
appear.   The most general Lagrangian built from SM fields with
operators of dimension up to dimension 4 --- that is, with
renormalizable interactions --- is the SM itself.   The Lagrangian of
SMEFT then takes the form
\beq
   {\cal L}_{\SMEFT} =   {\cal L}_{\SM} + \sum_{d = 5}^\infty \sum_j  { C^{(d)}_j
     \over \Lambda^{d-4} } {\cal O}_j^{(d)}   \ ,
  \eeq{basicSMEFT}
where $ {\cal O}_j^{(d)}$ is an operator of dimension $d$
invariant under  the $SU(3)_c\times SU(2)_L\times U(1)_Y$ gauge group of
the SM, $\Lambda$ is the mass scale of new particles, with the
power of $\Lambda$ in each term determined by dimensional analysis,
and $C^{(d)}_j$ is a dimensionless number, the ``Wilson coefficient''
of the operator.   For any new physics model satisfying the
hypothesis above, integrating out the new fields   produces a
Lagrangian with this structure.  If the model is weakly coupled at
the scale $\Lambda$, the Wilson coefficients can be determined
systematically by Feynman diagram calculations.

At each value $d$ of the operator dimensions, there is only a
finite number of non-redundant operators.  Thus, in principle, it is
possible to make a closed determination of the Wilson coefficients
for all operators up to some dimension $D$ 
from experiment, and these can be compared to the predictions of
specific models.

The goal of this paper is to present the current status of our
understanding of this experimental determination, to illustrate some
of its subtleties, and to prepare for the determination of the Wilson
coefficients in experiments at future accelerators.

The program
described in the previous two paragraphs is a very general one, but we
will need to make some simplifications to make progress.  The number
of operators appearing in each term in the sum over $d$ in
\leqn{basicSMEFT} increases very rapidly with  $d$.  For this reason,
we will restrict our study to the first relevant corrections to the
SM. 
The Lagrangian \leqn{basicSMEFT} contains 2 operators of dimension $d
= 5$, but these contribute only to the neutrino masses and are not
relevant to collider physics.  More generally, operators of odd
dimension require lepton- or baryon-number violation and will be
omitted from our study.   The first relevant corrections to the SM
then occur at dimension 6.  In this  paper, we will restrict ourselves
to dimension 6 effects, and we will consider their effects  only in linear
order. These contributions are of the order of $1/\Lambda^2$. 
 Effects proportional to the squares of the dimension 6
amplitudes are proportional to $1/\Lambda^4$ and therefore are on the
same footing as the (much more numerous and complex) dimension 8
contributions.

Even in this simplified contexts, further restrictions are needed.
At dimension 6, the SMEFT Lagrangian still contains a large number of
unknown coefficients.   The total number of dimension 6
gauge-invariant operators is 84 for 1 generation of fermions (76 if 
one restricts to baryon- and lepton-number conserving operators, 59 if one further restricts to CP-conserving operators) and
3045 (2499 without baryon- and lepton-number violation) for 3 generations~\cite{Henning:2015alf}.  Thus, it is necessary to make
physically motivated restrictions on the class of operators being
considered. We will discuss fits to subsets of data in which the 
total number of relevant operators is manageable.  
 Also, although many published SMEFT analyses consider the
effects of one operator at a time, a model-independent analysis
requires that the coefficients of all relevant operators be varied
simultaneously.   Otherwise, we cannot match the effects generated by an
arbitrary underlying new physics theory.  In fits with a large number
of free coefficients, one often finds ``flat directions'' that are not
constrained by the fit, each corresponding to a  linear combinations of
operators for which the collective effect is not measured by the
experimental inputs.  To address these cases, we need to add inputs or
find good reasons to further restrict the class of relevant
operators.  This issue of balancing the number of operators considered
in the fit with the available experimental inputs comes front and
center in formulating meaningful global fits using SMEFT.  Our main
goal in this paper is to understand how to make this balance in 
practical examples.

The outline of this paper is as follows:   In Section 2, we will
present the SMEFT Lagrangian at dimension 6, presenting a preferred
operator basis for our analysis and defining the various SMEFT Wilson
coefficients that will appear.   In Section 3, we will discuss the
relation between this operator basis and the one used in the recent
ECFA study of the capabilities of future facilities.   In Section 4,
we will review the input measurements for our global fit.  

 In Section
5, we will present the results of an analysis of a subset of the
global fit using Higgs and electroweak operators only, together with
relevant experimental inputs. We will discuss the current constraints,
and the constraints expected from future data from HL-LHC, from 
$e^+e^-$ Higgs factories and muon colliders. We will also address the impact of theory errors in this global fit.
In Section 6, we will present the
results of a subset of the
global fit focusing on 4-fermion operators, together with
relevant experimental inputs, both current and future.   These two analyses will be done with
CP-conserving operators only.  In Section 7, we will extend the fit of
Section 5 to CP-violating operators. 

 All of the analyses up to this
point will be done for fermions that can be considered massless in
collider physics.  Inclusion of the massive top quark brings in an
additional set of operators.  In Section 8, we will present an
analysis that constrains this operator set using current LHC data and
data from future facilities.

%In Section 9, we will bring these threads together into a truly global
%analysis and discuss some physics implications.
In Section 9, we will address the findings from above global fits and 
implications for future colliders. We will also put out some outlook about
potential studies that can future bring improvement to this work.

In this report, we will not address the question of distinguishing the
SMEFT from more general effective field theories of electroweak
symmetry breaking such as Higgs Effective Field Theory (HEFT).  
Distinguishing these models and demonstrating that SMEFT is not
sufficient requires measurements beyond the scope of this report, such
as measurements of multiple Higgs boson production.  Please see
\cite{Alonso:2016oah,Helset:2020yio,Cohen:2021ucp}
 for a detailed discussion.

%%%%%%%%%%%%%%%%%%%%%%%%%%%%%%%%%%
\section{The Standard Model Effective Field Theory Lagrangian\label{sec:SMEFT}}
%%%%%%%%%%%%%%%%%%%%%%%%%%%%%%%%%%
%[editor: Jorge]

As we have introduced above, we will study experimental 
constraints on the SMEFT Lagrangian, truncating the EFT 
expansion to dimension 6,
\beq
   {\cal L}_{\SMEFT} =   {\cal L}_{\SM} + \sum_j  { C^{(6)}_j
     \over \Lambda^{2} } {\cal O}_j^{(6)}  \ ,
  \eeq{SMEFT6}
 and including only the leading-order new physics effects in observables, i.e. the linear ${\cal O}(1/\Lambda^2)$ contributions.
 In what follows, as we will only consider dimension-6 operators, we drop the superscript ``$(6)$'' from the Wilson coefficients and operators.
 In this section, we will write out this Lagrangian explicitly in our preferred basis,
 the so-called {\it Warsaw basis} \cite{Grzadkowski:2010es}, for the cases of baryon- and lepton-number-conserving operators, % 
 giving a total of 59 dimension-6 operator coefficients for 1 generation (not counting hermitian conjugates separately).
  
We start first with those operators which, after electroweak symmetry breaking, modify the vertices already present in ${\cal L}_{\SM}$ (possibly introducing new tensor structures): 

\begin{equation}
\begin{split}
{\cal L}_{\mathrm{SMEFT}}^{d=6}\supset &\phantom{+}
\frac{C_\phi}{\Lambda^2}\left(\phi^{\dagger} \phi\right)^3 
+ \frac{C_{\phi\square}}{\Lambda^2} \left(\phi^{\dagger} \phi\right)\square
      \left(\phi^{\dagger} \phi\right) 
+ \frac{C_{\phi D} }{\Lambda^2} \left(\phi^{\dagger}D_\mu \phi\right)(\left(D^\mu \phi\right)^{\dagger}\phi) \\
&
+\frac{C_W }{\Lambda^2} \varepsilon_{abc} W^{a\,\nu}_\mu W^{b\,\rho}_\nu W^{c\,\mu}_\rho +\frac{C_G }{\Lambda^2} f_{ABC} G^{A\,\nu}_\mu G^{B\,\rho}_\nu G^{C\,\mu}_\rho \\%
&
+\frac{C_{\phi B}}{\Lambda^2} \phi^\dagger \phi B_{\mu\nu} B^{\mu\nu} 
+\frac{C_{\phi W}}{\Lambda^2} \phi^\dagger \phi W_{\mu\nu}^a W^{a\,\mu\nu}
+\frac{C_{\phi WB}}{\Lambda^2} \phi^\dagger \sigma_a \phi W^a_{\mu\nu} B^{\mu\nu}
+\frac{C_{\phi G} }{\Lambda^2} \phi^\dagger \phi G_{\mu\nu}^A G^{A\,\mu\nu} \\
&+\left(\frac{\left(C_{e\phi}\right)_{ij} }{\Lambda^2} \left(\phi^{\dagger} \phi\right)
      (\bar{l}_L^i \phi e_R^j)
      +\frac{\left(C_{d\phi}\right)_{ij}}{\Lambda^2} \left(\phi^{\dagger} \phi\right)
      (\bar{q}_L^i \phi d_R^j) 
       +\frac{\left(C_{u\phi}\right)_{ij}}{\Lambda^2} \left(\phi^{\dagger} \phi\right)
      (\bar{q}_L^i \tilde{\phi} u_R^j)  + \hc \right) \\
&+\left(\frac{\left(C_{eB}\right)_{ij} }{\Lambda^2} B^{\mu\nu}
      (\bar{l}_L^i \phi \sigma_{\mu\nu} e_R^j)
      +\frac{\left(C_{dB}\right)_{ij}}{\Lambda^2} B^{\mu\nu}
      (\bar{q}_L^i \phi \sigma_{\mu\nu} d_R^j) 
       +\frac{\left(C_{uB}\right)_{ij}}{\Lambda^2} B^{\mu\nu}
      (\bar{q}_L^i \tilde{\phi} \sigma_{\mu\nu} u_R^j)  + \hc \right)\\
&+\left(\frac{\left(C_{eW}\right)_{ij} }{\Lambda^2} W^{a~\!\mu\nu}
      (\bar{l}_L^i \phi \sigma_{\mu\nu} \sigma_a e_R^j)
      +\frac{\left(C_{dW}\right)_{ij}}{\Lambda^2} W^{a~\!\mu\nu}
      (\bar{q}_L^i \phi \sigma_{\mu\nu} \sigma_a d_R^j)\right.\\
      &\quad\quad\left.
       +\frac{\left(C_{uW}\right)_{ij}}{\Lambda^2} W^{a~\!\mu\nu}
      (\bar{q}_L^i \tilde{\phi} \sigma_{\mu\nu} \sigma_a u_R^j)  + \hc \right)\nonumber \\
\end{split} 
%\label{eq:LSMEFT}
\end{equation}
\begin{equation}
\begin{split}      
&+\left(\frac{\left(C_{dG}\right)_{ij}}{\Lambda^2} G^{A~\!\mu\nu}
      (\bar{q}_L^i \phi \sigma_{\mu\nu} T_A d_R^j) 
       +\frac{\left(C_{uG}\right)_{ij}}{\Lambda^2} G^{A~\!\mu\nu}
      (\bar{q}_L^i \tilde{\phi} \sigma_{\mu\nu} T_A u_R^j)  + \hc \right) \\
&
+\frac{\left(C_{\phi l}^{(1)}\right)_{ij}}{\Lambda^2} (\phi^{\dagger} i\overset{\leftrightarrow}{D}_\mu \phi)
      (\bar{l}_L^i \gamma^\mu l_L^j)
+\frac{\left(C_{\phi l}^{(3)}\right)_{ij}}{\Lambda^2} (\phi^{\dagger} i\lrDa_\mu \phi)
      (\bar{l}_L^i \gamma^\mu \sigma_a l_L^j)  \\
&
+\frac{\left(C_{\phi e}\right)_{ij}}{\Lambda^2}(\phi^{\dagger} i\overset{\leftrightarrow}{D}_\mu \phi)
      (\bar{e}_R^i \gamma^\mu e_R^j) \\
&
+\frac{\left(C_{\phi q}^{(1)}\right)_{ij}}{\Lambda^2} (\phi^{\dagger} i\overset{\leftrightarrow}{D}_\mu \phi)
      (\bar{q}_L^i \gamma^\mu q_L^j)
+\frac{\left(C_{\phi q}^{(3)}\right)_{ij}}{\Lambda^2} (\phi^{\dagger} i \lrDa_\mu \phi)
      (\bar{q}_L^i \gamma^\mu\sigma_a q_L^j) \\
&
+\frac{\left(C_{\phi u}\right)_{ij}}{\Lambda^2} (\phi^{\dagger} i\overset{\leftrightarrow}{D}_\mu \phi)
      (\bar{u}_R^i \gamma^\mu u_R^j)
+\frac{\left(C_{\phi d}\right)_{ij}}{\Lambda^2} (\phi^{\dagger} i\overset{\leftrightarrow}{D}_\mu \phi)
      (\bar{d}_R^i \gamma^\mu d_R^j) \\
&+\frac{\left(C_{\phi ud}\right)_{ij}}{\Lambda^2} (\tilde{\phi}^{\dagger} i\overset{\leftrightarrow}{D}_\mu \phi)
      (\bar{u}_R^i \gamma^\mu d_R^j).
\end{split} 
\label{eq:LSMEFT}
\end{equation}
The hermitian derivatives $\lrD$ and $\lrDa$ are defined as: 
$$\lrD_\mu \equiv \overset{\rightarrow}{D}_\mu - \overset{\leftarrow}{D}_\mu$$ 
and 
$$\lrDa_\mu \equiv \sigma_a \overset{\rightarrow}{D}_\mu - \overset{\leftarrow}{D}_\mu \sigma_a,$$
with $D_\mu=\partial_\mu - i g^\prime B_\mu Y  -i g W_\mu^a T_a -i g_s G_\mu^A {\rm T_A}$, and $Y$, $T_a$, ${\rm T}_A$ the hypercharge and $SU(2)_L$ and $SU(3)_c$ generators, respectively, and $\sigma_a$ the Pauli matrices. The symbols $B_{\mu\nu}$, $W_{\mu\nu}^a$ and $G_{\mu\nu}^A$ denote the corresponding SM gauge-boson field strengths. Finally, for the scalar doublet, $\tilde{\phi}=i\sigma_2 \phi^*$. In the fermionic operators, summation over the flavour indices is implicit. In practice, only diagonal entries will contribute to most of the observables we will consider.

The previous set of interactions do not include any purely bosonic CP-odd operator. These will be relevant for the discussion in Section~\ref{sec:CPodd} and are also listed here for completeness:
\begin{equation}
\begin{split}
{\cal L}_{\mathrm{SMEFT}}^{d=6, {\rm CP-odd, bos}} = &
\frac{C_{\widetilde{W}}}{\Lambda^2} \varepsilon_{abc} \widetilde{W}^{a\,\nu}_\mu W^{b\,\rho}_\nu W^{c\,\mu}_\rho +\frac{C_{\widetilde{G}} }{\Lambda^2} f_{ABC} \widetilde{G}^{A\,\nu}_\mu G^{B\,\rho}_\nu G^{C\,\mu}_\rho \\%
&
+\frac{C_{\phi \widetilde{B}}}{\Lambda^2} \phi^\dagger \phi \widetilde{B}_{\mu\nu} B^{\mu\nu} 
+\frac{C_{\phi \widetilde{W}}}{\Lambda^2} \phi^\dagger \phi \widetilde{W}_{\mu\nu}^a W^{a\,\mu\nu}
+\frac{C_{\phi \widetilde{W}B}}{\Lambda^2} \phi^\dagger \sigma_a \phi \widetilde{W}^a_{\mu\nu} B^{\mu\nu}\\%
&+\frac{C_{\phi \widetilde{G}} }{\Lambda^2} \phi^\dagger \phi \widetilde{G}_{\mu\nu}^A G^{A\,\mu\nu},
\end{split}
\label{eq:LSMEFT-CPodd-bos}
\end{equation}
with $\widetilde{X}_{\mu\nu}=\frac 12 \varepsilon_{\mu\nu\sigma\rho}X^{\sigma\rho}$ the Hodge dual of the corresponding field-strength tensors.

For the electroweak $Z$-pole and diboson observables, and most of the Higgs processes considered here, four-fermion operators do not contribute or
are expected to have negligible effects under the resonances. The exceptions are $t\bar{t}H$ and, if one chooses $G_F$ as part of the SM electroweak input parameters, as we will do here, the four-lepton operator $\left(\overline{l_L}\gamma^\mu l_L\right)\left(\overline{l_L}\gamma_\mu l_L\right)$. In the studies presented in this report, however, we will also consider the constraints induced by 2 to 2 fermion processes, in which case contact interactions between four fermions need to be considered. Those relevant at future lepton colliders are:
\begin{equation}
\begin{split}
{\cal L}_{\mathrm{SMEFT}}^{d=6, \ell^2 \psi^2}\supset 
&\phantom{+}
\frac{\left(C_{ll}\right)_{ijkl} }{\Lambda^2} \left(\overline{l_L^i}\gamma^\mu l_L^j\right)\left(\overline{l_L^k}\gamma_\mu l_L^l\right)
      \\
&+
\frac{\left(C_{lq}^{(1)}\right)_{ijkl} }{\Lambda^2} \left(\overline{l_L^i}\gamma^\mu l_L^j\right)\left(\overline{q_L^k}\gamma_\mu q_L^l\right)
      +\frac{\left(C_{lq}^{(3)}\right)_{ijkl} }{\Lambda^2} \left(\overline{l_L^i}\gamma^\mu\sigma_a l_L^j\right)\left(\overline{q_L^k}\gamma_\mu\sigma_a q_L^l\right) \\
&+
\frac{\left(C_{ee}\right)_{ijkl} }{\Lambda^2} \left(\overline{e_R^i}\gamma^\mu e_R^j\right)\left(\overline{e_R^k}\gamma_\mu e_R^l\right)\\
&+
\frac{\left(C_{eu}\right)_{ijkl} }{\Lambda^2} \left(\overline{e_R^i}\gamma^\mu e_R^j\right)\left(\overline{u_R^k}\gamma_\mu u_R^l\right)
+
\frac{\left(C_{ed}\right)_{ijkl} }{\Lambda^2} \left(\overline{e_R^i}\gamma^\mu e_R^j\right)\left(\overline{d_R^k}\gamma_\mu d_R^l\right)\\
&+
\frac{\left(C_{le}\right)_{ijkl} }{\Lambda^2} \left(\overline{l_L^i}\gamma^\mu l_L^j\right)\left(\overline{e_R^k}\gamma_\mu e_R^l\right)
      +\frac{\left(C_{qe}\right)_{ijkl} }{\Lambda^2} \left(\overline{q_L^i}\gamma^\mu q_L^j\right)\left(\overline{e_R^k}\gamma_\mu e_R^l\right) \\
&+
\frac{\left(C_{lu}\right)_{ijkl} }{\Lambda^2} \left(\overline{l_L^i}\gamma^\mu l_L^j\right)\left(\overline{u_R^k}\gamma_\mu u_R^l\right)+
\frac{\left(C_{ld}\right)_{ijkl} }{\Lambda^2} \left(\overline{l_L^i}\gamma^\mu l_L^j\right)\left(\overline{d_R^k}\gamma_\mu d_R^l\right) \\
&+
\frac{\left(C_{lequ}\right)_{ijkl} }{\Lambda^2} \left(\overline{l_L^i} e_R^j\right)i\sigma_2\left(\overline{q_L^k}^T{u_R^l}\right)+
\frac{\left(C_{ledu}^{(3)}\right)_{ijkl} }{\Lambda^2} \left(\overline{l_L^i} \sigma^{\mu\nu}e_R^j\right)i\sigma_2\left(\overline{q_L^k}^T\sigma_{\mu\nu}{u_R^l}\right)\\
&+\frac{\left(C_{ledq}\right)_{ijkl} }{\Lambda^2} \left(\overline{l_L^i}e_R^j\right)\left(\overline{d_R^j}{q_L^l}\right),
\end{split}
\label{eq:LSMEFT4flf}
\end{equation}
where $^T$ denotes the transpose in the $SU(2)_L$ indices and, again, summation over flavour indices is understood. 
We will also include results pertaining the limits that can be obtained from Top processes at the HL-LHC, so we also need to consider the following relevant four-quark operators
\begin{equation}
\begin{split}
{\cal L}_{\mathrm{SMEFT}}^{d=6, q^4}\supset 
&\phantom{+}
\frac{\left(C_{qq}^{(1)}\right)_{ijkl} }{\Lambda^2} \left(\overline{q_L^i}\gamma^\mu q_L^j\right)\left(\overline{q_L^k}\gamma_\mu q_L^l\right)
+\frac{\left(C_{qq}^{(3)}\right)_{ijkl} }{\Lambda^2} \left(\overline{q_L^i}\gamma^\mu\sigma_a q_L^j\right)\left(\overline{q_L^k}\gamma_\mu\sigma_a q_L^l\right) \\
&
+\frac{\left(C_{uu}\right)_{ijkl} }{\Lambda^2} \left(\overline{u_R^i}\gamma^\mu u_R^j\right)\left(\overline{u_R^k}\gamma_\mu u_R^l\right)
+\frac{\left(C_{ud}^{(8)}\right)_{ijkl} }{\Lambda^2} \left(\overline{u_R^i}\gamma^\mu T_A u_R^j\right)\left(\overline{d_R^k}\gamma_\mu T_A d_R^l\right)\\
&+
\frac{\left(C_{qu}^{(8)}\right)_{ijkl} }{\Lambda^2} \left(\overline{q_L^i}\gamma^\mu T_A q_L^j\right)\left(\overline{u_R^k}\gamma_\mu T_A u_R^l\right) 
+\frac{\left(C_{qd}^{(8)}\right)_{ijkl} }{\Lambda^2} \left(\overline{q_L^i}\gamma^\mu T_A q_L^j\right)\left(\overline{d_R^k}\gamma_\mu T_A d_R^l\right),
\end{split}
\label{eq:LSMEFT4ftop}
\end{equation}
where we only considered those that can interfere with the SM QCD contributions.
For the Top constraints we will follow the recommendations of \cite{Aguilar-Saavedra:2018ksv} and then express the
results in terms of the combinations of Wilson coefficients reported in Table~\ref{tab:WilsonDefined}. 
\begin{table}[h!]
\centering
{\footnotesize  
\begin{tabular}{ |c|c|c|c| } 
 \hline
 \multicolumn{4}{|c|}{\textbf{Coefficients fitted in the top-quark processes}} \\
 \hline
  \multirow{3}{*}{\!\!2-quark\!\!\!} & $C_{ t G}=(C_{uG})_{33}$  & $C_{\phi Q}^3=\left(C_{\phi q}^{(3)}\right)_{33}$  & $C_{\phi Q}^- = \left(C_{\phi q}^{(1)}\right)_{33}-\left(C_{\phi q}^{(3)}\right)_{33}$  \\ 
    &   $C_{\phi t} = (C_{\phi u})_{33}$ &  $C_{\phi b} = (C_{\phi d})_{33}$  &    $C_{t Z} = \cos{\theta_w} (C_{uW})_{33} -\sin{\theta_w} (C_{uB})_{33}$   \\ 
 
    &  -- &  $C_{t \phi } = (C_{u \phi })_{33}$ &   $C_{t W} = (C_{uW})_{33}$    \\ 
 \hline
 \hline
     \multirow{3}{*}{\!\!4-quark\!\!} &  $ C_{tu}^8 = \sum\limits_{\scaleto{i=1,2}{4pt}} 2\left(C_{uu}\right)_{i33i}$ &  $ C_{td}^8 = \sum\limits_{\scaleto{i=1,2,3}{4pt}} \left(C_{ud}^{(8)}\right)_{33ii}$ &  $C_{Qq}^{1,8} = \sum\limits_{\scaleto{i=1,2}{4pt}} \Big(\left(C_{qq}^{(1)}\right)_{i33i}+3\left(C_{qq}^{(3)}\right)_{i33i}\Big)$  \\

        & $ C_{Qu}^{8} = \sum\limits_{\scaleto{i=1,2}{4pt}} \left( C_{qu}^{(8)}\right)_{33ii}$ &  $ C_{Qd}^{8} =\sum\limits_{\scaleto{i=1,2,3}{4pt}} \left(C_{qd}^{(8)}\right)_{33ii}$  &   $C_{Qq}^{3,8} =\sum\limits_{\scaleto{i=1,2}{4pt}} \Big( \left( C_{qq}^{(1)}\right)_{i33i}-\left(C_{qq}^{(3)}\right)_{i33i}\Big)$  \\

    &    -- &   --  &  $C_{tq}^{8} = \sum\limits_{\scaleto{i=1,2}{4pt}} \left(C_{qu}^{(8)}\right)_{ii33}$  \\
 \hline
 \hline
 \multirow{3}{*}{\makecell{\!\!2-quark\!\!\\\!\!2-lepton\!\!}}  & $C_{eb}=\left(C_{ed}\right)_{1133}$ & $C_{et}= \left(C_{eu}\right)_{1133}$ & $ C_{ l Q}^+ = \left(C_{lq}^{(1)}\right)_{1133}+\left(C_{lq}^{(3)}\right)_{1133}$  \\
 
   & $C_{lb}=\left(C_{ld}\right)_{1133}$ &  $C_{lt}= \left(C_{lu}\right)_{1133}$ &  $C_{ l Q}^- = \left(C_{lq}^{(1)}\right)_{1133}-\left(C_{lq}^{(3)}\right)_{1133}$  \\
  
    &  -- & --  &   $C_{eQ} = \left(C_{qe}\right)_{3311}$ \\
   \hline
\end{tabular}
}
\caption{Here we present the Wilson coefficients that have been fitted in our top-quark analysis in terms of those of Eqs.~\eqref{eq:LSMEFT},\eqref{eq:LSMEFT4flf} and \eqref{eq:LSMEFT4ftop}. The first block are related with the 2-quark operators, the second block are related with the 4-quark operators and the last block is related with the 2-quark 2-lepton operators.}
\label{tab:WilsonDefined}
\end{table}

\subsection{Effective Lagrangian in the mass eigenstate basis\label{sec:SMEFTphysB}}

After electroweak symmetry breaking and upon writing the Lagrangian in the physical basis, the dimension-six operator introduced above give rise to both modifications of the SM interactions as well as to new terms not present in the SM Lagrangian. Following \cite{Falkowski:2015fla}, using the $\{\alpha, M_Z,G_F\}$ input scheme, putting our focus on the electroweak and Higgs interactions, and restricting to CP-even interactions for the moment, the effective Lagrangian including dimension-6 terms contain the following pieces:

\begin{itemize}

{\item {\bf Higgs couplings to vector bosons:}
\begin{eqnarray}
\label{eq:LhVV}
\Delta {\cal L}^{\rm hVV}_{6}\!\! &=\!\! & {\frac{h}{v}} \left [ 
\vphantom{\frac{1}{2}}
2 \delta c_W\,   m_W^2 W_\mu^+ W^{-\mu} +   \delta c_Z\,   m_Z^2 Z_\mu Z^\mu
\right . \nonumber\\ & & \left . 
+ c_{WW}\,  {\frac{g^2}{2}} W_{\mu \nu}^+  W^{-\mu\nu}  
+ c_{W \Box}\, g^2 \left (W^{-\mu} \partial^\nu W_{\mu \nu}^+ + {\rm h.c.} \right )  
\right . \nonumber\\ & & \left . 
+ c_{gg}\! {\frac{g_s^2}{4}} G_{\mu \nu}^A G^{A\mu \nu} \!  + c_{\gamma \gamma}\! {\frac{e^2}{4}} A_{\mu \nu} A^{\mu \nu} \!
+ c_{Z \gamma}\! {\frac{e \sqrt{g^2+g^{\prime~\!2}}}{2}} Z_{\mu \nu} A^{\mu\nu}\! + c_{ZZ}\, {\frac{g^2+g^{\prime~\!2}}{4}} Z_{\mu \nu} Z^{\mu\nu}
\right . \nonumber\\ & & \left .
+ c_{Z \Box}\, g^2 Z^\mu \partial^\nu Z_{\mu \nu} + c_{\gamma \Box}\, g g^\prime Z^\mu \partial^\nu A_{\mu \nu}
\vphantom{\frac{1}{2}}
\right ],
\end{eqnarray}
where only $c_{gg}, \  \delta c_Z,  \ c_{\gamma \gamma}, \ c_{Z \gamma},  \ c_{ZZ},  \ c_{Z \Box}$ are independent parameters:
\begin{eqnarray}
\label{eq:dep_pars}
\delta  c_{W} &=&  \delta c_Z + 4 \delta m , 
\nonumber\\
c_{WW} &=&  c_{ZZ} + 2 \sin^2{\theta_w} c_{Z \gamma} + \sin^4{\theta_w} c_{\gamma \gamma}, 
\nonumber\\
c_{W \Box}  &= & {\frac{1}{g^2 - g^{\prime~\!2}}} \left [ 
g^2 c_{Z \Box} + g^{\prime~\!2} c_{ZZ}  - e^2 \sin^2{\theta_w}   c_{\gamma \gamma}  -(g^2 - g^{\prime~\!2}) \sin^2{\theta_w}  c_{Z \gamma} 
\right ],  
\nonumber\\
  c_{\gamma \Box}  &= &  
  {\frac{1}{g^2 - g^{\prime~\!2}}} \left [ 
2 g^2 c_{Z \Box} + (g^2+ g^{\prime~\!2}) c_{ZZ}  - e^2  c_{\gamma \gamma}  -(g^2 - g^{\prime~\!2})   c_{Z \gamma} 
\right ], 
\end{eqnarray}
with $\theta_w$ the weak mixing angle and the parameter $\delta m$ contains the
dimension-6 contributions to $M_W$ with respect to the SM value,
\begin{equation}
\Delta {\cal L}_{W,Z}^{\rm mass}
=\frac{(g^2+g^{\prime~\!2})v^2}{8}Z_\mu Z^\mu + \frac{g^2v^2}{4}(1+\delta m)^2W_\mu^+W^{-\mu}.
\end{equation}
}
{\item {\bf Trilinear Gauge Couplings:}
  \begin{eqnarray}
   \label{eq:LaTGC}
\Delta {\cal L}^{\mathrm{aTGC}} &= & i e \delta \kappa_\gamma\, A^{\mu\nu} W_\mu^+ W_\nu^- +
i g  \cos{\theta_w} \left[ \delta g_{1,Z}\, (W_{\mu\nu}^+ W^{-\mu} - W_{\mu\nu}^- W^{+\mu})Z^\nu\right. \nonumber\\
&&\left.+(\delta g_{1,Z}-\frac{g^{\prime~\!2}}{g^2}\delta \kappa_\gamma)\, Z^{\mu\nu} W_\mu^+ W_\nu^- \right]
\nonumber\\
&&+\frac{ig \lambda_Z}{m_W^2}\left( \sin{\theta_w} W_{\mu}^{+\nu} W_{\nu}^{-\rho} A_{\rho}^{\mu} + \cos{\theta_w} W_{\mu}^{+\nu} W_{\nu}^{-\rho} Z_{\rho}^{\mu} \right),
  \end{eqnarray}
  where two of the three coefficients, $\delta g_{1,Z}$ and $ \delta \kappa_\gamma$ depend on $c_{gg}, \  \delta c_Z,  \ c_{\gamma \gamma}, \ c_{Z \gamma},  \ c_{ZZ},  \ c_{Z \Box}$:
\begin{eqnarray}
 \delta  g_{1,Z} &=& 
{\frac{1}{2} (g^2 - g^{\prime~\!2})} \left [   c_{\gamma\gamma} e^2 g^{\prime~\!2} + c_{Z \gamma} (g^2 - g^{\prime~\!2}) g^{\prime~\!2}  - c_{ZZ} (g^2 + g^{\prime~\!2}) g^{\prime~\!2}  - c_{Z\Box} (g^2 + g^{\prime~\!2}) g^2 \right ], 
 \nonumber\\
 \delta \kappa_\gamma  &=& - {\frac{g^2}{2}} \left ( c_{\gamma\gamma} {\frac{e^2 }{g^2 + g^{\prime~\!2}}}   + c_{Z\gamma} \frac{g^2  - g^{\prime~\!2}}{ g^2 + g^{\prime~\!2}} - c_{ZZ} \right ), 
 \end{eqnarray}  
while $\lambda_Z$ is  an independent parameter. Quartic gauge couplings also receive contributions in the effective Lagrangian but, to dimension 6, they are always connected to the trilinear ones. 
}
{\item {\bf Yukawa couplings:}
\begin{equation}
\label{eq:hff}
\Delta {\cal L}^{\rm hff}_{6}  =  - {\frac{h}{v}} \sum_{f \in u,d,e}    \hat \delta y_f \, m_f  \bar{f} f  + {\rm h.c.} , 
\end{equation}
where $\hat \delta y_f \, m_f$ should be thought as $3\times 3$ matrices in flavour space. FCNC are avoided when $\hat \delta y_f$ is diagonal in the same basis as $m_f$. Note that once we include dimension-6 contributions, the SM relation between the fermion masses and Yukawa interactions no longer holds and these are two sets of independent parameters. 
}
{\item {\bf Vector couplings to fermions:} while corrections to the QED and QCD vertices are protected by gauge invariance, the electroweak interactions of fermions $Vff$ ($V=Z,W$) are modified at dimension 6. These modifications are directly related to contact interactions of the form $hVff$:
  \begin{eqnarray} 
\label{eq:vff}
\Delta {\cal L}^{\rm Vff, hVff}_{6}\!\!&=&\!\! \frac{g}{\sqrt{2}} \left(1+2 \frac{h}{v}\right)   W_\mu^+  \!\! \left (
 \hat \Delta g^{\ell }_{W} \bar{\nu}_L \gamma^\mu  e_L
 +   \hat \Delta g^{q}_{W,L} \bar{u}_L \gamma^\mu  d_L
+ \hat \Delta g^{q}_{W,R}  \bar{u}_R  \gamma^\mu   d_R
 + \hc  \right )
\nonumber\\ 
&+&\!\! \sqrt{g^2 + g^{\prime~\!2}}  \left(1+2 \frac{h}{v} \right)  \!\! Z_\mu \!\!
\left [ \sum_{f = u,d,e,\nu} \hat  \Delta g^{f}_{Z,L}  \bar{f}_L \gamma^\mu f_L  + \!\!  
\sum_{f = u,d,e} \hat  \Delta g^{f}_{Z,R}  \bar{f}_R \gamma^\mu f_R  \right ].\nonumber\\
\end{eqnarray}
The $\hat \Delta g^{Y}_{X,L/R}$ are, again, $3$x$3$ matrices in flavor space and parameterize, in particular, {\it absolute} modifications of the EW couplings}. Also, not all terms in the previous equation are independent and the following relations hold to dimension 6:
\begin{equation} 
\hat \Delta g^{\ell }_{W} =\hat  \Delta g^{\nu}_{Z,L} - \hat \Delta g^{e}_{Z,L},~~~~~~~\hat \Delta g^{q}_{W,L} = \hat \Delta g^{u}_{Z,L} V_{\rm CKM} - V_{\rm CKM} \hat \Delta g^{d}_{Z,L}
,
\end{equation} 
with $V_{\rm CKM}$ the {\it Cabibbo-Kobayashi-Maskawa} (CKM) matrix which, unless otherwise is stated, we approximate to the identity matrix.
\end{itemize}

\subsection{Effective couplings\label{sec:SMEFTeffC}}
As done in \cite{deBlas:2019wgy,deBlas:2019rxi}, some of the results will be presented, not in terms of the Wilson coefficients of the manifestly gauge-invariant operators, but in terms of pseudo-observable quantities, referred to as {\it effective Higgs and electroweak couplings}, computed from physical observables and thus, independent of the basis one could have chosen for the dimension-6 Lagrangian. 
This is done by performing the fit {\it internally} in terms of the Wilson coefficients and then, from the posterior of the fit, compute the posterior prediction for the quantities
\begin{equation}
g_{HX}^{\mathrm{eff}~2}\equiv \frac{\Gamma_{H\to X}}{\Gamma_{H\to X}^{\SM}}.
\label{eq:gHeff}
\end{equation}
for the {\it Higgs effective couplings}, or the quantities $g_{Zff,L/R}^{\mathrm{eff}}$ for the {\it electroweak effective couplings}, defined from:
\begin{equation}
\Gamma_{Z\to e^+ e^-}=\frac{\alpha ~\! M_Z}{6 \sin^2{\theta_w} \cos^2{\theta_w}}(|g_{Zee,L}^{\mathrm{eff}}|^2 + |g_{Zee,R}^{\mathrm{eff}}|^2) ,\quad\quad A_e=\frac{|g_{Zee,L}^{\mathrm{eff}}|^2 - |g_{Zee,R}^{\mathrm{eff}}|^2}{|g_{Zee,L}^{\mathrm{eff}}|^2 + |g_{Zee,R}^{\mathrm{eff}}|^2}.
\label{eq:gZeff}
\end{equation}

Note that the definition in \autoref{eq:gHeff} is not phenomenologically possible for the top-Higgs coupling and the Higgs self-interaction. Being aware of this, for presentational purpose we will nevertheless still apply similar definition for $g_{Htt}^{\mathrm{eff}}$. To further connect with diboson processes, and even though they are technically not pseudo-observables, we will also use the aTGC $\delta g_{1,Z}$, $\delta \kappa_\gamma$ and $\lambda_Z$. Finally, we use $g_{HHH}\equiv \lambda_3 /\lambda_3^{\SM}$, to describe modifications of the Higgs self coupling.

In the results presented below, we will report the expected sensitivities to relative modifications of these effective couplings with respect to the SM values, whenever these are non-zero. Such relative shifts are always indicated by the symbol $\delta$, whereas absolute shifts will be indicated with $\Delta$, i.e., given a quantity $X$:
\begin{equation}
\Delta X\equiv X-X_{\SM},~~~~~~~\delta X\equiv \frac{\Delta X}{X_{\SM}}.
\end{equation}
For instance, in this notation, the new physics contributions to the effective couplings between fermions and electroweak bosons are given by:
\begin{equation}
\delta g_{V,L/R}^{ff} \equiv \frac{(\hat \Delta g_{V,L/R}^{f})_{ff}}{g_{V,L/R}^{f,\rm SM}}.
\end{equation}
Whenever a given quantity is zero in the SM, e.g. $\lambda_Z$ or any of the Wilson coefficients $C_i$, the sensitivity will be reported directly on the parameter.

%%%%%%%%%%%%%%%%%%%%%%%%%%%%%%%%%%
\section{Recap on SMEFT fits for ESG}
%%%%%%%%%%%%%%%%%%%%%%%%%%%%%%%%%%
%[editor: Christophe]

Global fits of the data expected at HL-LHC and future colliders have been carried out in the context of the 2020 European Strategy Update for Particle Physics~\cite{deBlas:2019rxi} with a special emphasis on the Higgs sector. One key question addressed was the sensitivity of the various colliders to the deformations of the Higgs couplings to the different SM particles compared to their values predicted robustly in SM itself. These fits relied on the measurements of the Higgs production cross section times its decay branching ratios in the different channels. Two different approaches, as model-independent as possible, were adopted. On the one hand, in the $\kappa$-framework, it is assumed that the structure of the Higgs interactions remain identical to the SM one. While rather simple and adequate to capture dominant effects in well-motivated New Physics scenarios like composite Higgs models, this approach lacks some generality and makes it difficult to fully exploit information collected away for the Higgs pole and is not easily amenable  to a full inclusion of quantum higher-order corrections. The more general Effective Field Theory approach aims to remedy these limitations and to fully capture all possible effects generated by new heavy degrees of freedom. 

The ESU fits made use of only inclusive cross section times branching ratio measurements (or even ratios thereof), omitting precious kinematic information, like Higgs transverse momentum distribution, which could reveal higher sensitivity to New Physics but requires more detailed estimates of the theoretical uncertainties. 

The resulting fits have been produced using the fitting framework of the HEPfit package~\cite{DeBlas:2019ehy}, a general tool to combine information from direct and indirect searches and test the Standard Model. It used the Markov-Chain Monte-Carlo implementation provided by the Bayesian Analysis Toolkit~\cite{Caldwell:2008fw,Caldwell:2010zz,Beaujean:2011zz}, to perform a Bayesian statistical analysis of the sensitivity to deformations from the SM at the different future collider projects. The experimental projections for the different observables included in the fits have been implemented in the likelihood assuming Gaussian distributions, with SM central values and standard deviations given by the corresponding projected uncertainties estimated by the different future collider projects. Finally, theory uncertainties, when included, were introduced via nuisance parameters with Gaussian priors. intrinsic theory uncertainties, arising from missing higher-order corrections, were not included, while parametric theory uncertainties arising from the propagation of experimental errors on SM parameters have been properly taken into account. Experimental uncertainties accounted for statistical uncertainties and the estimated experimental systematic uncertainties,  as well as background theory uncertainties and signal-acceptance related theory uncertainties.

\begin{figure}%[t]plots
    \includegraphics[width=\textwidth]{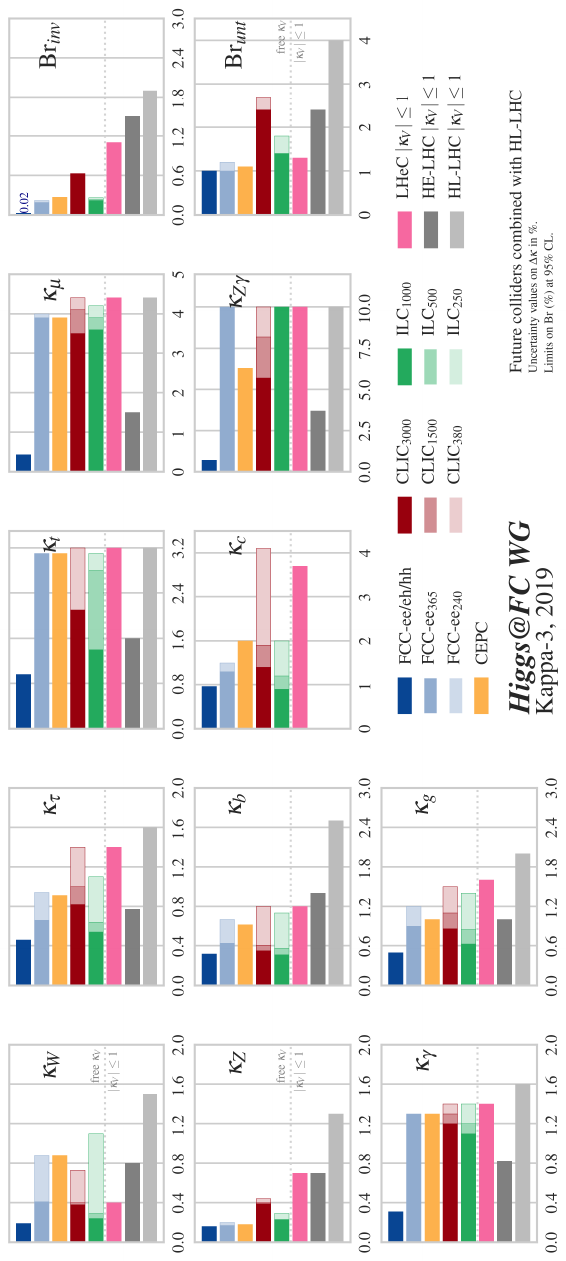}
    \caption{Expected relative precision (\%) of the $\kappa$ parameters in the kappa-3 scenario}
    \label{fig:ESU-kappa3}
\end{figure}

Four benchmark scenarios for the $\kappa$ analyses: the kappa-0 benchmark assumed that there exist no light BSM particles to which the Higgs boson can decay; the  kappa-1,2 benchmarks considered possible new BSM Higgs decays and explored their impact on the determination the Higgs width, i.e. the absolute normalisation of all the Higgs couplings. Finally, in the kappa-3 benchmark, the combination of the HL-LHC data with each of the future accelerators were studied. The results are summarised in Figure~\ref{fig:ESU-kappa3}.

For the EFT analyses, a set of assumptions were made to reduce the set of operators considered to 18 and 30 independent parameters for two specific flavour scenarios: (1) flavour universality, or linear minimal flavour violation, where the only sources of violation of the maximal flavour symmetry group originate from the Yukawa matrices; (2) neutral diagonality where all the new physics flavour interactions remain diagonal in the same basis as the Yukawa matrices. Additionally, it was considered that the vast subset of 4-fermion operators, with the only exception of the one that contributes to the muon decay and thus directly affects the Fermi constant, could be more strongly constrained by other processes and were thus omitted. It was further argued that all the dipole operators should exhibit the same chiral suppression of the Yukawa couplings and could be safely ignored, at least for the light quarks. For the sake of simplicity, the top quark dipole operators were not considered, even though their effects could be relevant.

\begin{figure}%[t]
    \includegraphics[width=\textwidth]{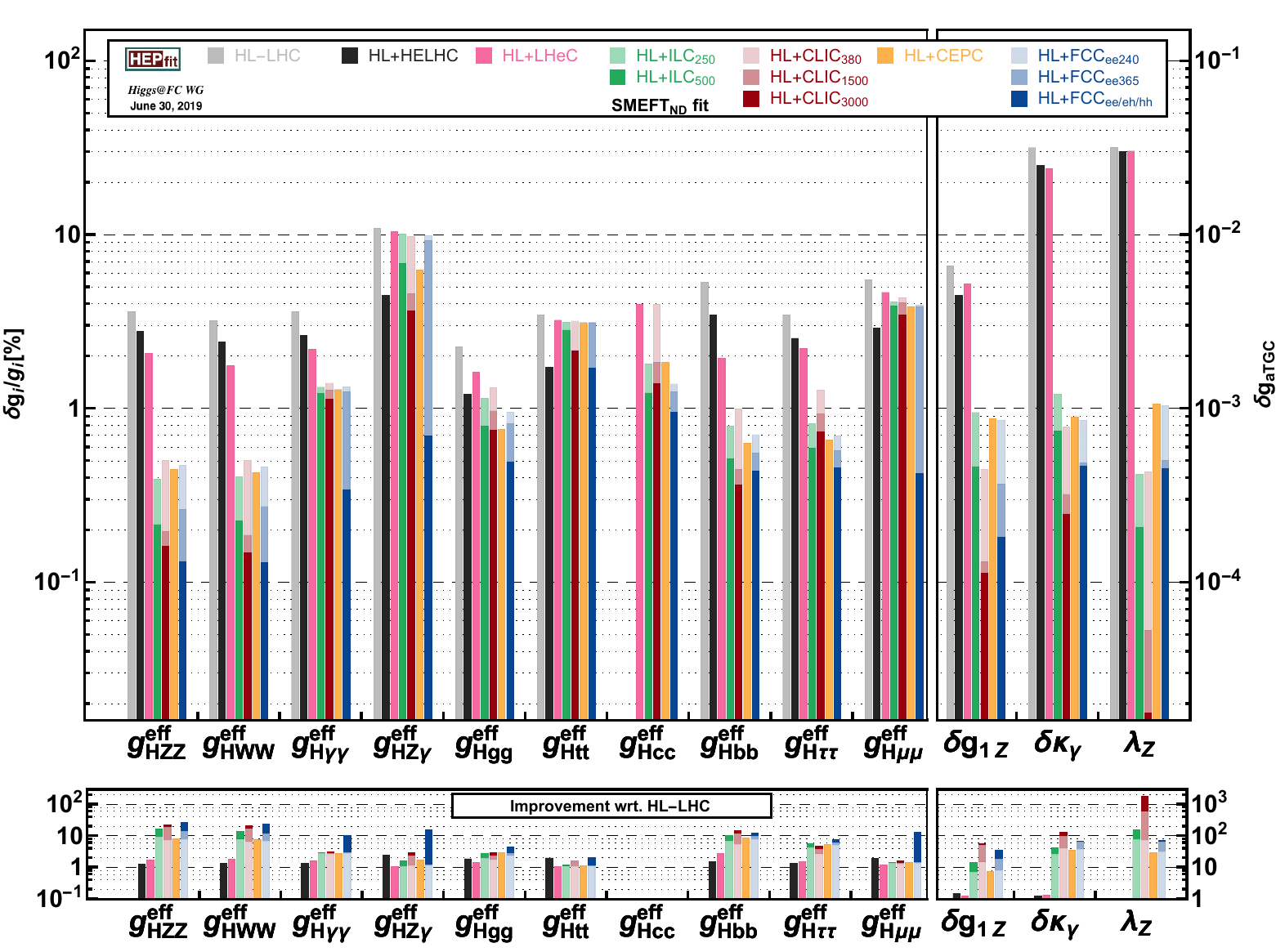}
    \caption{Sensitivity at 68\% probability to deviations in the different effective Higgs couplings and aTGC from a global fit to the projections available at each future collider project.}
    \label{fig:ESU-SMEFT}
\end{figure}

To assess the New Physics deformations with respect to the SM in a operator-basis independent way, the results of the SMEFT fit were projected onto a set of Higgs effective couplings capturing the on-shell properties of the Higgs boson, defined exactly as those presented in Section~\ref{sec:SMEFTeffC}. Detailed results are reported in the report~\cite{deBlas:2019rxi}. The results for the more general neutral diagonality flavour scenario are also shown in Figure~\ref{fig:ESU-SMEFT}  where the results are compared across colliders, emphasising the relative improvement compared to the HL-LHC results.

%

%%%%%%%%%%%%%%%%%%%%%%%%%%%%%%%%%%
\section{Input measurements}\label{sec:input}
%%%%%%%%%%%%%%%%%%%%%%%%%%%%%%%%%%
%[editor: Junping; EF01, 03, 04: HL-LHC/LHeC/ILC/CLIC/CEPC/FCC-ee/FCC-eh/FCC-hh]

The inputs observables and their measurement uncertainties that will be needed in following global fits are summarized in this section. The projections at future colliders
are taken mainly from corresponding collider collaborations (HL-LHC/ILC/CLIC/FCC-ee/CEPC/MuC) as well as the reports compiled by Energy Frontier Topical Groups (EF01/03/04)~\cite{Dawson:2022zbb,Agashe:2022plx,Belloni:2022due}. Collider scenarios considered in this work are summarized in Tab.~\ref{tab:epem_setup}. In a few places where the needed inputs are missing, we carried out our own analysis to give consistent projections for all colliders. The details are explained in the following.

\begin{table}[]
    \centering
    \begin{tabular}{c|c|c|c|c}
    \hline
       Machine & Pol. ($e^-,e^+$) & Energy & Luminosity & Reference\\\hline
        HL-LHC & Unpolarised & 14~TeV & 3 ab$^{-1}$ & \cite{Cepeda:2019klc} \\ \hline
        \multirow{4}{*}{ILC} & \multirow{3}{*}{$(\mp80\%,\,\pm30\%)$}  & 250 GeV & 2 \iab{} & \multirow{4}{*}{\cite{ILCInternationalDevelopmentTeam:2022izu}} \\
         &  & 350 GeV & 0.2 \iab{} & \\
         &  & 500 GeV & 4 \iab{} & \\
         & ($\mp80\%,\,\pm20\%)$ & 1 TeV & 8 \iab{} & \\\hline
        \multirow{3}{*}{CLIC} & \multirow{3}{*}{$(\pm80\%,\,0\%)$} &  380 GeV & 1 \iab{} & \multirow{4}{*}{\cite{Robson:2018zje}}\\
         &  & 1.5 TeV & 2.5 \iab{} & \\
         &  & 3 TeV & 5 \iab{} & \\\hline
         \multirow{5}{*}{FCC-$ee$} & \multirow{5}{*}{Unpolarised} &  Z-pole & 150 \iab{} & \multirow{5}{*}{\cite{Bernardi:2022hny}}\\
         &  & 2$m_W$  & 10 \iab{} \\
         &  & 240 GeV & 5 \iab{} & \\
         &  & 350 GeV & 0.2 \iab{} & \\
         &  & 365 GeV & 1.5 \iab{} & \\\hline
          \multirow{5}{*}{CEPC} & \multirow{5}{*}{Unpolarised} &  Z-pole & 100 \iab{} & \multirow{5}{*}{\cite{Cheng:2022zyy}}\\
         &  & 2$m_W$  & 6 \iab{} \\
         &  & 240 GeV & 20 \iab{} & \\
         &  & 350 GeV & 0.2 \iab{} & \\
         &  & 360 GeV & 1 \iab{} & \\\hline
         \multirow{3}{*}{MuC} & \multirow{3}{*}{Unpolarised} &  125 GeV & 0.02 \iab{} & \multirow{3}{*}{\cite{Forslund:2022xjq,deBlas:2022aow}}\\
         &  & 3 TeV & 3 \iab{} & \\
         &  & 10 TeV & 10 \iab{} & \\\hline
    \end{tabular}
    \caption{Future collider scenarios considered in this work.}
    \label{tab:epem_setup}
\end{table}

\subsection{Electroweak Precision Measurements}

The observables related to precision measurements of $Z$ properties include following: mass ($m_Z$), total width ($\Gamma_Z$), left-right asymmetry ($A_f$) as
defined in Eq.~\ref{eq:gZeff} for $f=b,c,e,\mu,\tau$, partial decay width relative to the total hadronic width ($R_f$), and the total hadronic cross section of 
$e^+e^-\to Z\to hadrons$ ($\sigma_{had}^0$). The projections of their uncertainties now at future $e^+e^-$ are listed in Tab.~\ref{tab:EWPO}, broken down into statistical error and experimental systematic error in most cases. 
The numbers for FCC-ee, CEPC and ILC-GigaZ are based on a dedicated $Z$-pole run, while the ones for ILC250 and CLIC380 are based on radiative return events available at 250 GeV and 380 GeV run respectively. Note that the consistency on
common systematic errors has improved significantly comparing to the numbers 
for ESG; see more details in the EF04 report. The observables $W$
mass ($m_W$), $W$ total width ($\Gamma_W$), Higgs mass ($m_H$) and fine-structure constant ($\alpha(m_Z)^{-1}$) are also listed in 
Tab.~\ref{tab:EWPO}. The other useful observables related to $W$ branching ratios are implicitly included in the optimal observables that will be explained in Sec.~\ref{sec:diboson}. Additional electroweak observables above or below $Z$-pole that are relevant to 4-fermion interactions will be explained in Sec.~\ref{sec:4f}.

\begin{table}[tb]
  \begin{center}
  \begin{adjustbox}{max width=\textwidth}
  \begin{tabular}{|c|c|c|c|c|c|c|}
    \hline
    Quantity       & current  &    ILC250      & ILC-GigaZ   &      FCC-ee            &  CEPC      &  CLIC380     \\
    \hline
    $\Delta\alpha(\mz)^{-1}\;(\times 10^3)$ &    17.8$^*$  & 17.8$^*$  &             &    3.8 (1.2)  & 17.8$^*$ &           \\
    $\Delta\mw$ (MeV)     & 12$^*$     &    0.5 (2.4)  &             &    0.25 (0.3)   & 0.35 (0.3)   &           \\
    $\Delta\mz$ (MeV)    & 2.1$^*$      &    0.7 (0.2)  &  0.2           &    0.004 (0.1)   & 0.005 (0.1)   &    2.1$^*$       \\
    $\Delta\mh$ (MeV)   & 170$^*$       &    14  &             &    2.5 (2)   & 5.9   &    78       \\
    $\Delta\Gamma_W$ (MeV)     & 42$^*$ &    2   &             &    1.2 (0.3)   & 1.8 (0.9)   &           \\    
    $\Delta\gz$ (MeV)     & 2.3$^*$ &    1.5 (0.2)  &  0.12    &    0.004 (0.025)   & 0.005 (0.025)   &    2.3$^*$       \\ 
    %\hdashline
    \cdashline{1-7}
    $\Delta A_e\;(\times 10^5)$   & 190$^*$ &    14 (4.5)  &   1.5 (8)  &    0.7 (2)   & 1.5    &    64       \\ 
    $\Delta A_\mu\;(\times10^5)$  & 1500$^*$ &    82 (4.5)  &   3 (8)  &    2.3 (2.2)   & 3.0 (1.8)   &    400       \\    
    $\Delta A_\tau\;(\times10^5)$ & 400$^*$ &   86 (4.5)  &   3 (8)  &    0.5 ({20})   & 1.2 ({6.9})   &    570       \\        
    $\Delta A_b\;(\times10^5)$   & 2000$^*$ &   53 (35)  &   9 (50)  &    2.4 (21)   & 3 (21)  &    380       \\
    $\Delta A_c\;(\times10^5)$  & 2700$^*$  &  140 (25)  &  20 (37)  &    20 (15)   & 6 ({30})   &    200       \\ 
    %\hdashline
    \cdashline{1-7}
    $\Delta \sigma_{\rm had}^0$ (pb)  & 37$^*$ &           &          &    0.035 (4)      & 0.05 (2)     &    37$^{*}$       \\     
    $\delta R_e\;(\times10^3)$    & 2.4$^*$ &    0.5 (1.0)    &   0.2 (0.5)  &    0.004 (0.3)   & 0.003 (0.2)   &    2.7       \\
    $\delta R_\mu\;(\times10^3)$  &  1.6$^*$  & 0.5 (1.0)    &   0.2 (0.2)  &    0.003 (0.05)  & 0.003 (0.1)   &    2.7       \\
    $\delta R_\tau\;(\times10^3)$ &  2.2$^*$ &  0.6 (1.0)    &   0.2 (0.4)  &    0.003 (0.1)   & 0.003 (0.1)   &    6       \\
    $\delta R_b\;(\times10^3)$    & 3.0$^*$ &   0.4 (1.0)    &   0.04 (0.7)  &   0.0014 ($<0.3$)   & 0.005 (0.2)   &    1.8       \\
    $\delta R_c(\times10^3)$    &  17$^*$ &  0.6 (5.0)    &   0.2 (3.0)  &    0.015 (1.5)   & 0.02 (1)   &    5.6       \\
    \hline    
  \end{tabular}
  \end{adjustbox}
  \end{center}
\vspace{-2ex}
\caption{EWPOs at future $e^+e^-$: statistical error (experimental systematic error). $\Delta$ ($\delta$) stands for absolute (relative) uncertainty, while * indicates inputs taken from current data \cite{pdg2020}. See Refs.~\cite{deBlas:2019rxi,DeBlas:2019qco,LCCPhysicsWorkingGroup:2019fvj,Blondel:2021ema,ILCInternationalDevelopmentTeam:2022izu,Cheng:2022zyy}.}
\label{tab:EWPO}
\end{table}

\subsection{Higgs Measurements}
The observables related to measurement of Higgs properties include mostly
the production cross section times decay branching ratio ($\sigma\times BR_X$) 
for various production and decay channels. One exception is the inclusive production cross section measurement for $e^+e^-\to ZH$, which is enabled by recoil mass technique at lepton colliders. 
Observables for differential cross sections are not included.
The projections of Higgs measurements are listed in Tab.~\ref{tab:LHC_HiggsSxBR} for HL-LHC, Tab.~\ref{tab:ILC_Higgs}-\ref{tab:CLIC3000_Higgs} for future $e^+e^-$, and Tab.~\ref{tab:MuC3000_Higgs}-\ref{tab:MuC125_Higgs} for muon colliders.
Note that in some cases two sets of numbers are provided for each observable.
Numbers without parentheses are directly provided by collider collaborations, which the final fit results will be based on. However there are often subtle inconsistency between different projections for similar observables due to different level of realism that was adopted in the relevant simulation analyses. Apparently those types of difference should not bring bias to the true capabilities of different future colliders. In order to allow us to isolate out those differences, we also provide uncertainties in parentheses that are 
extrapolated from a same set of analyses, which are mostly from 
ILC full detector simulation studies\footnote{This doesn't mean the extrapolated uncertainties are more accurate, but provides a way of comparing the capabilities of future colliders in a more equal footing.}. This can provide us a clue to understand the difference in final results.  In fact one can see from the tables that two sets of inputs are rather consistent in most cases. 
It often happens that the list of input observables directly provided by collaborations is not complete. 
Whenever that happens we try to fill out the missing inputs by extrapolations.
This further helps isolate out certain baises in the comparison. 
One such example which plays a quantitatively important role is the branching ratio of $H\to\gamma Z$. 

\begin{table}[h!]
  \centering
  \begin{tabular}{|c|c|c|c|c|c|}
    \hline
    HL-LHC    &   \multicolumn{5}{|c|}{3 ab$^{-1} ~ $ATLAS$+$CMS} \\
    \hline
    Prod.     &    $ggH$     &   VBF   &    $WH$   &    $ZH$   & $ttH$ \\
    $\sigma$  & - & -  & - & - & - \\
    $\sigma\times BR_{bb}$  &   19.1  &  -  & 8.3  & 4.6  & 10.7 \\
    $\sigma\times BR_{cc}$  &   -     &  -  & -    & -    & - \\   
    $\sigma\times BR_{gg}$  &   -     &  -  & -    & -    & - \\     
    $\sigma\times BR_{ZZ}$  &   2.5   &  9.5  & 32.1  & 58.3  & 15.2 \\
    $\sigma\times BR_{WW}$  &   2.5   &  5.5  & 9.9   & 12.8  & 6.6 \\    
    $\sigma\times BR_{\tau\tau}$  &   4.5   &  3.9  & -  & -  & 10.2 \\    
    $\sigma\times BR_{\gamma\gamma}$  &   2.5   &  7.9  & 9.9  & 13.2  & 5.9 \\    
    $\sigma\times BR_{\gamma Z}$  &   24.4   &  51.2  & -  & -  & - \\    
    $\sigma\times BR_{\mu\mu}$  &   11.1   &  30.7  & -  & -  & - \\    
    $\sigma\times BR_{inv.}$  &   -   &  2.5  & -  & -  & - \\ 
    $\Delta m_H$  & 10-20 MeV & - & - & - & - \\
    \hline    
  \end{tabular}
\caption{Projected uncertainties of Higgs observables at HL-LHC: numbers by default in \%.
}
\label{tab:LHC_HiggsSxBR}
\end{table}

\begin{table}[h!]
  \centering
  \begin{tabular}{|c|c|c|c|c|}
    \hline
    ILC250    &   \multicolumn{2}{c|}{0.9ab$^{-1}$~(-0.8,+0.3)} & \multicolumn{2}{c|}{0.9ab$^{-1}$~(+0.8,-0.3)}\\
    \hline
    Prod.     &    $ZH$     &   $\nu\nu H$   &    $ZH$   &    $\nu\nu H$ \\
    $\sigma$  & 1.07 & -  & 1.07 & - \\
    $\sigma\times BR_{bb}$  &   0.714  &  4.27  & 0.714  & 17.4 \\
    $\sigma\times BR_{cc}$  &   4.38     &  -  & 4.38    & -    \\   
    $\sigma\times BR_{gg}$  &   3.69     &  -  & 3.69    & -    \\     
    $\sigma\times BR_{ZZ}$  &   9.49   &  -  & 9.49  & -  \\
    $\sigma\times BR_{WW}$  &   2.43   &  -  & 2.43   & -  \\    
    $\sigma\times BR_{\tau\tau}$  &   1.7   &  -  & 1.7  & - \\    
    $\sigma\times BR_{\gamma\gamma}$  &   17.9   &  -  & 17.9  & - \\    
    $\sigma\times BR_{\gamma Z}$  &   63   &  -  & 59  & - \\    
    $\sigma\times BR_{\mu\mu}$  &   37.9   &  -  & 37.9  & - \\    
    $\sigma\times BR_{inv.}$  &  0.336   &  -  & 0.277  & - \\ 
%    $\Delta m_H$  & 10-20 MeV & - & - & - & - \\    
 \hline    
  \end{tabular}
\caption{Projected uncertainties of Higgs observables at ILC250: numbers by default in \%.}
\label{tab:ILC_Higgs}
\end{table}

\begin{table}[h!]
  \centering
  \begin{tabular}{|c|c|c|c|c|}
    \hline
        &   \multicolumn{2}{c|}{FCCee240  5ab$^{-1}$} & \multicolumn{2}{c|}{CEPC240  20ab$^{-1}$}\\
    \hline
    Prod.     &    $ZH$     &   $\nu\nu H$   &    $ZH$   &    $\nu\nu H$ \\
    $\sigma$  & 0.5(0.537) & -  & 0.26 & - \\
    $\sigma\times BR_{bb}$  &   0.3(0.380)  &  3.1(2.78)  & 0.14  & 1.59 \\
    $\sigma\times BR_{cc}$  &   2.2(2.08)     &  -  & 2.02    & -    \\   
    $\sigma\times BR_{gg}$  &   1.9(1.75)     &  -  & 0.81    & -    \\     
    $\sigma\times BR_{ZZ}$  &   4.4(4.49)   &  -  & 4.17  & -  \\
    $\sigma\times BR_{WW}$  &   1.2(1.16)   &  -  & 0.53   & -  \\    
    $\sigma\times BR_{\tau\tau}$  &   0.9(0.822)   &  -  & 0.42  & - \\    
    $\sigma\times BR_{\gamma\gamma}$  &   9(8.47)   &  -  & 3.02  & - \\    
    $\sigma\times BR_{\gamma Z}$  &   (17$^*$) & -  & 8.5  & - \\    
    $\sigma\times BR_{\mu\mu}$  &   19(17.9)   &  -  & 6.36  & - \\    
    $\sigma\times BR_{inv.}$  &  0.3(0.226)   &  -  & 0.07  & - \\ 
%    $\Delta m_H$  & 10-20 MeV & - & - & - & - \\    
 \hline    
  \end{tabular}
\caption{Projected uncertainties of Higgs observables at FCCee240 and CEPC240: numbers by default in \%.}
\label{tab:FCC_Higgs}
\end{table}

\begin{table}[h!]
  \centering
  \begin{tabular}{|c|c|c|c|c|}
    \hline
        CLIC380 &   \multicolumn{2}{c|}{0.5 ab$^{-1}$~(-0.8,0)} & \multicolumn{2}{c|}{0.5 ab$^{-1}$~(+0.8,0)}\\
    \hline
    Prod.     &    $ZH$     &   $\nu\nu H$   &    $ZH$   &    $\nu\nu H$ \\
    $\sigma$  & 1.5(1.43) & -  & 1.8(1.43) & - \\
    $\sigma\times BR_{bb}$  &   0.81(1.2)  &  1.4(1.47)  & 0.92(1.2)  & 4.1(4.4) \\
    $\sigma\times BR_{cc}$  &   13(8.7)     &  19(15.3)  &  15(8.7)   & 24(46)  \\ 
    $\sigma\times BR_{gg}$  &   5.7(6.6)     &  3.3(6.2)  & 6.5(6.6)  & 20(18.8)\\  
    $\sigma\times BR_{ZZ}$  &   (19.7)   &  (16.1)  & (19.7)  & (46)  \\
    $\sigma\times BR_{WW}$  &   5.1(4.4)   &  (4.6)  & (4.4)   & (14)  \\    
    $\sigma\times BR_{\tau\tau}$  &   5.9(3.2)   &  (12.9)  & 6.6(3.2)  & (39) \\   
    $\sigma\times BR_{\gamma\gamma}$  &   (31)   &  (36)  & (31)  & (108) \\    
    %$\sigma\times BR_{\gamma Z}$  &   () & -  & ()  & - \\    
    $\sigma\times BR_{\mu\mu}$  &   (69)   &  (129)  & (69)  & (129) \\    
    $\sigma\times BR_{inv.}$  &  0.57(0.68)   &  -  & 0.64(0.64)  & - \\ 
%    $\Delta m_H$  & 10-20 MeV & - & - & - & - \\    
 \hline    
  \end{tabular}
\caption{Projected uncertainties of Higgs observables at CLIC380: numbers by default in \%; numbers in parentheses are extrapolated from ILC350.}
\label{tab:CLIC380_Higgs}
\end{table}

\begin{table}[h!]
  \centering
  \begin{tabular}{|c|c|c|c|c|}
    \hline
        ILC350 &   \multicolumn{2}{c|}{0.135 ab$^{-1}$~(-0.8,+0.3)} & \multicolumn{2}{c|}{0.045 ab$^{-1}$~(+0.8,-0.3)}\\
    \hline
    Prod.     &    $ZH$     &   $\nu\nu H$   &    $ZH$   &    $\nu\nu H$ \\
    $\sigma$  & 2.46 & -  & 4.3 & - \\
    $\sigma\times BR_{bb}$  &   2.05  &  2.46  & 3.5  & 17.7 \\
    $\sigma\times BR_{cc}$  &   15     &  25.9  &  25.9   & 186  \\ 
    $\sigma\times BR_{gg}$  &   11.4     &  10.5  & 19.8  & 75 \\  
    $\sigma\times BR_{ZZ}$  &   34   &  27.2  & 59  & 191  \\
    $\sigma\times BR_{WW}$  &   7.6   &  7.8  & 13.2   & 57  \\    
    $\sigma\times BR_{\tau\tau}$  &   5.5   &  21.8  & 9.4  & 156 \\   
    $\sigma\times BR_{\gamma\gamma}$  &  53   &  61  & 92  & 424 \\    
    %$\sigma\times BR_{\gamma Z}$  &   () & -  & ()  & - \\    
    $\sigma\times BR_{\mu\mu}$  &   118   &  218  & 205  & 1580 \\    
    $\sigma\times BR_{inv.}$  &  1.15   &  -  & 1.83  & - \\ 
%    $\Delta m_H$  & 10-20 MeV & - & - & - & - \\    
 \hline    
  \end{tabular}
\caption{Projected uncertainties of Higgs observables at ILC350: numbers by default in \%.}
\label{tab:ILC350_Higgs}
\end{table}

\begin{table}[h!]
  \centering
  \begin{tabular}{|c|c|c|c|c|}
    \hline
         &   \multicolumn{2}{c|}{1.5 ab$^{-1}$~FCC-ee365} & \multicolumn{2}{c|}{1.0 ab$^{-1}$~CEPC360}\\
    \hline
    Prod.     &    $ZH$     &   $\nu\nu H$   &    $ZH$   &    $\nu\nu H$ \\
    $\sigma$  & 0.9(0.84) & -  & 1.4(1.02) & - \\
    $\sigma\times BR_{bb}$  &   0.5(0.71)  &  0.9(1.14)  & 0.90(0.86)  & 1.1(1.39) \\
    $\sigma\times BR_{cc}$  &   6.5(5.0)     &  10(11.9)  &  8.8(6.1)   & 16(14.5)  \\ 
    $\sigma\times BR_{gg}$  &   3.5(3.8)    &  4.5(4.8)  & 3.4(4.7)  & 4.5(5.9)\\  
    $\sigma\times BR_{ZZ}$  &   12(11.4)   &  10(12.5)  & 20(13.9)  & 21(15.3)  \\
    $\sigma\times BR_{WW}$  &   2.6(2.55)   &  (3.6)  & 2.8(3.12)   & 4.4(4.4)  \\    
    $\sigma\times BR_{\tau\tau}$  &  1.8(1.83) &  8(10)  & 2.1(2.24)  & 4.2(12.2) \\   
    $\sigma\times BR_{\gamma\gamma}$  & 18(17.7) & 22(28.1) & 11(21.7) & 16(34.4) \\   
    %$\sigma\times BR_{\gamma Z}$  &   () & -  & ()  & - \\    
    $\sigma\times BR_{\mu\mu}$  &  40(40)   &  (100)  & 41(48)  & 57(123) \\    
    $\sigma\times BR_{inv.}$  &  0.60(0.42)   &  -  & (0.49)  & - \\ 
%    $\Delta m_H$  & 10-20 MeV & - & - & - & - \\    
 \hline    
  \end{tabular}
\caption{Projected uncertainties of Higgs observables at FCC-ee365 and CEPC360: numbers by default in \%; numbers in parentheses are extrapolated from ILC350.}
\label{tab:CEPC360_Higgs}
\end{table}

\begin{table}[h!]
  \centering
  \begin{tabular}{|c|c|c|c|c|}
    \hline
        ILC500 &   \multicolumn{2}{c|}{1.6 ab$^{-1}$~(-0.8,+0.3)} & \multicolumn{2}{c|}{1.6 ab$^{-1}$~(+0.8,-0.3)}\\
    \hline
    Prod.     &    $ZH$     &   $\nu\nu H$   &    $ZH$   &    $\nu\nu H$ \\
    $\sigma$  & 1.67 & -  & 1.67 & - \\
    $\sigma\times BR_{bb}$  &   1.01  &  0.42  & 1.01  & 1.52 \\
    $\sigma\times BR_{cc}$  &   7.1   &  3.48  & 7.1   & 14.2  \\ 
    $\sigma\times BR_{gg}$  &   5.9   &  2.3   & 5.9   & 9.5 \\  
    $\sigma\times BR_{ZZ}$  &   13.8  &  4.8   & 13.8  & 19  \\
    $\sigma\times BR_{WW}$  &   3.1   &  1.36  & 3.1   & 5.5  \\    
    $\sigma\times BR_{\tau\tau}$      &  2.42   &  3.9   & 2.42  & 15.8 \\   
    $\sigma\times BR_{\gamma\gamma}$  &  18.6   &  10.7  & 18.6  & 44 \\    
    %$\sigma\times BR_{\gamma Z}$  &   () & -  & ()  & - \\    
    $\sigma\times BR_{\mu\mu}$ & 47     &  40  & 47  & 166 \\    
    $\sigma\times BR_{inv.}$   & 0.83  &  -   & 0.60  & - \\ 
%    $\Delta m_H$  & 10-20 MeV & - & - & - & - \\    
 \hline    
  \end{tabular}
\caption{Projected uncertainties of Higgs observables at ILC500: numbers by default in \%.}
\label{tab:ILC500_Higgs}
\end{table}

\begin{table}[h!]
  \centering
  \begin{tabular}{|c|c|c|}
    \hline
        ILC1000 &   3.2 ab$^{-1}$~(-0.8,+0.2) & 3.2 ab$^{-1}$~(+0.8,-0.2)\\
    \hline
    Prod.     &      $\nu\nu H$     &    $\nu\nu H$ \\
    %$\sigma$  & -  & - \\
    $\sigma\times BR_{bb}$  &   0.32  & 1.0 \\
    $\sigma\times BR_{cc}$  &   1.7   & 6.4  \\ 
    $\sigma\times BR_{gg}$  &   1.3   & 4.7 \\  
    $\sigma\times BR_{ZZ}$  &   2.3   & 8.4  \\
    $\sigma\times BR_{WW}$  &   0.91  & 3.3  \\    
    $\sigma\times BR_{\tau\tau}$      &  1.7  & 6.4 \\   
    $\sigma\times BR_{\gamma\gamma}$  &  4.8  & 17 \\    
    %$\sigma\times BR_{\gamma Z}$  &   () & -  & ()  & - \\    
    $\sigma\times BR_{\mu\mu}$ & 17  & 64 \\    
    %$\sigma\times BR_{inv.}$   & 0.83  &  -   & 0.60  & - \\ 
%    $\Delta m_H$  & 10-20 MeV & - & - & - & - \\    
 \hline    
  \end{tabular}
\caption{Projected uncertainties of Higgs observables at ILC1000: numbers by default in \%.}
\label{tab:ILC1000_Higgs}
\end{table}

\begin{table}[h!]
  \centering
  \begin{tabular}{|c|c|c|}
    \hline
        CLIC1500 &   2 ab$^{-1}$~(-0.8,0) & 0.5 ab$^{-1}$~(+0.8,0)\\
    \hline
    Prod.     &      $\nu\nu H$     &    $\nu\nu H$ \\
    %$\sigma$  & -  & - \\
    $\sigma\times BR_{bb}$  &   0.25  & 1.5 \\
    $\sigma\times BR_{cc}$  &   3.9   & 24  \\ 
    $\sigma\times BR_{gg}$  &   3.3   & 20 \\  
    $\sigma\times BR_{ZZ}$  &   3.6   & 22  \\
    $\sigma\times BR_{WW}$  &   0.67  & 4.0  \\    
    $\sigma\times BR_{\tau\tau}$      &  2.8  & 17 \\   
    $\sigma\times BR_{\gamma\gamma}$  &  10  & 60 \\    
    $\sigma\times BR_{\gamma Z}$  &  28   & 170 \\    
    $\sigma\times BR_{\mu\mu}$ & 24  & 150 \\    
    %$\sigma\times BR_{inv.}$   & 0.83  &  -   & 0.60  & - \\ 
%    $\Delta m_H$  & 10-20 MeV & - & - & - & - \\    
 \hline    
  \end{tabular}
\caption{Projected uncertainties of Higgs observables at CLIC1500: numbers by default in \%.}
\label{tab:CLIC1500_Higgs}
\end{table}

\begin{table}[h!]
  \centering
  \begin{tabular}{|c|c|c|}
    \hline
        CLIC3000 &   4 ab$^{-1}$~(-0.8,0) & 1 ab$^{-1}$~(+0.8,0)\\
    \hline
    Prod.     &      $\nu\nu H$     &    $\nu\nu H$ \\
    %$\sigma$  & -  & - \\
    $\sigma\times BR_{bb}$  &   0.17  & 1.0 \\
    $\sigma\times BR_{cc}$  &   3.7   & 22  \\ 
    $\sigma\times BR_{gg}$  &   2.3   & 14 \\  
    $\sigma\times BR_{ZZ}$  &   2.1   & 13  \\
    $\sigma\times BR_{WW}$  &   0.33  & 2.0  \\    
    $\sigma\times BR_{\tau\tau}$      &  2.3  & 14 \\   
    $\sigma\times BR_{\gamma\gamma}$  &  5.0  & 30 \\    
    $\sigma\times BR_{\gamma Z}$  &  16   & 95 \\    
    $\sigma\times BR_{\mu\mu}$ & 13  & 80 \\    
    %$\sigma\times BR_{inv.}$   & 0.83  &  -   & 0.60  & - \\ 
%    $\Delta m_H$  & 10-20 MeV & - & - & - & - \\    
 \hline    
  \end{tabular}
\caption{Projected uncertainties of Higgs observables at CLIC3000: numbers by default in \%.}
\label{tab:CLIC3000_Higgs}
\end{table}

\begin{table}[h!]
  \centering
  \begin{tabular}{|c|c|c|}
    \hline
        MuC3000 &   \multicolumn{2}{|c|}{3 ab$^{-1}$}\\
    \hline
    Prod.     &      $\nu\nu H$     &    $\mu\mu H$ \\
    %$\sigma$  & -  & - \\
    $\sigma\times BR_{bb}$  &   0.8  & 2.6 \\
    $\sigma\times BR_{cc}$  &   12   & 72  \\ 
    $\sigma\times BR_{gg}$  &   2.8   & 14 \\  
    $\sigma\times BR_{ZZ}$  &   11   & 34  \\
    $\sigma\times BR_{WW}$  &   1.5  & 7.5  \\    
    $\sigma\times BR_{\tau\tau}$      &  3.8  & 21 \\   
    $\sigma\times BR_{\gamma\gamma}$  &  6.4  & 23 \\    
    $\sigma\times BR_{\gamma Z}$  &  45   & - \\    
    $\sigma\times BR_{\mu\mu}$ & 28  & - \\    
 \hline    
  \end{tabular}
\caption{Projected uncertainties of Higgs observables at 3 TeV muon collider: numbers by default in \%.
}
\label{tab:MuC3000_Higgs}
\end{table}

\begin{table}[h!]
  \centering
  \begin{tabular}{|c|c|c|}
    \hline
        MuC10000 &   \multicolumn{2}{|c|}{10 ab$^{-1}$}\\
    \hline
    Prod.     &      $\nu\nu H$     &    $\mu\mu H$ \\
    %$\sigma$  & -  & - \\
    $\sigma\times BR_{bb}$  &   0.22  & 0.77 \\
    $\sigma\times BR_{cc}$  &   3.6   & 17  \\ 
    $\sigma\times BR_{gg}$  &   0.79   & 3.3 \\  
    $\sigma\times BR_{ZZ}$  &   3.2   & 11  \\
    $\sigma\times BR_{WW}$  &   0.40  & 1.8  \\    
    $\sigma\times BR_{\tau\tau}$      &  1.1  & 4.8 \\   
    $\sigma\times BR_{\gamma\gamma}$  &  1.7  & 4.8 \\    
    $\sigma\times BR_{\gamma Z}$  &  12   & - \\    
    $\sigma\times BR_{\mu\mu}$ & 5.7  & - \\    
 \hline    
  \end{tabular}
\caption{Projected uncertainties of Higgs observables at 10 TeV muon collider: numbers by default in \%.
}
\label{tab:MuC10000_Higgs}
\end{table}

\begin{table}[h!]
  \centering
  \begin{tabular}{|c|c|}
    \hline
        MuC125 &   20 fb$^{-1}$\\
    \hline
    Prod.     &      $\mu\mu\to H$     \\
    %$\sigma$  & -  & - \\
    $\sigma\times BR_{bb}$  &   0.49  \\
    $\sigma\times BR_{cc}$  &   12  \\ 
    $\sigma\times BR_{gg}$  &   5.3   \\  
    $\sigma\times BR_{ZZ}$  &   2.9   \\
    $\sigma\times BR_{WW}$  &   0.67  \\    
    $\sigma\times BR_{\tau\tau}$      &  2.4  \\   
    $\sigma\times BR_{\gamma\gamma}$  &  94  \\    
    $\sigma\times BR_{\gamma Z}$  &  -   \\    
    $\sigma\times BR_{\mu\mu}$ & -  \\    
 \hline    
  \end{tabular}
\caption{Projected uncertainties of Higgs observables at 125 GeV muon collider: numbers by default in \%.}
\label{tab:MuC125_Higgs}
\end{table}

\subsection{Light fermion pair measurements}
The input observables for $e^-e^+\to f\bar{f}\,(f=e,\mu,\tau,c,b)$ at $\sqrt{s}>>m_Z$ are summarized in Tables \ref{tab:light_fermion_pair_ILCee}-\ref{tab:light_fermion_pair_CEPC}, these are later used in \autoref{sec:4f} for the 4-fermion fit. The uncertainties for the total cross sections $\sigma_f$ and the forward-backward asymmetries $A_{FB}^f$ are obtained from a common
analysis using optimal observable method for all
futue $e^+e^-$. The efficiencies used in the analysis
were taken from ILD full simulation studies~\cite{ILCInternationalDevelopmentTeam:2022izu} all for double-tagged events.
However due to insufficient input about systematic errors for all channels and the fact that the technical
implementation of systematic errors in the optimal
observable method is unclear at this moment,
the uncertainties given in above tables are only statistical. For the ILC uncertainties, the systematics
would play a minor role. But for CEPC or FCC-ee, the
uncertainties are likely significantly underestimated
in particular in the lepton channels. It's worth pointing out one source of the contradictory inputs is about systematic error for luminosity measurement at similar $\sqrt{s}$. It is assumed to be 0.01\% at FCCee240 while 0.1\% at ILC250, one order of magnitude difference. This contradictory systematic errors couldn't get resolved in time. After all we left out
systematic errors for 2-fermion observables off the $Z$-pole.
%Systematical errors are more subtle to be implemented and are thus left out in the current fit. The total selection efficiency is obtained based on full ILD simulations, using the double-tagged events only.\footnote{We thank Adrian Irles for providing us these numbers.}

%%%%%%%%%%%%%%%%% ILC
\begin{table}[]
\centering
\begin{adjustbox}{max width=1\textwidth, max height=0.49\textheight}
\begin{tabular}{c|c|c|c|c|c|c}
\hline
{\textbf{ILC}} $\sqrt{s}$ [GeV] & Pol. ($e^-,e^+$) & $\mathcal{L}\,\rm[fb^{-1}]$ & $\sigma_e$ [fb] & $A_{FB}^e$ & $[c_\theta^{\rm min},c_\theta^{\rm max}]$ & $\epsilon$\\\hline\hline
\multirow{4}{*}{250} & $(-80\%,\,-30\%)$    & 100&  64510.2$\pm$ 25.29& 0.956$\pm$ 0.0001156& [-0.9, 0.9]& 0.98\\
                                & $(-80\%,\,+30\%)$    & 900& 68282.6$\pm$ 8.69& 0.962$\pm$ 0.0000348& [-0.9, 0.9]& 0.98\\
                                & $(+80\%,\,-30\%)$    & 900& 66455.6$\pm$ 8.28& 0.999$\pm$ 0.00000595 & [-0.9, 0.9]& 0.98\\
                                & $(+80\%,\,+30\%)$     & 100& 86359.7$\pm$ 28.93& 0.933$\pm$ 0.0001202& [-0.9, 0.9]& 0.98\\
\hline
\multirow{4}{*}{500} & $(-80\%,\,-30\%)$    & 400& 15566.2$\pm$ 6.22& 0.956$\pm$ 0.0001176& [-0.9, 0.9]& 0.98\\
                                & $(-80\%,\,+30\%)$    & 1600& 19081.8$\pm$ 3.45& 0.965$\pm$ 0.0000474& [-0.9, 0.9]& 0.98\\
                                & $(+80\%,\,-30\%)$    & 1600& 16326.7$\pm$ 3.1& 0.982$\pm$ 0.0000362& [-0.9, 0.9]& 0.98\\
                                & $(+80\%,\,+30\%)$     & 400& 23477.5$\pm$ 7.57& 0.929$\pm$ 0.0001196& [-0.9, 0.9]& 0.98\\
\hline
\multirow{4}{*}{1000} & $(-80\%,\,-20\%)$    & 800& 4084.86$\pm$ 2.253& 0.958$\pm$ 0.0001582& [-0.9, 0.9]& 0.98\\
                                & $(-80\%,\,+20\%)$    & 3200& 4922.72$\pm$ 1.238& 0.966$\pm$ 0.0000654& [-0.9, 0.9]& 0.98\\
                                & $(+80\%,\,-20\%)$    & 3200& 4429.19$\pm$ 1.15& 0.963$\pm$ 0.0000701& [-0.9, 0.9]& 0.98\\
                                & $(+80\%,\,+20\%)$     & 800& 5828.42$\pm$ 2.67& 0.934$\pm$ 0.000164& [-0.9, 0.9]& 0.98\\
\hline
    \end{tabular}
    \end{adjustbox}
    \caption{Projections for $e^-e^+\to e^-e^+$ at ILC, where the last column is the total selection efficiency.}
    \label{tab:light_fermion_pair_ILCee}
\end{table}

\begin{table}[]
\centering
\begin{adjustbox}{max width=1\textwidth, max height=0.49\textheight}
\begin{tabular}{c|c|c|c|c|c|c}
\hline
{\textbf{ILC}} $\sqrt{s}$ [GeV] & Pol. ($e^-,e^+$) & $\mathcal{L}\,\rm[fb^{-1}]$ & $\sigma_\mu$ [fb] & $A_{FB}^\mu$ & $[c_\theta^{\rm min},c_\theta^{\rm max}]$ & $\epsilon$\\\hline\hline
\multirow{4}{*}{250} & $(-80\%,\,-30\%)$    & 100&  1396.06$\pm$3.74& 0.53$\pm$0.00227& [-0.95, 0.95]& 0.98\\
                                & $(-80\%,\,+30\%)$    & 900& 2329.5$\pm$1.61& 0.535$\pm$0.000584& [-0.95, 0.95]& 0.98\\
                                & $(+80\%,\,-30\%)$    & 900& 1929.12$\pm$1.464& 0.494$\pm$0.00066& [-0.95, 0.95]& 0.98\\
                                & $(+80\%,\,+30\%)$    & 100& 1214.07$\pm$3.484& 0.5$\pm$0.002485& [-0.95, 0.95]& 0.98\\
\hline
\multirow{4}{*}{500} & $(-80\%,\,-30\%)$    & 400& 336.28$\pm$0.917& 0.492$\pm$0.002374& [-0.95, 0.95]& 0.98\\
                                & $(-80\%,\,+30\%)$    & 1600& 559.91$\pm$0.592& 0.597$\pm$0.000917& [-0.95, 0.95]& 0.98\\
                                & $(+80\%,\,-30\%)$    & 1600& 472.88$\pm$0.544& 0.4535$\pm$0.001025& [-0.95, 0.95]& 0.98\\
                                & $(+80\%,\,+30\%)$    & 400& 296.72$\pm$0.861& 0.46$\pm$0.00258& [-0.95, 0.95]& 0.98\\
\hline
\multirow{4}{*}{1000} & $(-80\%,\,-20\%)$    & 800& 92.65$\pm$0.34& 0.484$\pm$0.003214& [-0.95, 0.95]& 0.98\\
                                & $(-80\%,\,+20\%)$     & 3200& 129.58$\pm$0.2012& 0.487$\pm$0.001356& [-0.95, 0.95]& 0.98\\
                                & $(+80\%,\,-20\%)$     & 3200& 110.43$\pm$0.1858& 0.445$\pm$0.001507& [-0.95, 0.95]& 0.98\\
                                & $(+80\%,\,+20\%)$    & 800& 81.16$\pm$0.3185& 0.449$\pm$0.00351& [-0.95, 0.95]& 0.98\\
\hline
    \end{tabular}
    \end{adjustbox}
    \caption{Projections for $e^-e^+\to \mu^-\mu^+$ at ILC, where the last column is the total selection efficiency.}
    \label{tab:light_fermion_pair_ILCmumu}
\end{table}

\begin{table}[]
\centering
\begin{adjustbox}{max width=1\textwidth, max height=0.49\textheight}
\begin{tabular}{c|c|c|c|c|c|c}
\hline
{\textbf{ILC}} $\sqrt{s}$ [GeV] & Pol. ($e^-,e^+$) & $\mathcal{L}\,\rm[fb^{-1}]$ & $\sigma_\tau$ [fb] & $A_{FB}^\tau$ & $[c_\theta^{\rm min},c_\theta^{\rm max}]$ & $\epsilon$\\\hline\hline
\multirow{4}{*}{250} & $(-80\%,\,-30\%)$    & 100& 1185.87$\pm$3.444& 0.515$\pm$0.00249& [-0.9, 0.9]& 0.9\\
                                & $(-80\%,\,+30\%)$    & 900& 1978.78$\pm$1.483& 0.519$\pm$0.00064& [-0.9, 0.9]& 0.9\\
                                & $(+80\%,\,-30\%)$    & 900& 1638.57$\pm$1.35& 0.48$\pm$0.000723& [-0.9, 0.9]& 0.9\\
                                & $(+80\%,\,+30\%)$   & 100& 1031.22$\pm$3.21& 0.4855$\pm$0.00272& [-0.9, 0.9]& 0.9\\
\hline
\multirow{4}{*}{500} & $(-80\%,\,-30\%)$    & 400& 285.63$\pm$0.845& 0.477$\pm$0.0026& [-0.9, 0.9]& 0.9\\
                                & $(-80\%,\,+30\%)$    & 1600& 475.59$\pm$0.545& 0.482$\pm$0.001004& [-0.9, 0.9]& 0.9\\
                                & $(+80\%,\,-30\%)$    & 1600& 401.64$\pm$0.501& 0.44$\pm$0.00112& [-0.9, 0.9]& 0.9\\
                                & $(+80\%,\,+30\%)$    & 400& 252.02$\pm$0.794& 0.446$\pm$0.00282& [-0.9, 0.9]& 0.9\\
\hline
\multirow{4}{*}{1000} & $(-80\%,\,-20\%)$    & 800& 78.69$\pm$0.3136& 0.47$\pm$0.00352& [-0.9, 0.9]& 0.9\\
                                & $(-80\%,\,+20\%)$    & 3200& 110.07$\pm$0.1855& 0.473$\pm$0.001485& [-0.9, 0.9]& 0.9\\
                                & $(+80\%,\,-20\%)$    & 3200& 93.8$\pm$0.1712& 0.432$\pm$0.001646& [-0.9, 0.9]& 0.9\\
                                & $(+80\%,\,+20\%)$    & 800& 68.93$\pm$0.2935& 0.436$\pm$0.00383& [-0.9, 0.9]& 0.9\\
\hline
    \end{tabular}
    \end{adjustbox}
    \caption{Projections for $e^-e^+\to \tau^-\tau^+$ at ILC, where the last column is the total selection efficiency.}
    \label{tab:light_fermion_pair_ILCtautau}
\end{table}

\begin{table}[]
\centering
\begin{adjustbox}{max width=1\textwidth, max height=0.49\textheight}
\begin{tabular}{c|c|c|c|c|c|c}
\hline
{\textbf{ILC}} $\sqrt{s}$ [GeV] & Pol. ($e^-,e^+$) & $\mathcal{L}\,\rm[fb^{-1}]$ & $\sigma_c$ [fb] & $A_{FB}^c$ & $[c_\theta^{\rm min},c_\theta^{\rm max}]$ & $\epsilon$\\\hline\hline
\multirow{4}{*}{250} & $(-80\%,\,-30\%)$    & 100&  81.79$\pm$0.904& 0.599$\pm$0.00886& [-0.9, 0.9]& 0.03\\
                                & $(-80\%,\,+30\%)$    & 900&  143.08$\pm$0.399& 0.594$\pm$0.00224& [-0.9, 0.9]& 0.03\\
                                & $(+80\%,\,-30\%)$    & 900& 68.5$\pm$0.276& 0.662$\pm$0.00302& [-0.9, 0.9]& 0.03\\
                                & $(+80\%,\,+30\%)$   & 100& 47.89$\pm$0.692& 0.646$\pm$0.01103& [-0.9, 0.9]& 0.03\\
\hline
\multirow{4}{*}{500} & $(-80\%,\,-30\%)$    & 400&  18.88$\pm$0.2173& 0.57$\pm$0.00946& [-0.9, 0.9]& 0.03\\
                                & $(-80\%,\,+30\%)$    & 1600& 32.93$\pm$0.1435& 0.565$\pm$0.003594& [-0.9, 0.9]& 0.03\\
                                & $(+80\%,\,-30\%)$    & 1600& 16.52$\pm$0.1016& 0.629$\pm$0.00478& [-0.9, 0.9]& 0.03\\
                                & $(+80\%,\,+30\%)$    & 400& 11.42$\pm$0.169& 0.614$\pm$0.01168& [-0.9, 0.9]& 0.03\\
\hline
\multirow{4}{*}{1000} & $(-80\%,\,-20\%)$    & 800& 5.21$\pm$0.0807& 0.561$\pm$0.01282& [-0.9, 0.9]& 0.03\\
                                & $(-80\%,\,+20\%)$    & 3200& 7.5$\pm$0.0484& 0.559$\pm$0.00535& [-0.9, 0.9]& 0.03\\
                                & $(+80\%,\,-20\%)$    & 3200& 3.88$\pm$0.03484& 0.618$\pm$0.00705& [-0.9, 0.9]& 0.03\\
                                & $(+80\%,\,+20\%)$    & 800& 3.037$\pm$0.0616& 0.609$\pm$0.0161& [-0.9, 0.9]& 0.03\\
\hline
    \end{tabular}
    \end{adjustbox}
    \caption{Projections for $e^-e^+\to c\bar{c}$ at ILC, where the last column is the total selection efficiency.}
    \label{tab:light_fermion_pair_ILCcc}
\end{table}

\begin{table}[]
\centering
\begin{adjustbox}{max width=1\textwidth, max height=0.49\textheight}
\begin{tabular}{c|c|c|c|c|c|c}
\hline
{\textbf{ILC}} $\sqrt{s}$ [GeV] & Pol. ($e^-,e^+$) & $\mathcal{L}\,\rm[fb^{-1}]$ & $\sigma_b$ [fb] & $A_{FB}^b$ & $[c_\theta^{\rm min},c_\theta^{\rm max}]$ & $\epsilon$\\\hline\hline
\multirow{4}{*}{250} & $(-80\%,\,-30\%)$    & 100&  268.77$\pm$1.64& 0.648$\pm$0.00464& [-0.9, 0.9]& 0.15\\
                                & $(-80\%,\,+30\%)$    & 900&  483.56$\pm$0.733& 0.66$\pm$0.001138& [-0.9, 0.9]& 0.15\\
                                & $(+80\%,\,-30\%)$    & 900&  134.94$\pm$0.387& 0.351$\pm$0.002687& [-0.9, 0.9]& 0.15\\
                                & $(+80\%,\,+30\%)$   & 100&  110.32$\pm$1.05& 0.4585$\pm$0.00846& [-0.9, 0.9]& 0.15\\
\hline
\multirow{4}{*}{500} & $(-80\%,\,-30\%)$    & 400&  58.18$\pm$0.3814& 0.641$\pm$0.00503& [-0.9, 0.9]& 0.15\\
                                & $(-80\%,\,+30\%)$    & 1600&  104.77$\pm$0.256& 0.649$\pm$0.00186& [-0.9, 0.9]& 0.15\\
                                & $(+80\%,\,-30\%)$    & 1600&  28.58$\pm$0.1336& 0.446$\pm$0.00419& [-0.9, 0.9]& 0.15\\
                                & $(+80\%,\,+30\%)$    & 400&  23.55$\pm$0.2426& 0.517$\pm$0.00882& [-0.9, 0.9]& 0.15\\
\hline
\multirow{4}{*}{1000} & $(-80\%,\,-20\%)$    & 800& 15.94$\pm$0.141& 0.64$\pm$0.0068& [-0.9, 0.9]& 0.15\\
                                & $(-80\%,\,+20\%)$    & 3200& 23.44$\pm$0.0856& 0.645$\pm$0.00279& [-0.9, 0.9]& 0.15\\
                                & $(+80\%,\,-20\%)$    & 3200& 6.68$\pm$0.0457& 0.4755$\pm$0.00602& [-0.9, 0.9]& 0.15\\
                                & $(+80\%,\,+20\%)$    & 800& 5.88$\pm$0.0857& 0.518$\pm$0.01248& [-0.9, 0.9]& 0.15\\
\hline
    \end{tabular}
    \end{adjustbox}
    \caption{Projections for $e^-e^+\to b\bar{b}$ at ILC, where the last column is the total selection efficiency.}
    \label{tab:light_fermion_pair_ILCbb}
\end{table}

%%%%%%%%%%%%%%%%% CLIC

\begin{table}[]
\centering
\begin{adjustbox}{max width=1\textwidth, max height=0.49\textheight}
\begin{tabular}{c|c|c|c|c|c|c}
\hline
{\textbf{CLIC}} $\sqrt{s}$ [GeV] & Pol. ($e^-,e^+$) & $\mathcal{L}\,\rm[fb^{-1}]$ & $\sigma_e$ [fb] & $A_{FB}^e$ & $[c_\theta^{\rm min},c_\theta^{\rm max}]$ & $\epsilon$\\\hline\hline
\multirow{2}{*}{380} & $(-80\%,\,0\%)$     &  500& 29422.4$\pm$7.65& 0.96$\pm$0.0000727&[ -0.9, 0.9]& 0.98\\
                                & $(+80\%,\,0\%)$    &  500& 33886.8$\pm$8.06& 0.954$\pm$0.0000713&[ -0.9, 0.9]& 0.98\\
\hline
\multirow{2}{*}{1500} & $(-80\%,\,0\%)$    &  2000& 2024.97$\pm$1.004& 0.963$\pm$0.0001345&[ -0.9, 0.9]& 0.98\\
                                  & $(+80\%,\,0\%)$   &  500& 2298.87$\pm$2.11& 0.945$\pm$0.000299&[ -0.9, 0.9]& 0.98\\
\hline
\multirow{2}{*}{3000} & $(-80\%,\,0\%)$    &  4000& 510.3$\pm$0.3565& 0.963$\pm$0.0001888&[ -0.9, 0.9]& 0.98\\
                                  & $(+80\%,\,0\%)$   &  1000& 578.04$\pm$0.749& 0.945$\pm$0.000424&[ -0.9, 0.9]& 0.98\\
\hline
    \end{tabular}
    \end{adjustbox}
    \caption{Projections for $e^-e^+\to e^-e^+$ at CLIC, where the last column is the total selection efficiency.}
    \label{tab:light_fermion_pair_CLICee}
\end{table}

\begin{table}[]
\centering
\begin{adjustbox}{max width=1\textwidth, max height=0.49\textheight}
\begin{tabular}{c|c|c|c|c|c|c}
\hline
{\textbf{CLIC}} $\sqrt{s}$ [GeV] & Pol. ($e^-,e^+$) & $\mathcal{L}\,\rm[fb^{-1}]$ & $\sigma_\mu$ [fb] & $A_{FB}^\mu$ & $[c_\theta^{\rm min},c_\theta^{\rm max}]$ & $\epsilon$\\\hline\hline
\multirow{2}{*}{380} & $(-80\%,\,0\%)$     &  500& 782.38$\pm$1.25& 0.504$\pm$0.00138& [-0.95, 0.95]& 0.98\\
                                & $(+80\%,\,0\%)$    &  500& 669.18$\pm$1.157& 0.465$\pm$0.00153& [-0.95, 0.95]& 0.98\\
\hline
\multirow{2}{*}{1500} & $(-80\%,\,0\%)$    &  2000& 49.31$\pm$0.157& 0.484$\pm$0.002786& [-0.95, 0.95]& 0.98\\
                                  & $(+80\%,\,0\%)$   &  500& 42.54$\pm$0.2917& 0.445$\pm$0.00614& [-0.95, 0.95]& 0.98\\
\hline
\multirow{2}{*}{3000} & $(-80\%,\,0\%)$    &  4000& 12.32$\pm$0.0555& 0.483$\pm$0.003945& [-0.95, 0.95]& 0.98\\
                                  & $(+80\%,\,0\%)$   &  1000& 10.63$\pm$0.1031& 0.444$\pm$0.00869& [-0.95, 0.95]& 0.98\\
\hline
    \end{tabular}
    \end{adjustbox}
    \caption{Projections for $e^-e^+\to \mu^-\mu^+$ at CLIC, where the last column is the total selection efficiency.}
    \label{tab:light_fermion_pair_CLICmumu}
\end{table}

\begin{table}[]
\centering
\begin{adjustbox}{max width=1\textwidth, max height=0.49\textheight}
\begin{tabular}{c|c|c|c|c|c|c}
\hline
{\textbf{CLIC}} $\sqrt{s}$ [GeV] & Pol. ($e^-,e^+$) & $\mathcal{L}\,\rm[fb^{-1}]$ & $\sigma_\tau$ [fb] & $A_{FB}^\tau$ & $[c_\theta^{\rm min},c_\theta^{\rm max}]$ & $\epsilon$\\\hline\hline
\multirow{2}{*}{380} & $(-80\%,\,0\%)$     &  500& 664.57$\pm$1.153& 0.489$\pm$0.001513& [-0.9, 0.9]& 0.9\\
                                & $(+80\%,\,0\%)$    &  500& 568.38$\pm$1.066& 0.4515$\pm$0.001674& [-0.9, 0.9]& 0.9\\
\hline
\multirow{2}{*}{1500} & $(-80\%,\,0\%)$    &  2000& 41.89$\pm$0.1447& 0.47$\pm$0.00305& [-0.9, 0.9]& 0.9\\
                                  & $(+80\%,\,0\%)$   &  500& 36.13$\pm$0.269& 0.432$\pm$0.00671& [-0.9, 0.9]& 0.9\\
\hline
\multirow{2}{*}{3000} & $(-80\%,\,0\%)$    &  4000& 10.46$\pm$0.0511& 0.469$\pm$0.00432& [-0.9, 0.9]& 0.9\\
                                  & $(+80\%,\,0\%)$   &  1000& 9.03$\pm$0.095& 0.431$\pm$0.0095& [-0.9, 0.9]& 0.9\\
\hline
    \end{tabular}
    \end{adjustbox}
    \caption{Projections for $e^-e^+\to \tau^-\tau^+$ at CLIC, where the last column is the total selection efficiency.}
    \label{tab:light_fermion_pair_CLICtautau}
\end{table}

\begin{table}[]
\centering
\begin{adjustbox}{max width=1\textwidth, max height=0.49\textheight}
\begin{tabular}{c|c|c|c|c|c|c}
\hline
{\textbf{CLIC}} $\sqrt{s}$ [GeV] & Pol. ($e^-,e^+$) & $\mathcal{L}\,\rm[fb^{-1}]$ & $\sigma_c$ [fb] & $A_{FB}^c$ & $[c_\theta^{\rm min},c_\theta^{\rm max}]$ & $\epsilon$\\\hline\hline
\multirow{2}{*}{380} & $(-80\%,\,0\%)$     &  500& 45.7$\pm$0.3023& 0.574$\pm$0.00542& [-0.9, 0.9]& 0.03\\
                                & $(+80\%,\,0\%)$    &  500& 24.41$\pm$0.221& 0.631$\pm$0.00702& [-0.9, 0.9]& 0.03\\
\hline
\multirow{2}{*}{1500} & $(-80\%,\,0\%)$    &  2000& 2.81$\pm$0.0375& 0.558$\pm$0.01106& [-0.9, 0.9]& 0.03\\
                                  & $(+80\%,\,0\%)$   &  500& 1.534$\pm$0.0554& 0.613$\pm$0.02854& [-0.9, 0.9]& 0.03\\
\hline
\multirow{2}{*}{3000} & $(-80\%,\,0\%)$    &  4000& 0.7$\pm$0.01323& 0.558$\pm$0.01568& [-0.9, 0.9]& 0.03\\
                                  & $(+80\%,\,0\%)$   &  1000& 0.383$\pm$0.01956& 0.612$\pm$0.0404& [-0.9, 0.9]& 0.03\\
\hline
    \end{tabular}
    \end{adjustbox}
    \caption{Projections for $e^-e^+\to c\bar{c}$ at CLIC, where the last column is the total selection efficiency.}
    \label{tab:light_fermion_pair_CLICcc}
\end{table}

\begin{table}[]
\centering
\begin{adjustbox}{max width=1\textwidth, max height=0.49\textheight}
\begin{tabular}{c|c|c|c|c|c|c}
\hline
{\textbf{CLIC}} $\sqrt{s}$ [GeV] & Pol. ($e^-,e^+$) & $\mathcal{L}\,\rm[fb^{-1}]$ & $\sigma_b$ [fb] & $A_{FB}^b$ & $[c_\theta^{\rm min},c_\theta^{\rm max}]$ & $\epsilon$\\\hline\hline
\multirow{2}{*}{380} & $(-80\%,\,0\%)$     &  500& 145.83$\pm$0.54& 0.649$\pm$0.00282& [-0.9, 0.9]& 0.15\\
                                & $(+80\%,\,0\%)$    &  500& 46.81$\pm$0.306& 0.46$\pm$0.0058&[ -0.9, 0.9]& 0.15\\
\hline
\multirow{2}{*}{1500} & $(-80\%,\,0\%)$    &  2000& 8.69$\pm$0.0659& 0.643$\pm$0.00581& [-0.9, 0.9]& 0.15\\
                                  & $(+80\%,\,0\%)$   &  500& 2.77$\pm$0.0744& 0.498$\pm$0.0233& [-0.9, 0.9]& 0.15\\
\hline
\multirow{2}{*}{3000} & $(-80\%,\,0\%)$    &  4000& 2.16$\pm$0.02325& 0.642$\pm$0.00824& [-0.9, 0.9]& 0.15\\
                                  & $(+80\%,\,0\%)$   &  1000& 0.689$\pm$0.02625& 0.5$\pm$0.033& [-0.9, 0.9]& 0.15\\
\hline
    \end{tabular}
    \end{adjustbox}
    \caption{Projections for $e^-e^+\to b\bar{b}$ at CLIC, where the last column is the total selection efficiency.}
    \label{tab:light_fermion_pair_CLICbb}
\end{table}

%%%%%%%%%%%%%%%%% FCC-ee

\begin{table}[]
\centering
\begin{adjustbox}{max width=1\textwidth, max height=0.49\textheight}
\begin{tabular}{c|c|c|c|c|c|c}
\hline
{\textbf{FCC-ee}} $\sqrt{s}$ [GeV] & Final state & $\mathcal{L}\,\rm[fb^{-1}]$ & $\sigma$ [fb] & $A_{FB}$ & $[c_\theta^{\rm min},c_\theta^{\rm max}]$ & $\epsilon$\\\hline\hline
\multirow{5}{*}{240} & $e^-e^+$            &  \multirow{5}{*}{5000} & 77330.4$\pm$3.87& 0.96$\pm$0.00001388& [-0.9, 0.9]& 0.98\\
                                & $\mu^-\mu^+$    &                                    & 1870.84$\pm$0.612& 0.521$\pm$0.000279& [-0.95, 0.95]& 0.98\\
                                & $\tau^-\tau^+$    &                                    & 1589.15$\pm$0.564& 0.506$\pm$0.000306& [-0.9, 0.9]& 0.9\\
                                & $c\bar{c}$          &                                    & 93.38$\pm$0.1367& 0.62$\pm$0.00115& [-0.9, 0.9]& 0.03\\
                                & $b\bar{b}$          &                                    & 275.64$\pm$0.235& 0.592$\pm$0.000687& [-0.9, 0.9]& 0.15\\
\hline
\multirow{5}{*}{365} & $e^-e^+$            &  \multirow{5}{*}{1500} & 34221.5$\pm$4.72& 0.957$\pm$0.0000399& [-0.9, 0.9]& 0.98\\
                                & $\mu^-\mu^+$    &                                    & 787.74$\pm$0.725& 0.488$\pm$0.000803& [-0.95, 0.95]& 0.98\\
                                & $\tau^-\tau^+$    &                                    & 669.11$\pm$0.668& 0.473$\pm$0.00088& [-0.9, 0.9]& 0.9\\
                                & $c\bar{c}$          &                                    & 38.11$\pm$0.1594& 0.595$\pm$0.00336& [-0.9, 0.9]& 0.03\\
                                & $b\bar{b}$          &                                    & 105.12$\pm$0.2647& 0.603$\pm$0.00201& [-0.9, 0.9]& 0.15\\
\hline
    \end{tabular}
    \end{adjustbox}
    \caption{Projections for $e^-e^+\to f\bar{f}$ at FCC-ee, where the last column is the total selection efficiency.}
    \label{tab:light_fermion_pair_FCCee}
\end{table}

%%%%%%%%%%%%%%%%% CEPC

\begin{table}[]
\centering
\begin{adjustbox}{max width=1\textwidth, max height=0.49\textheight}
\begin{tabular}{c|c|c|c|c|c|c}
\hline
{\textbf{CEPC}} $\sqrt{s}$ [GeV] & Final state & $\mathcal{L}\,\rm[fb^{-1}]$ & $\sigma$ [fb] & $A_{FB}$ & $[c_\theta^{\rm min},c_\theta^{\rm max}]$ & $\epsilon$\\\hline\hline
\multirow{5}{*}{240} & $e^-e^+$            &  \multirow{5}{*}{20000} & 77330.4$\pm$1.937& 0.96$\pm$0.00000694& [-0.9, 0.9]& 0.98\\
                                & $\mu^-\mu^+$    &                                    & 1870.84$\pm$0.306& 0.521$\pm$0.0001395& [-0.95, 0.95]& 0.98\\
                                & $\tau^-\tau^+$    &                                    & 1589.15$\pm$0.282& 0.506$\pm$0.000153& [-0.9, 0.9]& 0.9\\
                                & $c\bar{c}$          &                                    & 93.38$\pm$0.0683& 0.62$\pm$0.000574& [-0.9, 0.9]& 0.03\\
                                & $b\bar{b}$          &                                    & 275.64$\pm$0.1174& 0.592$\pm$0.0003434& [-0.9, 0.9]& 0.15\\
\hline
\multirow{5}{*}{360} & $e^-e^+$            &  \multirow{5}{*}{1000} & 35147.9$\pm$5.85& 0.957$\pm$0.0000482& [-0.9, 0.9]& 0.98\\
                                & $\mu^-\mu^+$    &                                    & 810.18$\pm$0.9& 0.4885$\pm$0.00097& [-0.95, 0.95]& 0.98\\
                                & $\tau^-\tau^+$    &                                    & 688.17$\pm$0.83& 0.474$\pm$0.001061& [-0.9, 0.9]& 0.9\\
                                & $c\bar{c}$          &                                    & 39.22$\pm$0.198& 0.596$\pm$0.004056& [-0.9, 0.9]& 0.03\\
                                & $b\bar{b}$          &                                    & 108.33$\pm$0.329& 0.602$\pm$0.002425& [-0.9, 0.9]& 0.15\\
\hline
    \end{tabular}
    \end{adjustbox}
    \caption{Projections for $e^-e^+\to f\bar{f}$ at CEPC, where the last column is the total selection efficiency.}
    \label{tab:light_fermion_pair_CEPC}
\end{table}

\subsection{Top-quark measurements}
Input observables related to top-quark sector are explained in Sec.~\ref{sec:topfit}.
%\subsection{Low-energy measurements}
%Input observables from low-energy measurements and their uncertainties are currently listed in Section 6.

%remove the clearpage below in the end
\clearpage

\subsection{Diboson measurements}
\label{sec:diboson}

The diboson ($\eeww$) measurements provide important constraints on a set of operator coefficients that are essential to the Higgs + EW fit. 
Conventionally, the new physics effects are parameterized in terms of three CP-even anomalous triple gauge couplings (aTGCs).  This was for instance done by the LEP collaboration~\cite{ALEPH:2013dgf} and also in the ILC analysis~\cite{Marchesini:2011aka}.  Considering the tree-level contributions of SMEFT CP-even dimension-6 operators, and omitting those that only contribute to the $W$-boson decay rate, a total number of 7 independent parameters contribute to the $e^+e^- \to W^+W^-$ process. 
Among them, one degree of freedom can be associated with the modification of the $W$-boson mass, which we discard here due to the strong constraints from the $W$-mass measurements.  In the language of the effective Lagrangian in \autoref{sec:SMEFTeffC}, the remaining 6 parameters are\footnote{Here we assume that the initial particles are $e^+e^-$.  For  $\mu^+\mu^-$, one needs to replace the last three parameters by the corresponding muon couplings.}
\begin{equation}
    \delta g_{1,Z},~~\delta \kappa_{\gamma},~~\lambda_Z,~~\delta g^{ee}_{Z,L},~~\delta g^{ee}_{Z,R}~~{\rm and}~~\delta g^{e\nu}_{W} \,,  \label{eq:wwpara6} 
\end{equation}
where $\delta g_{1,Z}$, $\delta \kappa_{\gamma}$, $\lambda_Z$ are the familiar aTGCs, and $\delta g^{ee}_{Z,L}$, $\delta g^{ee}_{Z,R}$, $\delta g^{e\nu}_{W}$ correspond to modifications in the $Ze_L\bar{e}_L$, $Ze_R\bar{e}_R$ and $W e\nu$ couplings.  The latter are particularly relevant if the measurement precision of the diboson process is comparable or even better than those of Z-pole measurements.   
All 6 parameters are included in our global SMEFT analysis.  

While several studies of the diboson measurements already exist from various collider collaborations (such as the ILC one~\cite{Marchesini:2011aka}), not all projections for future lepton colliders are available, and many of the available ones uses the 3-aTGC framework which is not directly applicable in the global SMEFT framework.  Furthermore, these parameters are very sensitive to the multi-dimensional differential distribution of the $\eeww$ process, which is practically difficult to be fully utilized in an analysis with binned-distributions.  To efficiently extract information from the measurements, and for a consistent treatment among various different colliders, we perform a simplified phenomenological analysis based on Optimal Observables~\cite{Diehl:1993br,Gunion:1996vv} for all collider scenarios to extract the likelihood (or $\chi^2$) of the six parameters in \autoref{eq:wwpara6}. 
The optimal observable analysis assumes that the new physics contributions enter observables only at the linear level. 
This is a good approximation for the very precise diboson measurements that are expected at future lepton colliders, and is also consistent with the SMEFT treatment in our analysis.  
More explicitly, the differential cross section is parameterized as
\begin{equation}
 \frac{d \sigma}{d \Omega} = S_0 +\underset{i}{\sum} S_{1,i} \,g_i \,,   
\end{equation}
where $g_{i=1,...,6}$ are the six parameters in \autoref{eq:wwpara6}, and $S_0$, $S_{1,i}$ are functions of the differential observables.  In the narrow width approximation, each event can be described by five independent observables, which are the production polar angle $\theta$ and two decay angles for each $W$. $S_0 =\frac{d\sigma_{\rm SM}}{d\Omega}$ is the SM differential cross section.  It can be shown that the best possible reaches on the $g_i$ are given by the inverse covariance matrix 
\begin{equation}
c^{-1}_{ij} = \int d\Omega \frac{S_{1,i}S_{1,j} }{S_0} \cdot \mathcal{L} \,, 
\end{equation}
where $\mathcal{L}$ is the total integrated luminosity.  The $c^{-1}_{ij}$ can be obtained by measuring the optimal observables, defined as $\mathcal{O}_i = \frac{S_{1,i}}{S_0}$, 
and is simply given by the covariance matrix $V_{ij}$ of the $O_i$, $c^{-1}_{ij} = n V_{ij}$ where $n$ is the number of events. 

It should be noted that the diboson process also receives  contributions from (four) additional operators that modify the $W$ branching ratios.  While an optimal observable analysis could be done for each diboson decay channel including these additional parameters, it is more convenient to separate the information in the rate measurements of $e^+e^- \to W^+W^-$ from the differential ones, as the latter depend only on the parameters in  \autoref{eq:wwpara6}.   
It is straightforward to subtract from $c^{-1}_{ij}$ the contribution from the total rate measurement, which we treat separately with possible modifications in the $W$ branching ratios.  For the differential analysis, several assumptions are made:  We consider only the statistical uncertainties of the signal and assume a conservative selection efficiency of $45\%$ in all $WW$ events, chosen to agree with the results from the ILC full simulation analysis of the semi-leptonic channel at 500 GeV. 
We include all decay channels of the $WW$ pairs.  For the hadronic decay, one could not distinguish the two quarks, and the corresponding angular distributions are ``folded.''  This effect is implemented in the optimal observables.  The $\tau$ channel is treated in the same way as the lepton channels, assuming a good $\tau$ reconstruction can be achieved at the future lepton colliders.  For the dilepton channels, the momenta of the two missing neutrinos cannot be directly reconstructed.  They could be obtained by imposing the conditions of the $W$ on-shell mass and the center of mass energies, which gives a set of quadratic equations that can be solved.  We assume that the correct solution is always chosen for each event.  The optimal observable analysis is performed for each $WW$ channel, and the resultant likelihoods are combined in the end.    
For jets and leptons we use detector acceptance cuts on the polar angle of $\left|\cos{\theta}\right|<$0.9  and 0.95, respectively.
We have checked that the effects of detector acceptance and smearing have only a small impact on the results, reducing the precision reaches by at most $\sim 10$-$15\%$.
It was also shown in Ref.~\cite{Marchesini:2011aka} that, with appropriate selection cuts, most backgrounds can be removed with a signal efficiency of around $70\%$.  

For the rate measurements of the $WW$ process, the following treatment is implemented.  The decay of a single $W$ can be separated into four channels, $e\nu$, $\mu\nu$, $\tau\nu$ and $jj$.  For each possible decay channel of the $WW$ pair, we estimate the precision of the rate measurement (in terms of $\sigma_{\eeww} \times {\rm BR}_{W^+} \times {\rm BR}_{W^-}$), considering only the statistical uncertainties of the signal and the above-mentioned selection efficiency of $45\%$ in all $WW$ events.
We also assume that the $W$ has no exotic decay, so that the relation 
\begin{equation}
    {\rm BR}_{W^- \to e\nu} + {\rm BR}_{W^- \to \mu\nu} + {\rm BR}_{W^- \to \tau\nu} + {\rm BR}_{W^- \to jj} =1 \,, \label{eq:WBR}
\end{equation}
is imposed.  With this condition, one could extract the precision of the total cross section $\sigma_{WW}$ and the 
$W$ branching ratios from the measurements of all the $WW$ channels.  
Additional measurements of the W-boson width ($\it e.g.$ the ones from threshold scan) as listed in \autoref{tab:EWPO}, as well as the projected reach of the HL-LHC are also included in the global analysis.  With these measurements, 
the total cross section and branching ratios can be determined even without imposing \autoref{eq:WBR}. 
From the projections on total cross section $\sigma_{WW}$ and the $W$ branching ratios, the likelihood ($\chi^2$) of the operator coefficients can be obtained, which is combined with the one from the differential analysis.
%

%%%%%%%%%%%%%%%%%%%%%%%%%%%%%%%%%%
\section{Higgs + EW fit}
\label{sec:higgsew}
%%%%%%%%%%%%%%%%%%%%%%%%%%%%%%%%%%
%[editor: Jiayin, Jorge, Michael]

In this section we report the results of a global analysis of the Higgs and EW measurements in the dimension-6 SMEFT framework. 
The following measurements are included in the analysis:
\begin{itemize}

    \item The Higgs rate measurements listed in \autoref{tab:LHC_HiggsSxBR}-\ref{tab:CLIC3000_Higgs}. In particular, the results for all future colliders are assumed to be combined with the HL-LHC Higgs measurements. 
    
    \item  The electroweak precision observables in \autoref{tab:EWPO}. Here we assume as a baseline 
    the current set of precisions for EWPO, but with SM central values.~\footnote{We exclude the recent measurement of the $W$ mass from CDF.} This is combined with all future collider scenarios.
    
    \item  The diboson measurements in \autoref{sec:diboson}.  For lepton colliders, this include the $W$ branching ratio measurements and the differential analysis using optimal observables; for the HL-LHC, we implement the results from Ref.~\cite{Grojean:2018dqj}.
    
    \item  For the high energy muon collider only, we also consider the $\gamma\gamma \to W^+W^-$ process for the measurements of the W branching ratios. The cross section~\cite{Han:2021kes} for this process is much larger than the one of $\mu^+\mu^- \to  W^+W^-$ at very high energy.   This mainly improves the reach on the operator coefficients that modify the W branching ratios. 
    
\end{itemize}

\begin{figure}%[t]
    \includegraphics[width=\textwidth]{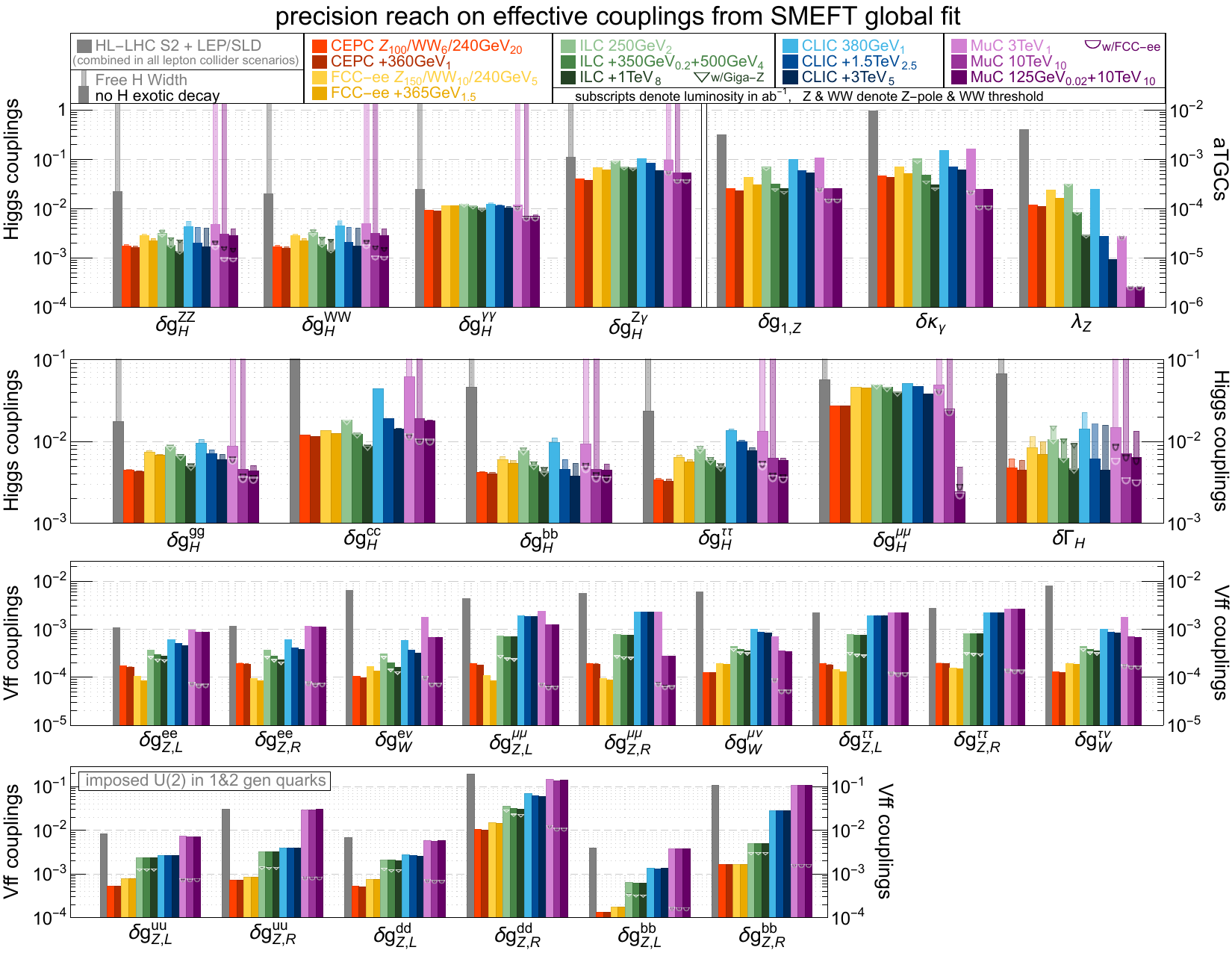}
    \caption{
    Precision reach on effective couplings from a SMEFT global analysis of the Higgs and EW measurements at various future colliders listed in \autoref{tab:epem_setup}.  The wide (narrow) bars correspond to the results  
    from the constrained-$\Gamma_H$ (free-$\Gamma_H$) fit. The HL-LHC and LEP/SLD measurements are combined with all lepton collider scenarios.  For $e^+e^-$ colliders, the high energy runs are always combined with the low energy ones.  For the ILC scenarios, the (upper edge of the) triangle mark shows the results for which a Giga-Z run is also included.  For the muon collider, 3 separate scenarios are considered.  The subscripts in the collider scenarios denote the corresponding integrated luminosity of the run in ${\rm ab}^{-1}$.
    Note the Higgs total width measurement from the off-shell Higgs processes at the HL-LHC is not included in the global fit.
    }
    \label{fig:fit1}
\end{figure}

For the study, we perform a series of fits of the dimension-6 SMEFT to the above-mentioned measurements for the different colliders scenarios in \autoref{tab:epem_setup}. 
These fits were performed with the {\tt HEPfit} code~\cite{DeBlas:2019ehy} and using a Bayesian approach. An independent cross check of the results was performed using a $\chi^2$ fit constructed with all the relevant measurements. In \autoref{fig:fit1} and \autoref{tab:fit1}, we present the result of the fits in terms of the 68\% probability sensitivity\footnote{This is estimated as the square root of the variance of the posterior predictive distribution of the corresponding parameter from the fits.} to modifications to the effective couplings introduced in \autoref{sec:SMEFTeffC}~\cite{Barklow:2017suo, deBlas:2019rxi, DeBlas:2019qco},
\begin{equation}
\delta g_{X}^{Y}=\frac{g_{XY}^{\rm eff}}{g_{XY}^{\rm eff, SM}} -1,
\end{equation}
for the various collider scenarios listed in \autoref{tab:epem_setup}.  For the $e^+e^-$ colliders, the runs are considered to be staged, i.e. the high energy runs are always combined with the low energy ones.  For the muon collider, three separate scenarios are considered: operating at 3 TeV, at 10 TeV, and the latter combined with a run at 125 GeV. (In \autoref{fig:fit1} we also show results of these three scenarios in combination with the information of FCCee.)  

\begin{figure}%[t]
    \includegraphics[width=\textwidth]{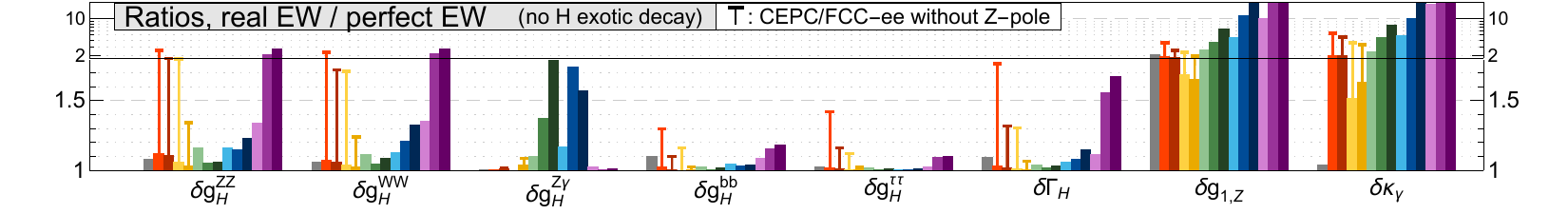}
    \caption{
    Ratios of the measurement precision (shown in \autoref{fig:fit1}) to the one assuming perfect EW measurements 
    in the constrained-$\Gamma_H$ fit. See text for details. Results are only shown for Higgs couplings and aTGCs with ratios noticeably larger than one.  For CEPC/FCC-ee, we also show (with the thin ``T'' lines) the results without the improved measurements of the EWPO that would be possible at the future Z-pole runs.
    }
    \label{fig:fit1-ratio}
\end{figure}
Two sets of results are shown for each scenario: one assumes that the Higgs decay channels are only the ones in the SM; the other assumes the total Higgs width is not constrained by the previous condition, and the Higgs can decay into non-SM states. (This is modeled in the fits by introducing a new parameter, BR$_{\rm Exo}\geq 0$, for the non-SM branching ratio.)  These two scenarios are represented by the wide and narrow bars, respectively.  They will be denoted as the constrained-$\Gamma_H$ fit and the free-$\Gamma_H$ fit later on.  For the ILC results, we consider an additional scenario with a Giga-Z run included, which is illustrated by the triangle marks. 
We also impose a $U(2)$ symmetry for the electroweak gauge couplings of the first two generation quarks, which can be written explicitly as $\delta g^{uu}_{Z,L/R} = \delta g^{cc}_{Z,L/R}$ and $\delta g^{dd}_{Z,L/R} = \delta g^{ss}_{Z,L/R}$.  As such, the results for $\delta g^{cc}_{Z,L/R}$ and $\delta g^{ss}_{Z,L/R}$ are not explicitly shown.  This assumption is necessary in our framework to remove a flat direction among these parameters, as the asymmetry observables using jet-charge could not be measured for the first generation quarks at lepton colliders\footnote{The measurement of $R_{uc}$ as included in the next section, which can measure the electric-charge asymmetry using final state photon radiation effects, can help to relax this assumption.}.  Note that the LHC could measure a similar asymmetry observable and lift this flat direction~\cite{Breso-Pla:2021qoe}, in which case our $U(2)$ assumption could be removed. Since there is no official HL-LHC projection for this specific measurement, we do not consider it in our analysis. We will however illustrate the impact of these asymmetry observables later in \autoref{sec:noU2}.
In the leptonic sector, on the other hand, we do not impose any ``universality'' condition, and couplings to electrons, muons and taus are assumed to be independent. Higgs couplings are also assumed to be diagonal but independent for the different fermion families. With the exception of the possibility of having possible non-SM decays of the Higgs boson, and thus $\Gamma_H$ effectively as an independent parameter, this is similar to the {SMEFT$_{\rm ND}$ fit scenario considered in \cite{deBlas:2019rxi}. As in that reference, dimension-6 SMEFT contributions are including at leading order and SM predictions are computed including the future projected uncertainties associated to the SM input parameters in the $\{\alpha, M_Z, G_F, m_t, m_H\}$ scheme. See \autoref{sec:SMuncFit1} for more details on the latter and a discussion on the impact of other SM uncertainties.

For the constrained-$\Gamma_H$ fit, the outcome of this analysis is similar to that presented in Ref.~\cite{DeBlas:2019qco}, with the exception of the CEPC results where one observes the expected improvement in the sensitivity to Higgs couplings derived from the increase in the luminosity at 240 GeV, together with the addition of the new set of measurements that would be possible at 360 GeV. 
The sensitivity to the aTGC via the optimal observable analysis presented in \autoref{sec:diboson} is also different compared to Ref.~\cite{DeBlas:2019qco}, as we now use all $W$ decay channels (as opposed to only the semi-leptonic channel), but we also use a slightly more conservative selection efficiency, consider cuts not included in \cite{DeBlas:2019qco}, and account for systematic effects associated to the knowledge of the effective beam polarization or the luminosity.

For the free-$\Gamma_H$ fit, it is essential to have a model independent determination of the Higgs width, without which the Higgs couplings could not be constrained.  Clearly, the $e^+e^-$ colliders have the advantage of the inclusive $HZ$ measurements, while a 125\,GeV muon collider is able to directly measure the Higgs width with a threshold scan.
There is a potential at the HL-LHC to determine the Higgs total width using off-shell Higgs measurements~\cite{Caola:2013yja,Campbell:2013una} with an uncertainty of 0.75 MeV~\cite{ATLAS:2022hsp,CMS:2022ley}\footnote{This uncertainty is likely to be improved once the $WW$ channel is employed in addition to the current $ZZ$ analyses.}. This piece of input has not been included in the global fit since the full EFT treatment for this measurement is not yet available \cite{Azatov:2022kbs}. 

It is worth noting that, in a global SMEFT framework, the EW  measurements are also relevant for the Higgs coupling determination, since they constrain many EW parameters that could also enter the Higgs processes.  To illustrate this, we show in \autoref{fig:fit1-ratio} the ratios of the measurement precision to the one obtained assuming {\it perfect EW} measurements 
for the Higgs couplings and aTGCs. This {\it perfect EW} scenario assumes that the experimental precision
of the EWPO sensitive to the $Z$ and $W$ couplings of fermions is so high that these can be assumed to be exactly SM like, i.e. $\delta g_{Z,L/R}^f, \delta g_{W}^{ff\prime}\equiv 0$.
The results are only shown for the constrained-$\Gamma_H$ fit, and for those with this ratio significantly larger than one.  This ratio is generally very close to one for the Higgs couplings at CEPC and FCC-ee, which benefit from the future Z-pole runs.  For comparison, we also show the results for CEPC and FCC-ee with such future Z-pole measurements removed, and much larger ratios are observed for many of the couplings.  On the other hand, the lack of better EW measurements could be a limiting factor for the determination of the $HWW$ and $HZZ$ couplings at a muon collider.
For the aTGCs the situation is different, and we observe a deterioration by a factor $\sim 2$ in  $\delta g_{1,Z}$ and $\delta \kappa_\gamma$ at the 240/250 GeV $e^+e^-$ Higgs/EW factories, which gets significantly worse at high energy lepton colliders ($e^+ e^-$ and $\mu^-\mu^-$). 
Interestingly, our results also suggest that the determination of the $HZ\gamma$ coupling at the linear colliders at high energies could be significantly improved with better EW measurements. 
This is due to its capability of probing the $HZ\gamma$ coupling via the Higgstrahlung process with polarized beams.\footnote{For unpolarized beams, the contribution of the $HZ\gamma$ coupling to the Higgstrahlung process is accidentally suppressed. See {\it e.g.} Ref.~\cite{Durieux:2017rsg}.}

\begin{table}%[]
  \centering
  \begin{adjustbox}{max width=\textwidth}
  \begin{tabular}{|c||c||c|c||c|c||cc|cc|cc||c|c|c||c|c|c|}\hline
in & HL- & \multicolumn{2}{c||}{CEPC} & \multicolumn{2}{c||}{FCC-ee}  & \multicolumn{6}{c||}{ILC} & \multicolumn{3}{c||}{CLIC}  & \multicolumn{3}{c|}{muon-collider} \\ \cline{3-18} %\hline
\% & LHC & 240 & +360 &  240 & +365 & \multicolumn{2}{c|}{250} & \multicolumn{2}{c|}{+500} & \multicolumn{2}{c||}{+1TeV} & 380 & +1.5TeV & +3TeV & 3TeV & 10TeV & 10TeV \\ 
&  & +Z/WW &  &  +Z/WW &  &  & Giga-Z &  & Giga-Z &  & Giga-Z &  &  &  &  &  & +125 \\ \hline \hline
$\delta g^{ZZ}_H$ & 2.2 & 0.17 & 0.16 & 0.28 & 0.22 & 0.31 & 0.29 & 0.18 & 0.18 & 0.13 & 0.13 & 0.43 & 0.19 & 0.16 & 0.48 & 0.31 & 0.28 \\ 
 \text{} & \text{--} & 0.19 & 0.17 & 0.31 & 0.25 & 0.37 & 0.35 & 0.26 & 0.25 & 0.23 & 0.23 & 0.56 & 0.41 & 0.4 & \text{--} & \text{--} & 0.39 \\
 \hline
$\delta g^{WW}_H$ & 2. & 0.17 & 0.15 & 0.28 & 0.22 & 0.32 & 0.31 & 0.19 & 0.18 & 0.14 & 0.14 & 0.44 & 0.21 & 0.17 & 0.49 & 0.31 & 0.28 \\ 
 \text{} & \text{--} & 0.18 & 0.17 & 0.31 & 0.25 & 0.37 & 0.36 & 0.26 & 0.26 & 0.24 & 0.23 & 0.56 & 0.42 & 0.41 & \text{--} & \text{--} & 0.39 \\
 \hline
$\delta g^{\gamma\gamma}_H $ & 2.5 & 0.91 & 0.89 & 1.2 & 1.1 & 1.2 & 1.2 & 1.1 & 1.1 & 0.98 & 0.97 & 1.2 & 1.1 & 1. & 1.2 & 0.7 & 0.69 \\ 
 \text{} & \text{--} & 0.91 & 0.9 & 1.2 & 1.1 & 1.2 & 1.2 & 1.1 & 1.1 & 1. & 1. & 1.3 & 1.2 & 1.1 & \text{--} & \text{--} & 0.74 \\
 \hline
$\delta g^{Z\gamma}_H $ & 11. & 4. & 3.8 & 6.7 & 6.1 & 9.3 & 9.1 & 7. & 6.8 & 6.7 & 6.6 & 10. & 8.3 & 5.8 & 9.7 & 5.2 & 5.2 \\ 
 \text{} & \text{--} & 4. & 3.8 & 6.7 & 6.1 & 9.3 & 9.1 & 7. & 6.8 & 6.7 & 6.6 & 10. & 8.3 & 5.8 & \text{--} & \text{--} & 5.2 \\
 \hline
$\delta g_{1,Z} $ & 0.31 & 0.025 & 0.023 & 0.044 & 0.03 & 0.069 & 0.067 & 0.031 & 0.025 & 0.025 & 0.022 & 0.1 & 0.06 & 0.052 & 0.1 & 0.025 & 0.025 \\ 
 \text{} & 0.31 & 0.025 & 0.023 & 0.043 & 0.03 & 0.069 & 0.067 & 0.031 & 0.025 & 0.025 & 0.022 & 0.1 & 0.06 & 0.052 & 0.1 & 0.025 & 0.025 \\
 \hline
$\delta \kappa_\gamma$ & 0.97 & 0.046 & 0.042 & 0.069 & 0.05 & 0.1 & 0.092 & 0.047 & 0.036 & 0.031 & 0.026 & 0.15 & 0.071 & 0.06 & 0.16 & 0.025 & 0.024 \\ 
 \text{} & 0.97 & 0.046 & 0.043 & 0.069 & 0.05 & 0.1 & 0.092 & 0.047 & 0.036 & 0.031 & 0.026 & 0.15 & 0.071 & 0.061 & 0.16 & 0.025 & 0.025 \\
 \hline
$\lambda_Z $ & 0.4 & 0.012 & 0.011 & 0.023 & 0.016 & 0.031 & 0.031 & 0.0082 & 0.0082 & 0.0028 & 0.0028 & 0.025 & 0.0028 & 0.00092 & 0.0027 & 0.00026 & 0.00025 \\ 
 \text{} & 0.4 & 0.012 & 0.011 & 0.023 & 0.016 & 0.031 & 0.031 & 0.0083 & 0.0082 & 0.0028 & 0.0028 & 0.025 & 0.0028 & 0.00092 & 0.0027 & 0.00026 & 0.00026 \\
 \hline
$\delta g^{gg}_H $ & 1.8 & 0.44 & 0.43 & 0.74 & 0.68 & 0.85 & 0.85 & 0.66 & 0.66 & 0.49 & 0.49 & 0.94 & 0.71 & 0.59 & 0.87 & 0.46 & 0.43 \\ 
 \text{} & \text{--} & 0.45 & 0.44 & 0.77 & 0.69 & 0.9 & 0.89 & 0.69 & 0.69 & 0.53 & 0.53 & 1.1 & 0.79 & 0.69 & \text{--} & \text{--} & 0.51 \\
 \hline
$\delta g^{cc}_H $ & \text{--} & 1.2 & 1.1 & 1.3 & 1.2 & 1.8 & 1.8 & 1.2 & 1.2 & 0.87 & 0.87 & 4.3 & 1.9 & 1.4 & 6.2 & 1.9 & 1.8 \\ 
 \text{} & \text{--} & 1.2 & 1.1 & 1.4 & 1.3 & 1.8 & 1.8 & 1.2 & 1.2 & 0.9 & 0.9 & 4.3 & 1.9 & 1.5 & \text{--} & \text{--} & 1.8 \\
 \hline
$\delta g^{bb}_H $ & 4.5 & 0.41 & 0.4 & 0.6 & 0.53 & 0.77 & 0.77 & 0.5 & 0.51 & 0.42 & 0.42 & 0.96 & 0.46 & 0.37 & 0.92 & 0.46 & 0.44 \\ 
 \text{} & \text{--} & 0.43 & 0.42 & 0.66 & 0.58 & 0.83 & 0.83 & 0.56 & 0.56 & 0.48 & 0.47 & 1.1 & 0.6 & 0.54 & \text{--} & \text{--} & 0.53 \\
 \hline
$\delta g^{\tau\tau}_H $ & 2.3 & 0.34 & 0.32 & 0.64 & 0.56 & 0.8 & 0.8 & 0.58 & 0.58 & 0.49 & 0.48 & 1.4 & 0.98 & 0.76 & 1.3 & 0.62 & 0.58 \\ 
 \text{} & \text{--} & 0.36 & 0.34 & 0.68 & 0.6 & 0.87 & 0.86 & 0.63 & 0.63 & 0.53 & 0.53 & 1.4 & 1. & 0.84 & \text{--} & \text{--} & 0.63 \\
 \hline
$\delta g^{\mu\mu}_H $ & 5.6 & 2.7 & 2.7 & 4.6 & 4.5 & 4.9 & 4.9 & 4.5 & 4.5 & 4. & 4. & 5.1 & 4.7 & 3.8 & 4.9 & 2.5 & 0.24 \\ 
 \text{} & \text{--} & 2.7 & 2.7 & 4.6 & 4.5 & 4.9 & 4.9 & 4.5 & 4.5 & 4. & 4. & 5.1 & 4.7 & 3.8 & \text{--} & \text{--} & 0.49 \\
 \hline
$\delta \Gamma_H $ & 6.7 & 0.47 & 0.44 & 0.82 & 0.69 & 1.1 & 1. & 0.62 & 0.62 & 0.46 & 0.46 & 1.4 & 0.6 & 0.45 & 1.5 & 0.7 & 0.63 \\ 
 \text{} & \text{--} & 0.61 & 0.59 & 1.1 & 0.98 & 1.5 & 1.5 & 1.1 & 1.1 & 0.94 & 0.93 & 2.3 & 1.6 & 1.6 & \text{--} & \text{--} & 1.3 \\
 \hline
$\delta g^{ee}_{Z,L} $ & 0.11 & 0.017 & 0.016 & 0.01 & 0.0083 & 0.036 & 0.027 & 0.03 & 0.023 & 0.028 & 0.023 & 0.061 & 0.051 & 0.046 & 0.095 & 0.085 & 0.085 \\ 
 \text{} & 0.11 & 0.017 & 0.016 & 0.01 & 0.0083 & 0.036 & 0.027 & 0.03 & 0.024 & 0.028 & 0.023 & 0.061 & 0.051 & 0.046 & 0.095 & 0.085 & 0.086 \\
 \hline
$\delta g^{ee}_{Z,R} $ & 0.12 & 0.019 & 0.019 & 0.0092 & 0.0085 & 0.036 & 0.027 & 0.028 & 0.023 & 0.023 & 0.02 & 0.06 & 0.041 & 0.037 & 0.11 & 0.11 & 0.11 \\ 
 \text{} & 0.12 & 0.02 & 0.019 & 0.0092 & 0.0085 & 0.036 & 0.027 & 0.028 & 0.023 & 0.023 & 0.02 & 0.06 & 0.041 & 0.038 & 0.11 & 0.11 & 0.11 \\
 \hline
$\delta g^{e\nu}_{W} $ & 0.65 & 0.01 & 0.0097 & 0.016 & 0.013 & 0.031 & 0.027 & 0.02 & 0.015 & 0.016 & 0.013 & 0.058 & 0.036 & 0.032 & 0.17 & 0.068 & 0.068 \\ 
 \text{} & 0.65 & 0.01 & 0.0097 & 0.016 & 0.013 & 0.031 & 0.027 & 0.02 & 0.015 & 0.016 & 0.013 & 0.058 & 0.036 & 0.032 & 0.18 & 0.068 & 0.068 \\
 \hline
$\delta g^{\mu\mu}_{Z,L} $ & 0.42 & 0.019 & 0.018 & 0.011 & 0.0085 & 0.071 & 0.028 & 0.07 & 0.025 & 0.07 & 0.024 & 0.19 & 0.19 & 0.19 & 0.23 & 0.12 & 0.12 \\ 
 \text{} & 0.42 & 0.019 & 0.018 & 0.011 & 0.0085 & 0.071 & 0.028 & 0.07 & 0.025 & 0.07 & 0.024 & 0.19 & 0.19 & 0.19 & 0.23 & 0.12 & 0.12 \\
 \hline
$\delta g^{\mu\mu}_{Z,R} $ & 0.55 & 0.019 & 0.019 & 0.0093 & 0.0086 & 0.076 & 0.028 & 0.075 & 0.026 & 0.075 & 0.026 & 0.23 & 0.23 & 0.23 & 0.23 & 0.027 & 0.027 \\ 
 \text{} & 0.55 & 0.019 & 0.019 & 0.0091 & 0.0086 & 0.076 & 0.028 & 0.075 & 0.026 & 0.075 & 0.026 & 0.23 & 0.23 & 0.23 & 0.23 & 0.027 & 0.027 \\
 \hline
$\delta g^{\mu\nu}_{W} $ & 0.6 & 0.013 & 0.012 & 0.019 & 0.018 & 0.044 & 0.039 & 0.038 & 0.033 & 0.035 & 0.032 & 0.1 & 0.087 & 0.083 & 0.068 & 0.035 & 0.034 \\ 
 \text{} & 0.6 & 0.013 & 0.012 & 0.019 & 0.018 & 0.044 & 0.039 & 0.038 & 0.033 & 0.035 & 0.032 & 0.1 & 0.087 & 0.083 & 0.069 & 0.035 & 0.035 \\
 \hline
$\delta g^{\tau\tau}_{Z,L} $ & 0.22 & 0.019 & 0.018 & 0.015 & 0.013 & 0.076 & 0.032 & 0.075 & 0.03 & 0.074 & 0.029 & 0.19 & 0.19 & 0.19 & 0.22 & 0.22 & 0.22 \\ 
 \text{} & 0.22 & 0.019 & 0.018 & 0.014 & 0.013 & 0.076 & 0.033 & 0.075 & 0.03 & 0.075 & 0.029 & 0.19 & 0.19 & 0.19 & 0.22 & 0.22 & 0.22 \\
 \hline
$\delta g^{\tau\tau}_{Z,R} $ & 0.27 & 0.019 & 0.019 & 0.015 & 0.015 & 0.08 & 0.032 & 0.079 & 0.031 & 0.079 & 0.03 & 0.22 & 0.22 & 0.22 & 0.26 & 0.26 & 0.26 \\ 
 \text{} & 0.27 & 0.02 & 0.02 & 0.015 & 0.015 & 0.081 & 0.032 & 0.079 & 0.031 & 0.079 & 0.031 & 0.22 & 0.22 & 0.22 & 0.26 & 0.26 & 0.26 \\
 \hline
$\delta g^{\tau\nu}_{W} $ & 0.79 & 0.013 & 0.012 & 0.019 & 0.018 & 0.044 & 0.039 & 0.038 & 0.033 & 0.035 & 0.032 & 0.1 & 0.087 & 0.083 & 0.18 & 0.068 & 0.068 \\ 
 \text{} & 0.79 & 0.013 & 0.013 & 0.019 & 0.018 & 0.044 & 0.039 & 0.038 & 0.033 & 0.035 & 0.032 & 0.1 & 0.087 & 0.083 & 0.18 & 0.068 & 0.068 \\
 \hline
$\delta g^{uu}_{Z,L} $ & 0.82 & 0.052 & 0.052 & 0.077 & 0.076 & 0.24 & 0.13 & 0.24 & 0.13 & 0.24 & 0.13 & 0.26 & 0.26 & 0.26 & 0.73 & 0.7 & 0.7 \\ 
 \text{} & 0.83 & 0.052 & 0.052 & 0.077 & 0.076 & 0.24 & 0.13 & 0.24 & 0.13 & 0.24 & 0.13 & 0.26 & 0.26 & 0.26 & 0.73 & 0.7 & 0.7 \\
 \hline
$\delta g^{uu}_{Z,R} $ & 3. & 0.071 & 0.071 & 0.084 & 0.084 & 0.32 & 0.14 & 0.31 & 0.14 & 0.31 & 0.14 & 0.39 & 0.39 & 0.39 & 2.9 & 2.9 & 2.9 \\ 
 \text{} & 3. & 0.071 & 0.071 & 0.084 & 0.084 & 0.32 & 0.14 & 0.31 & 0.14 & 0.31 & 0.14 & 0.39 & 0.39 & 0.39 & 2.9 & 2.9 & 2.9 \\
 \hline
$\delta g^{dd}_{Z,L} $ & 0.66 & 0.051 & 0.051 & 0.075 & 0.074 & 0.21 & 0.13 & 0.2 & 0.12 & 0.2 & 0.12 & 0.28 & 0.26 & 0.25 & 0.56 & 0.56 & 0.56 \\ 
 \text{} & 0.66 & 0.051 & 0.051 & 0.075 & 0.074 & 0.21 & 0.13 & 0.2 & 0.12 & 0.2 & 0.12 & 0.28 & 0.26 & 0.25 & 0.56 & 0.56 & 0.56 \\
 \hline
$\delta g^{dd}_{Z,R} $ & 19. & 1. & 1. & 1.5 & 1.4 & 3.6 & 2.9 & 3.1 & 2.3 & 3. & 2.2 & 6.8 & 6. & 5.8 & 15. & 14. & 14. \\ 
 \text{} & 19. & 1. & 1. & 1.5 & 1.4 & 3.6 & 2.9 & 3.1 & 2.3 & 3. & 2.2 & 6.8 & 6. & 5.8 & 15. & 14. & 14. \\
 \hline
$\delta g^{bb}_{Z,L} $ & 0.38 & 0.013 & 0.013 & 0.017 & 0.017 & 0.063 & 0.034 & 0.062 & 0.033 & 0.062 & 0.033 & 0.13 & 0.13 & 0.13 & 0.38 & 0.37 & 0.37 \\ 
 \text{} & 0.38 & 0.013 & 0.013 & 0.017 & 0.017 & 0.063 & 0.034 & 0.062 & 0.033 & 0.062 & 0.033 & 0.13 & 0.13 & 0.13 & 0.38 & 0.37 & 0.37 \\
 \hline
$\delta g^{bb}_{Z,R} $ & 11. & 0.16 & 0.16 & 0.16 & 0.16 & 0.49 & 0.3 & 0.49 & 0.3 & 0.49 & 0.3 & 2.8 & 2.8 & 2.8 & 11. & 10. & 10. \\ 
 \text{} & 11. & 0.16 & 0.16 & 0.16 & 0.16 & 0.49 & 0.3 & 0.49 & 0.3 & 0.49 & 0.3 & 2.8 & 2.8 & 2.8 & 11. & 10. & 10. \\ \hline
  \end{tabular}
  \end{adjustbox}
\caption{%\jb{UPDATED Oct 4}
Precision reach (in percentage) on effective couplings from a SMEFT global analysis of the Higgs and EW measurements at various future colliders listed in \autoref{tab:epem_setup}. For each coupling, the first (second) row shows the results from the constrained-$\Gamma_H$ (free-$\Gamma_H$) fit.  The results match those in \autoref{fig:fit1}.
}
\label{tab:fit1}
\end{table}

\subsection{Impact of the Standard Model uncertainties on the results\label{sec:SMuncFit1}}

The previous results have been obtained from a fit to the experimental predictions presented in Section~\ref{sec:input}. On the theory side, we considered the observable predictions from the SM, complemented with the dimension-6 SMEFT operators. For the SM predictions, and as it was done in \cite{deBlas:2019rxi}, we include the effects associated to the projected experimental uncertainties on the SM input parameters, the so-called {\it parametric uncertainties}. Apart from these, one needs to take into account that the current precision of SM theory calculations, e.g. known in general at the 2-loop level for the EWPO, may not be enough to match the projected experimental precision of the different future measurements, i.e. the uncertainty associated to missing higher-order corrections, typically referred as {\it intrinsic theory uncertainties} may be a limiting factor. A lot of work has been dedicated to establish the theory requirements that would be needed so the theory calculations do not hinder the interpretation of the different precision measurements at a future $e^+ e^-$ EW/Higgs factories. (See, e.g., \cite{Blondel:2019qlh}.) In this subsection we quantify the impact of such intrinsic uncertainties from both {\it current} calculations and those projected to be available by the time a future $e^+ e^-$ collider operates.

A summary of the current and future intrinsic uncertainties for EWPO are given in Table~\ref{tab:EWPOUnc}. These estimates have been obtained in \cite{Awramik:2003rn,Dubovyk:2019szj,Freitas:2019bre}. Parametric errors are also shown in that table for a benchmark of precision for the SM inputs. In the actual fits, these are assigned to each observable at each collider according to the uncertainty of the corresponding SM input in Table~\ref{tab:EWPO}. For the strong coupling constant at the $Z$ pole and the Top mass, both missing in that table, we assume the following: a) an independent determination of $\alpha_s(M_Z)$ from lattice QCD will bring up a determination with an uncertainty $\sim 0.0002 $; b) the HL-LHC will be able to measure $m_t$ with an uncertainty of the order of 400 MeV, which would be reduced at a future $e^+e^-$ factory running at the $t\bar{t}$ threshold down to $\sim 20$ MeV.

For single Higgs production, following \cite{Freitas:2019bre}, we assume the current theory uncertainty for $e^+e^- \to ZH$ and $e^+ e^- \to \bar{\nu}\nu H$ via $W$ boson fusion is of $O(1\%)$, due to the missing 2-loop effects.\footnote{The two-loop corrections to $e^+e^- \to ZH$ have been recently computed in \cite{Freitas:2022hyp,Chen:2022mre}.} With the full 2-loop calculation for the $ZH$ process, the uncertainty is expected to be reduced to $\lesssim 0.3\%$, whereas in the more complicated case of $W$ boson fusion, a partial result could bring the uncertainty below $1\%$. The uncertainties for Higgs decays, also from \cite{Freitas:2019bre}, are summarized in table~\ref{tab:widthsUnc}. 

Finally, for the $e^+ e^- \to W^+W^-$ process, we used our own projections obtained via the optimal observable method. Unfortunately, there are no estimates available for the SM theory uncertainties in this case.

\begin{table}[ht]
\caption{\label{tab:EWPOUnc} 
Current and future (absolute) theory uncertainties in the SM predictions for different EWPO. Future parametric uncertainties correspond to $\Delta m_H=10$ MeV, $\Delta m_t=20$ MeV, $\Delta \alpha_s(m_Z)=0.0002$, $\Delta m_Z=0.1$ MeV and two uncertainties for $\Delta \alpha(m_Z)^{-1}=17.8/3.2$. The latter has a particular impact in the uncertainties of the $W$ mass and the effective weak mixing angle. Current parametric uncertainties from \cite{deBlas:2021wap}.
}
\begin{center}
\begin{tabular}{ c|c c|c c}
\toprule
EWPO & \multicolumn{2}{c|}{current unc. $\Delta O$} & \multicolumn{2}{c}{future unc. $\Delta O$}\\
    & Th$_\mtxt{Intr}$ 
    & Th$_\mtxt{Par}$ %& Th$_\mtxt{Par}^{\alpha_s}$ & Th$_\mtxt{Par}^{m_H}$
    & Th$_\mtxt{Intr}$ 
    & Th$_\mtxt{Par}$ %& Th$_\mtxt{Par}^{\alpha_s}$ & Th$_\mtxt{Par}^{m_H}$ 
    \\     
\midrule
$M_W$ [MeV] & $4$  & $4.2$ %& $$ & $$
& $1$  & $2.4/0.6$ %& $$ & $$
\\
$\sin^2{\theta_W}$ & $5\cdot 10^{-5}$ & $4\cdot 10^{-3}$ %& $$ & $$
& $1.5\cdot 10^{-5}$ & $4.5\cdot 10^{-5}/ 10^{-5}$ %& $$ & $$
\\
$\Gamma_Z$ [MeV] & $0.4$ & $0.6$ %& $$ & $$
& $0.15$  & $0.16/0.1$ %& $$ & $$
\\
$\sigma_{\mtxt{had}}^0$ [pb] & $6$  & $5.3$ %& $$ & $$ 
& n/a  & $1/1$ %& $$ & $$
\\
$R_\ell^0$ & $6\cdot 10^{-3}$  & $6.3\cdot 10^{-3}$ %& $$ & $$ 
& $1.5\cdot 10^{-3}$  & $1.5\cdot 10^{-3}/1.2\cdot 10^{-3}$ %& $$ & $$
\\
$R_c^0$ & $5\cdot 10^{-5}$  & $2\cdot 10^{-5}$ %& $$ & $$
& n/a  & $4.7\cdot 10^{-6}/3.9\cdot 10^{-6}$ %& $$ & $$
\\
$R_b^0$ & $11\cdot 10^{-5}$  & $2\cdot 10^{-5}$ %& $$ & $$
& $5\cdot 10^{-5}$  & $2.8\cdot 10^{-6}/2.3\cdot 10^{-6}$ %& $$ & $$
\\
\bottomrule
\end{tabular}
\end{center}
\end{table}
\begin{table}[ht]
\caption{\label{tab:widthsUnc} 
Current and future (relative) uncertainties in the SM predictions for the different Higgs decay channels. The future parametric uncertainties correspond to an assumed precision of $\Delta m_b=13$~MeV, $\Delta m_c=7$~MeV, $\Delta m_t=50$~MeV, $\Delta \alpha_s=0.0002$ and $\Delta m_H=10$~MeV. 
}
\begin{center}
\begin{tabular}{ c|cccc|cccc}
\toprule
Decay & \multicolumn{4}{c|}{current unc. $\delta \Gamma$ [\%]} & \multicolumn{4}{c}{future unc. $\delta \Gamma$ [\%]}\\
    & Th$_\mtxt{Intr}$ & Th$_\mtxt{Par}^{m_q}$ & Th$_\mtxt{Par}^{\alpha_s}$ & Th$_\mtxt{Par}^{m_H}$& Th$_\mtxt{Intr}$ & Th$_\mtxt{Par}^{m_q}$ & Th$_\mtxt{Par}^{\alpha_s}$ & Th$_\mtxt{Par}^{m_H}$ \\     
\midrule
$H\to b\bar{b}$ & $<0.4$ & $1.4$ & $0.4$ & $-$ & $0.2$ & $0.6$ & $<0.1$ & $-$\\
$H\to \tau^+\tau^-$ & $<0.3$ & $-$ & $-$ & $-$& $<0.1$ & $-$ & $-$ & $-$\\\
$H\to  c\bar{c} $ & $<0.4$ & $4.0$ & $0.4$ & $-$& $0.2$ & $1.0$ & $<0.1$ & $-$\\
$H\to \mu^+\mu^-$ & $<0.3$ & $-$ & $-$ & $-$ & $<0.1$ & $-$ & $-$ & $-$ \\
$H\to W^+W^-$ & $0.5$ & $-$ & $-$ & $2.6$ & $0.3$ & $-$ & $-$ & $0.1$\\
$H\to gg$ & $3.2$ & $<0.2$ & $3.7$ & $-$ & $1.0$ & $-$ & $0.5$ & $-$\\
$H\to ZZ$ & $0.5$ & $-$ & $-$ & $3.0$ & $0.3$ & $-$ & $-$ & $0.1$\\
$H\to \gamma\gamma$ & $<1.0$ & $<0.2$ & $-$ & $-$ & $<1.0$ & $-$ & $-$ & $-$\\
$H\to Z\gamma$ & $5.0$ & $-$ & $-$ & 2.1 & $1.0$ & $-$ & $-$ & $0.1$\\
\bottomrule
\end{tabular}
\end{center}
\end{table}

In order to quantify the impact of these SM uncertainties we consider the results derived from a series of fits analogous to the one presented above in different scenarios:  
\begin{enumerate}[label=(\alph*)]
\item{including only the SM parametric uncertainties, denoted as SM$_{\rm Param.}$;}
\item{ including the SM parametric and {\it future} intrinsic uncertainties, denoted as SM$_{\rm Full(Future)}$;}
\item{ including the SM parametric and {\it current} intrinsic uncertainties, denoted as SM$_{\rm Full(Current)}$.}
\end{enumerate}
These are to be compared with another scenario:
\begin{enumerate}[label=(\alph*)]
  \setcounter{enumi}{3}
\item{ {\it ignoring all} the SM theory uncertainties, which we denote as the {\it No Error} scenario.}
\end{enumerate}

In Figure~\ref{fig:THunc}, we show the deterioration of the results with respect to last scenario without any SM uncertainty, by presenting the ratios of the uncertainties $\delta g^{(a,b,c)}/\delta g_{\rm No~Error}$ for the different Higgs and electroweak couplings, respectively. Given the lack of estimates for the SM theory uncertainties in $e^+ e^- \to W^+ W^-$ we chose not to show the results for the aTGC. One must note that, due to the connection between aTGC and the Higgs in the dimension-6 SMEFT formalism, the results for the determination of the Higgs couplings to vector bosons may also be affected by extra uncertainties in the $W^+ W^-$ process, and therefore, the numbers presented here should be taken as an optimistic estimate of the effect of the theory uncertainties.

\begin{figure}%[t]
\centering
    \includegraphics[width=\textwidth]{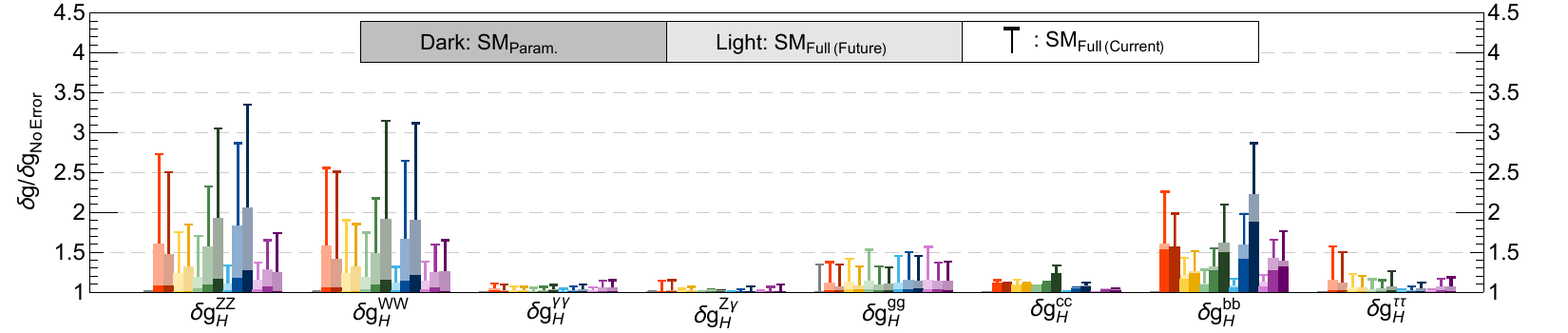}\\[1cm]
    \includegraphics[width=\textwidth]{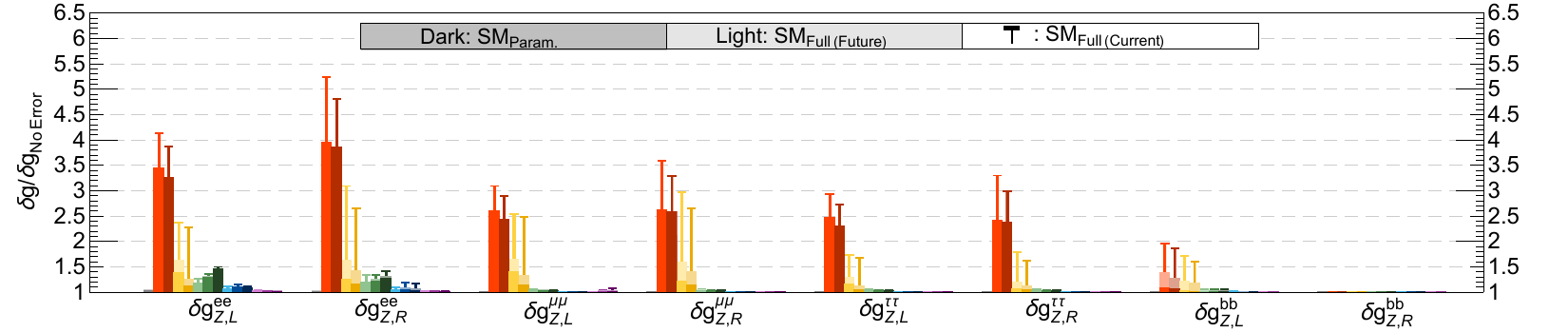}
    \caption{
    Impact of the parametric and intrinsic theory errors on the determination of the effective Higgs and EW couplings from the SMEFT fit. The impact is plotted in terms of the ratio between the uncertainties, $\delta g$, obtained when a given source of SM theory errors is included in the global fit and the ones derived in the case when these are not included, $\delta g_{\rm No~Error}$. The results indicated by the dark bars assume only parametric errors are included; in the light bars both parametric and intrinsic errors as projected in the future are included; finally, the thin ``T'' lines denote the case when future parametric errors and {\it current} intrinsic theory uncertainties are included.}
    \label{fig:THunc}
\end{figure}

As it is apparent from the results in the top panel of Figure \ref{fig:THunc}, and comparing with the uncertainties in Table~\ref{tab:fit1} assuming the future theory calculations would be sufficient for a determination of the couplings to vector bosons at the level of 0.2-0.3 percent. Below this level of precision, the determination of the $HZZ$ and $HWW$ couplings would be somewhat limited by the theory uncertainties. As also noted in \cite{deBlas:2019rxi}, the effect of parametric errors is relatively small, with perhaps the exception of $m_b$.  Focusing the attention in the bottom panel of Figure \ref{fig:THunc}, and looking at the case of circular colliders, we observe that, in particular, the inclusion of the SM uncertainties has a strong impact on the determination of the electroweak couplings of leptons, especially if no progress in the theory calculations is made. This is due to the higher precision that is expected for these couplings coming from the measurements at the future $Z$-pole runs, which then also requires of higher theoretical precision to take advantage of such experimental measurements. The second thing that is noticeable is the large impact of parametric uncertainties on the CEPC results compared to the FCCee ones. In this case, this is mostly attributed to the different expected precision for $\alpha(m_Z)$, where the FCCee plans to run slightly below and above the $Z$ pole to determine this parameter with improved accuracy~\cite{Janot:2015gjr}. Assuming similar measurements are performed at the CEPC would reduce significantly the impact of the parametric errors in the determination of the leptonic electroweak couplings.

\subsection{Relaxing $U(2)$ assumption \label{sec:noU2}}

As mentioned at the beginning of this section, using the future projections for the EWPO available in the literature, there is very little discriminating power between contributions from the electroweak couplings of the first family of quarks. Hence, in the fits presented here we assumed a $U(2)$ symmetry that effectively sets the first and second family quark couplings to the same value, allowing to close the fits without any flat direction. 
Indeed, as shown in \cite{Breso-Pla:2021qoe}, even though current $Z$-pole observables provide measurements that can be sensitive to the charm and even strange quark couplings separately, the traditional LEP/SLD EWPO are blind to a particular combination of up and down- quark couplings. 
To break this assumption we need to use an observable measurement that can be particularly sensitive to contributions from the lighter quark families. (Note also that there are currently no future $Z$ pole projections for strange quark observables, which complicates giving sensible estimates for the sensitivity to modifications of the corresponding couplings.)
In \cite{Breso-Pla:2021qoe} it was shown that such observable can be provided by the LHC measurement of the forward-backward asymmetry in the Drell-Yan process around the $Z$ pole, as a function of the rapidity of the dilepton system. This by itself cannot constrain all the $Zqq$ interactions but is able to do so in combination with $Z$-pole measurements.

HL-LHC projections for measurements using the Drell-Yan process are available as estimates for the sensitivity to the effective weak mixing angle $\sin^2{\theta_{\rm Eff}^{\rm lept}}$ \cite{ATLAS:2018qvs,CMS:2017vxj}, in which the projections of the forward-backward asymmetry ($A_{\rm FB}$) measurements are also presented. Here we use the ATLAS projections \cite{ATLAS:2018qvs} of the $A_{\rm FB}$ measurements in the $e^+e^-$ channel, which are divided into three sets with different rapidity ($|Y_{ee}|=[0.8-1.2]$, $[2.4-2.8]$, $[3.2-3.6]$), each presented for an invariant mass ($m_{ee}$) range of 60-200\,GeV.   We include all three rapidity sets in our analysis, and use only the two invariant mass bins of 80-90\,GeV and 90-100\,GeV for each set, in order to focus on the Z-pole effects and avoid large contamination from possible 4-fermion interactions.  For simplicity, we consider only the contributions from $g_{Z,L/R}^{uu,dd}$, assuming all other contributions of the dimension-6 operators are sufficiently well constrained by the Z-pole measurements at lepton colliders, following the treatment in \cite{Breso-Pla:2021qoe}. 

In \autoref{fig:afb1} we illustrate the impact of this hadron-collider measurement on the determination of the up and down quark coupling by showing the $\Delta \chi^2=1$ bounds on different planes of the $g_{Z,L/R}^{uu,dd}$ space from the FCC-ee fit, with and without our estimate for the sensitivity to the forward-backward asymmetry at the HL-LHC. Indeed, in the FCC-ee alone fit one observes the presence of the flat direction. Using the $A_{\rm FB}$ measurement across 3 rapidity sets and 2 invariant mass bins, one can close this in the global fit.  Furthermore, given the different correlations of the $Z$-pole flat direction and the $A_{FB}$ constraints, the combined bound can significantly improve the sensitivity beyond that which would be possible using the asymmetry alone, as shown in the upper-left panel. As discussed at the beginning, however, we prefer to leave this observable outside of the main fit presented in this section until proper HL-LHC estimates from ATLAS and/or CMS are available.

\begin{figure}%[t]
\centering
    \includegraphics[width=0.8\textwidth]{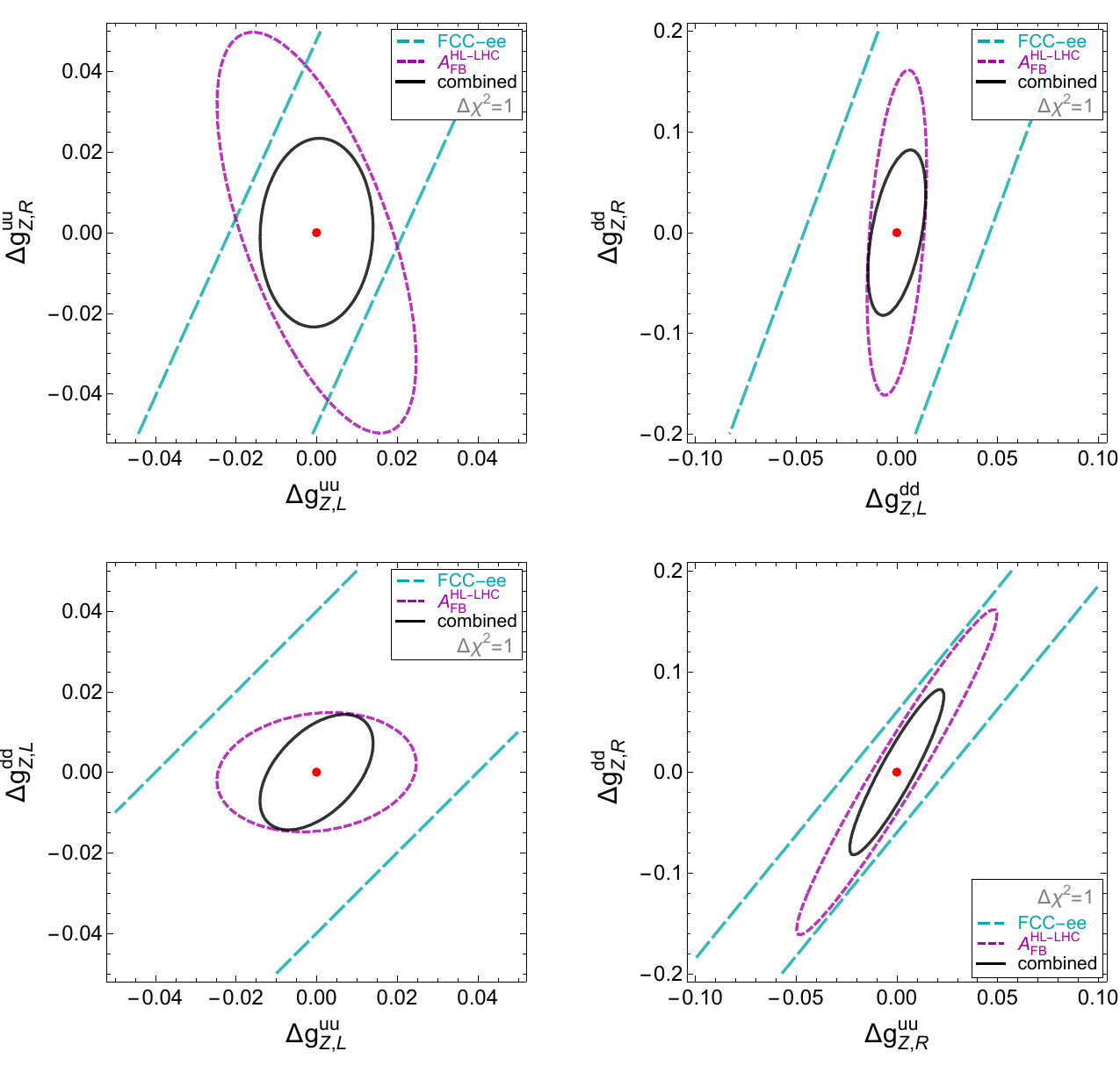}
    \caption{
    Marginalized $\Delta \chi^2=1$ bounds for $\Delta g_{Z,L/R}^{uu,dd}$ from the $A_{\rm FB}$ measurements at HL-LHC with ATLAS projection \cite{ATLAS:2018qvs} and the FCC-ee-only Higgs + EW fit.  Note that the couplings are not normalized to SM values. 
    }
    \label{fig:afb1}
\end{figure}

%================================================================
%!!!!!!!!!!!!!!!!!!!!!!!!!!!!!!!!!!!!!!!!!!!!!!!!!!!!!!!!!!!!!!!
%================================================================

%%%%%%%%%%%%%%%%%%%%%%%%%%%%%%%%%%
\section{Four-fermion operators}
%%%%%%%%%%%%%%%%%%%%%%%%%%%%%%%%%%
\label{sec:4f}
%[editor: Yong, Michael]

In this section we focus our attention on the sector of electroweak interactions, but extend the fit beyond electroweak precision measurements, considering $2 \to 2$ fermion processes above the pole as well as measurements at low energy experiments.
In the SMEFT, describing such a set of observables requires to consider not only the operators modifying the fermion electroweak couplings, as in the previous section, but also the corresponding set of four-fermion operators modifying these processes. For the different (projected) measurements, described in what follows, we will need to include  the leptonic and semi-leptonic four-fermion operators in Eq.~(\ref{eq:LSMEFT4flf}) and summarized again in Table~\ref{tab:4fope}.

\begin{table}
\centering{
\begin{tabular}{l|l}
\hline\hline
$2\ell2q$ operators ($p,r = 1,2,3$) & $4\ell$ operators ($p < r =1,2,3$)\\
\hline
\hspace{1.4cm}Chirality conserving  &  \hspace{1.4cm}Two flavors\\ 
\hline 
$ [\mathcal{O}_{\ell q}]_{pprr} = (\bar \ell_p\bar \sigma_\mu \ell_p)  (\bar q_r \bar \sigma^\mu q_r)$
 &  $ [\mathcal{O}_{\ell \ell}]_{pprr}  =  (\bar \ell_p\bar \sigma_\mu \ell_p)  (\bar \ell_r \bar \sigma^\mu \ell_r) $ \\ 
$ [O^{(3)}_{\ell q}]_{pprr} = (\bar \ell_p \bar \sigma_\mu \sigma^i \ell_p)  (\bar q_r \bar \sigma^\mu \sigma^i q_r)$ 
&  $[\mathcal{O}_{\ell \ell}]_{prrp} = (\bar \ell_p \bar \sigma_\mu \ell_r)  (\bar \ell_r \bar \sigma^\mu \ell_p)  $\\   
$ [\mathcal{O}_{\ell u}]_{pprr} = (\bar \ell_p\bar \sigma_\mu \ell_p)  (u^c_r \sigma^\mu \bar u^c_r)$ 
& $ [\mathcal{O}_{\ell e}]_{pprr}  =  (\bar \ell_p\bar \sigma_\mu \ell_p)  (e_r^c  \sigma^\mu \bar e_r^c)$\\
$ [\mathcal{O}_{\ell d}]_{pprr} = (\bar \ell_p\bar \sigma_\mu \ell_p)  (d^c_r \sigma^\mu \bar d^c_r)$ 
& $[\mathcal{O}_{\ell e}]_{rrpp}  =  (\bar \ell_r \bar \sigma_\mu \ell_r)  (e_p^c  \sigma^\mu \bar e_p^c)$ \\
$ [\mathcal{O}_{e q}]_{pprr} = (e^c_p \sigma_\mu \bar e^c_p)  (\bar q_r \bar \sigma^\mu q_r)$ 
& $[\mathcal{O}_{\ell e}]_{prrp}  =  (\bar \ell_p \bar \sigma_\mu \ell_r)  (e_r^c  \sigma^\mu \bar e_p^c)$ \\ 
$ [\mathcal{O}_{e u}]_{pprr} = (e^c_p \sigma_\mu \bar e^c_p)  (u^c_r \sigma^\mu \bar u^c_r)$ 
& $ [\mathcal{O}_{e e}]_{pprr}  =   (e_p^c  \sigma_\mu \bar e_p^c)   (e_r^c  \sigma^\mu \bar e_r^c) $\\ 
$ [\mathcal{O}_{e d}]_{pprr} = (e^c_p \sigma_\mu \bar e^c_p)  (d^c_r \sigma^\mu \bar d^c_r)$ & \\ 
\hline
\hspace{1.4cm}Chirality violating  & \hspace{1.4cm}One flavor\\
\hline
$ [\mathcal{O}_{\ell e q u}]_{pprr}  =  (\bar \ell^j_p \bar e_p^c) \epsilon_{jk} (\bar q^k_r \bar u^c_r) $ & $ [\mathcal{O}_{\ell \ell}]_{pppp} = {1\over 2} (\bar \ell_p\bar \sigma_\mu \ell_p)  (\bar \ell_p \bar \sigma^\mu \ell_p)$ \\
$ [O^{(3)}_{\ell e q u}]_{pprr}  =  (\bar \ell^j_p \bar \sigma_{\mu \nu} \bar e_p^c) \epsilon_{jk} (\bar q^k_r  \bar \sigma_{\mu \nu} \bar u^c_r) $ & $ [\mathcal{O}_{\ell e}]_{pppp} =  (\bar \ell_p\bar \sigma_\mu \ell_p)  (e_p^c  \sigma^\mu \bar e_p^c) $ \\
$ [\mathcal{O}_{\ell e d q}]_{pprr}  =  (\bar \ell^j_p \bar e_p^c)  (d^c_r q^j_r) $  &  $ [\mathcal{O}_{e e}]_{pppp} =   {1\over 2} (e_p^c  \sigma_\mu \bar e_p^c)   (e_p^c  \sigma^\mu \bar e_p^c) $ \\
\hline\hline
\end{tabular}
}
\caption{4-fermion operators included in this study. We use the notations in \cite{Grzadkowski:2010es} but rename $\mathcal{O}_{qe}$ as $\mathcal{O}_{eq}$ so the lepton bilinear product always appears first. In each operator, $p,r$ represent the flavor indices and $j,k$ the isospin indices.}\label{tab:4fope}
\end{table}

%%%%%%%%%%%%%%%%%%%%%%%
\subsection{Observables}
%%%%%%%%%%%%%%%%%%%%%%%

%%%%%%%%%%
\subsubsection{$Z$- and $W$-pole observables}
%%%%%%%%%%
The $Z$- and $W$-pole observables that are sensitive to the electroweak vertex corrections in \autoref{eq:vff} have been studied in \cite{Efrati:2015eaa} without any flavor assumption. We summarize the numerical results for these pole observables in \autoref{tab:zwpoleobs} with all the SM predictions taken from the best fit values from GFitter \cite{Baak:2014ora} except that the leptonic branching ratios for $W$ are taken from \cite{ALEPH:2013dgf}. For the experimental results, we use the data and the corresponding correlation matrices, whenever possible, from LEP-1 \cite{ALEPH:2005ab} for the $Z$-pole observables. We also include $A_s$ from SLD \cite{SLD:2000jop} and the very precise observable $R_{uc}$ from PDG \cite{ParticleDataGroup:2012pjm} in our fit. While for the $W$-pole observables, we use the data from LEP-2 \cite{ALEPH:2013dgf} for the leptonic $W$ branching ratios. The $W$ mass is taken from \cite{Group:2012gb} by CDF and D0, and with averaged $W$ boson decay width from \cite{ParticleDataGroup:2012pjm}. For $W$ couplings to quarks, we also include $R_{Wc}$ from \cite{ParticleDataGroup:2012pjm} and $R_\sigma$ from the CMS collaboration \cite{CMS:2014mgj}. In particular, the latter observable can be used to constrain ${\delta} g_{Z,L}^{tt}$, while ${\delta} g_{Z,R}^{tt}$ would remain free until top data is included as done in \autoref{sec:topfit}. In addition, two well-known flat directions exist for a global fit using only these pole observables. These flat directions can be lifted by less well-measured observables. To that end, we choose the model-independent $Z$-quark coupling measurements from D0\cite{D0:2011baz} as well as the forward-backward asymmetry measurements (projection) from the LHC (HL-LHC).

\begin{table}
\begin{center}
\begin{tabular}{|c|c|c|c|}
\hline
{Observable} & {Experimental value}   &   {Ref.}   &  {SM prediction}
  \\  \hline
$\Gamma_{Z}$ [GeV]  & $2.4952 \pm 0.0023$ & \cite{ALEPH:2005ab} & $ 2.4950$
 \\  \hline
$\sigma_{\rm had}$ [nb]  & $41.541\pm 0.037$ &\cite{ALEPH:2005ab} &  $41.484$
  \\  \hline
 $R_{e}$  & $20.804\pm 0.050$ & \cite{ALEPH:2005ab} &  $20.743$
   \\  \hline
 $R_{\mu}$  & $20.785 \pm 0.033$ & \cite{ALEPH:2005ab} &  $20.743$
   \\  \hline
 $R_{\tau}$  & $20.764\pm 0.045$ & \cite{ALEPH:2005ab} &  $20.743$
 \\  \hline
 $A_{\rm FB}^{0,e}$ & $0.0145\pm 0.0025$ &\cite{ALEPH:2005ab} &  $0.0163$
  \\  \hline
 $A_{\rm FB}^{0,\mu}$ & $0.0169\pm 0.0013$ &\cite{ALEPH:2005ab} &  $0.0163$
  \\  \hline
 $A_{\rm FB}^{0,\tau}$ & $0.0188\pm 0.0017$ &\cite{ALEPH:2005ab} &  $0.0163$
  \\ \hline
$R_b$ & $0.21629\pm0.00066$ & \cite{ALEPH:2005ab} & $0.21578$
   \\  \hline
$R_c$ & $0.1721\pm0.0030$  & \cite{ALEPH:2005ab}  & $0.17226$
\\  \hline
$A_{b}^{\rm FB}$ & $0.0992\pm 0.0016$ & \cite{ALEPH:2005ab}  & $0.1032$
 \\  \hline
 $A_{c}^{\rm FB}$ & $0.0707\pm 0.0035$  & \cite{ALEPH:2005ab} &  $0.0738$
  \\ \hline
 $A_e$ & $0.1516 \pm 0.0021$ &\cite{ALEPH:2005ab} &  $0.1472$
 \\ \hline
  $A_\mu$ & $0.142 \pm 0.015$ &\cite{ALEPH:2005ab} &  $0.1472$
 \\ \hline
 $A_\tau$ & $0.136 \pm 0.015$ &\cite{ALEPH:2005ab} &  $0.1472$
 \\ \hline
  $A_e$ & $0.1498 \pm 0.0049$ & \cite{ALEPH:2005ab} &  $0.1472$
 \\ \hline
 $A_\tau$ & $0.1439 \pm  0.0043$ & \cite{ALEPH:2005ab} &  $0.1472$
 \\ \hline
 $A_b$ & $0.923\pm 0.020$ & \cite{ALEPH:2005ab} & $0.935$
 \\  \hline
$A_c$ & $0.670 \pm 0.027$ & \cite{ALEPH:2005ab} & $0.668$
 \\ \hline
$A_s$ & $0.895 \pm 0.091$ & \cite{SLD:2000jop} & $0.935$
\\ \hline
$R_{uc}$ & $0.166 \pm 0.009$  & \cite{ParticleDataGroup:2012pjm}  & $0.1724$
\\ \hline
 $m_{W}$ [GeV]  & $80.385 \pm 0.015$ &\cite{Group:2012gb}    &  $80.364$
\\ \hline
$\Gamma_{W}$ [GeV]  & $ 2.085 \pm 0.042$  & \cite{ParticleDataGroup:2012pjm} &  $2.091$
\\ \hline
${\rm Br} (W \to e \nu)$ & $ 0.1071 \pm 0.0016$ &\cite{ALEPH:2013dgf} &  $0.1083$
\\ \hline
${\rm Br} (W \to \mu \nu)$ & $ 0.1063 \pm 0.0015$ &\cite{ALEPH:2013dgf} &  $0.1083$
\\ \hline
${\rm Br} (W \to \tau \nu)$ & $ 0.1138 \pm 0.0021$ &\cite{ALEPH:2013dgf} &  $0.1083$
 \\ \hline
 $R_{Wc}$ & $ 0.49 \pm 0.04$ & \cite{ParticleDataGroup:2012pjm}  &  $0.50$
 \\ \hline
 $R_{\sigma}$ & $0.998 \pm 0.041$  & \cite{CMS:2014mgj} & 1.000
  \\ \hline
\end{tabular}
\end{center}
\caption{$Z$ and $W$-pole observables used in the fit. $A_{e,\tau}$ appear twice in the above table: The first numbers are from SLC, and the second ones are from $\tau$ polarization at LEP-1.}
\label{tab:zwpoleobs}
 \end{table}

%%%%%%%%%%
\subsubsection{High-energy observables for 4-fermion operators}
%%%%%%%%%%
To go beyond the vertex shifts, observables off the $Z$ pole need to be included. LEP-2 \cite{ALEPH:2006bhb,ALEPH:2013dgf} reported fermion pair production above the $Z$ pole, with a center of mass energy $\sqrt{s}$ scan ranging from 130\,GeV to 209\,GeV. For the leptonic final state, the production cross sections and the forward-backward asymmetries were measured for $\mu^+\mu^-$ and $\tau^+\tau^-$ final states, and the differential cross section for the $e^+e^-$ final state was reported. In contrast, for the quark final state, only the total production cross sections and the forward-backward asymmetries were measured. Note that since the LEP-2 experiment was run below the top pair threshold, the $t\bar{t}$ final state would not be produced on-shell. For reference, all these observables are summarized in \autoref{tab:fpairobs}, where we generically use $f(s)$ to reflect the fact that the corresponding observable is $s$-dependent, and the angle $\theta$ for Bhabha is the scattering angle with respect to the incoming $e^-$. These channels receive contributions from both the vertex shifts discussed above, and the 4-fermion operators enumerated in \autoref{tab:4fope}. As noticed in \cite{Falkowski:2015krw,Falkowski:2017pss}, the energy scan of LEP-2 only provides 5 distinct observables, i.e, $\sum\limits_{q\ne t}\sigma(qq)$, $\sigma(b\bar{b})$, $\sigma(c\bar{c})$, ${\sigma_{\rm FB}(b\bar{b})}/{\sum\limits_{q\ne t}\sigma(q\bar{q})}$, ${\sigma_{\rm FB}(c\bar{c})}/{\sum\limits_{q\ne t}\sigma(q\bar{q})}$, which would allow to probe only four different combinations of the $2\ell2q$ operators. For this reason, less precise measurements of the total cross sections and the forward-backward asymmetries of $c\bar{c}$ and $b\bar{b}$ at $\sqrt{s}=58$\,GeV from VENUS\cite{VENUS:1993pob} and TOPAZ\cite{TOPAZ:2000evx} are also included in our fit.

\begin{table}
\begin{center}
\begin{adjustbox}{max width = \textwidth}
\begin{tabular}{|c|c|c|c|}
\hline
{Observable} & {Experimental value}   &   {Ref.}   &  {SM prediction}
 \\  \hline  \hline
$\sigma(\mu^+\mu^-)$ & $f(s)$ &\cite{ALEPH:2013dgf}    &  $f(s)$\cite{ALEPH:2013dgf}\\
\hline
$\sigma(\tau^+\tau^-)$ & $f(s)$ &\cite{ALEPH:2013dgf}    &  $f(s)$\cite{ALEPH:2013dgf}\\
\hline
$\sum\limits_{q\ne t}\sigma(q\bar{q})$ & $f(s)$ &\cite{VENUS:1993pob,TOPAZ:2000evx,ALEPH:2013dgf}    &  $f(s)$\cite{VENUS:1993pob,TOPAZ:2000evx,ALEPH:2013dgf}\\
\hline
$\sigma(b\bar{b})$ & $f(s)$ &\cite{VENUS:1993pob,TOPAZ:2000evx,ALEPH:2006bhb}    &  $f(s)$\cite{VENUS:1993pob,TOPAZ:2000evx,ALEPH:2006bhb}\\
\hline
$\sigma(c\bar{c})$ & $f(s)$ &\cite{VENUS:1993pob,TOPAZ:2000evx,ALEPH:2006bhb}    &  $f(s)$\cite{VENUS:1993pob,TOPAZ:2000evx,ALEPH:2006bhb}\\
\hline
$\frac{\sigma_{\rm FB}(b\bar{b})}{\sum\limits_{q\ne t}\sigma(q\bar{q})}$ & $f(s)$ &\cite{VENUS:1993pob,TOPAZ:2000evx,ALEPH:2006bhb}    &  $f(s)$\cite{VENUS:1993pob,TOPAZ:2000evx,ALEPH:2006bhb}\\
\hline
$\frac{\sigma_{\rm FB}(c\bar{c})}{\sum\limits_{q\ne t}\sigma(q\bar{q})}$ & $f(s)$ &\cite{VENUS:1993pob,TOPAZ:2000evx,ALEPH:2006bhb}    &  $f(s)$\cite{VENUS:1993pob,TOPAZ:2000evx,ALEPH:2006bhb}\\
\hline
$A_{\rm FB}(\mu^+\mu^-)$ & $f(s)$ &\cite{ALEPH:2013dgf}    &  $f(s)$\cite{ALEPH:2013dgf}\\
\hline
$A_{\rm FB}(\tau^+\tau^-)$ & $f(s)$ &\cite{ALEPH:2013dgf}    &  $f(s)$\cite{ALEPH:2013dgf}\\
\hline
$\frac{d\sigma}{d\cos\theta}(\rm Bhabha)$ & $f(s,\cos\theta)$ &\cite{ALEPH:2013dgf}    &  $f(s,\cos\theta)$\cite{ALEPH:2013dgf}
\\ \hline
\end{tabular}
\end{adjustbox}
\end{center}
\caption{Observables for fermion-pair production at lepton colliders. Here, we use $f(s)$ and $f(s,\cos\theta)$ to reflect the fact that the corresponding observables depend on the center of mass energy $\sqrt{s}$ and the scattering angle $\theta$ that is with respect to the incoming $e^-$.}
\label{tab:fpairobs}
\end{table}

%%%%%%%%%%
\subsubsection{Low-energy precision observables}
%%%%%%%%%%
\begin{table}
\begin{center}
\begin{adjustbox}{max width = \textwidth}
\begin{tabular}{|c|c|c|c|c|}
\hline
{Process} & {Observable} & {Experimental value}   &   {Ref.}   &  {SM prediction}
 \\  \hline  \hline
\multirow{2}{*}{$\stackrel{(-)}{\nu}_\mu-e^-$ scattering} & $g_{LV}^{\nu_\mu e}$ & $-0.035\pm0.017$ &\multirow{2}{*}{CHARM-II\cite{CHARM-II:1994dzw}}    &  $-0.0396$\cite{Erler:2013xha}\\
 & $g_{LA}^{\nu_\mu e}$ & $-0.503\pm0.017$  &   &  $-0.5064$\cite{Erler:2013xha}\\\hline
\multirow{2}{*}{$\tau$ decay} & $\frac{G_{\tau e}^2}{G_F^2}$ & $1.0029\pm0.0046$ & \multirow{2}{*}{PDG2014\cite{ParticleDataGroup:2014cgo}}    &  \multirow{2}{*}{$1$}\\ %PDG
 & $\frac{G_{\tau \mu}^2}{G_F^2}$ & $0.981\pm0.018$  &  & \\\hline
\multirow{6}{*}{Neutrino scattering} & $R_{\nu_\mu}$ & $0.3093\pm0.0031$ & \multirow{2}{*}{CHARM ($r=0.456$)\cite{CHARM:1987pwr}} & 0.3156\cite{CHARM:1987pwr} \\
 & $R_{\bar{\nu}_\mu}$ & $0.390\pm0.014$ & & 0.370\cite{CHARM:1987pwr} \\
 \cline{2-5}
 & $R_{\nu_\mu}$ & $0.3072\pm0.0033$ & \multirow{2}{*}{CDHS ($r=0.393$)\cite{Blondel:1989ev}}  & 0.3091\cite{Blondel:1989ev}\\
 & $R_{\bar{\nu}_\mu}$ & $0.382\pm0.016$ & & 0.380\cite{Blondel:1989ev}\\
 \cline{2-5}
  & $\kappa$ & $0.5820\pm0.0041$ & CCFR\cite{CCFR:1997zzq} & 0.5830\cite{CCFR:1997zzq}\\
 \cline{2-5}
  & $R_{\nu_e\bar{\nu}_e}$ & $0.406^{+0.145}_{-0.135}$ & CHARM\cite{CHARM:1986vuz} & 0.33\cite{ParticleDataGroup:2016lqr}\\
\hline
 \multirow{9}{*}{Parity-violating scattering} & $(s^2_w)^{\rm M\o ller}$ & $0.2397\pm0.0013$ &SLAC-E158\cite{SLACE158:2005uay}    &  $0.2381\pm0.0006$\cite{Czarnecki:1995fw}\\
 \cline{2-5}
 & $Q_W^{\rm Cs}(55,78)$ & $-72.62\pm0.43$ & PDG2016\cite{ParticleDataGroup:2016lqr} & $-73.25\pm0.02$\cite{ParticleDataGroup:2016lqr} \\
 \cline{2-5}
 & $Q_W^{\rm p}(1,0)$ & $0.064\pm0.012$ & QWEAK\cite{Qweak:2013zxf} & $0.0708\pm0.0003$\cite{ParticleDataGroup:2016lqr} \\
 \cline{2-5}
 & $A_1$ & $(-91.1\pm4.3)\times10^{-6}$ & \multirow{2}{*}{PVDIS\cite{PVDIS:2014cmd}} & $(-87.7\pm0.7)\times10^{-6}$\cite{PVDIS:2014cmd} \\
 & $A_2$ & $(-160.8\pm7.1)\times10^{-6}$ & & $(-158.9\pm1.0)\times10^{-6}$\cite{PVDIS:2014cmd}\\
 \cline{2-5}
 & \multirow{2}{*}{$g_{VA}^{eu}-g_{VA}^{ed}$} & $-0.042\pm0.057$ & SAMPLE ($\sqrt{Q^2}=200$\,MeV)\cite{Beise:2004py} & -0.0360\cite{ParticleDataGroup:2016lqr}\\
 & & $-0.12\pm0.074$ & SAMPLE ($\sqrt{Q^2}=125$\,MeV)\cite{Beise:2004py} & 0.0265\cite{ParticleDataGroup:2016lqr}\\
 \cline{2-5}
 & \multirow{2}{*}{$b_{\rm SPS}$} & $-(1.47\pm0.42)\times10^{-4}\rm\,GeV^{-2}$ & SPS $(\lambda=0.81)$\cite{Argento:1982tq} & $-1.56\times10^{-4}\rm\,GeV^{-2}$\cite{Argento:1982tq} \\
 & & $-(1.74\pm0.81)\times10^{-4}\rm\,GeV^{-2}$ & SPS $(\lambda=0.66)$\cite{Argento:1982tq} & $-1.57\times10^{-4}\rm\,GeV^{-2}$\cite{Argento:1982tq} \\
\hline
\multirow{2}{*}{$\tau$ polarization} & $\mathcal{P}_\tau$ & $0.012\pm0.058$ & \multirow{2}{*}{VENUS\cite{VENUS:1997cjg}} & 0.028\cite{VENUS:1997cjg} \\
  & $\mathcal{A}_{\mathcal{P}}$ & $0.029\pm0.057$ &  & 0.021\cite{VENUS:1997cjg} \\
\hline
{Neutrino trident production} & $\frac{\sigma}{\sigma^{\rm SM}}(\nu_\mu \gamma^*\to\nu_\mu\mu^+\mu^-)$ & $0.82\pm0.28$ & CCFR\cite{CHARM-II:1990dvf,CCFR:1991lpl,Altmannshofer:2014pba} & 1 \\
\hline
%\multirow{2}{*}{$d_j\to u_i\ell\bar{\nu}_\ell$} & $\tilde{V}_{ud}$ & \cite{Gonzalez-Alonso:2016etj} & \cite{Gonzalez-Alonso:2016etj} & 0.974353 \\
{$d_I\to u_J\ell\bar{\nu}_\ell(\gamma)$}& $\epsilon_{L,R,S,P,T}^{de_J}$ & See text & \cite{Gonzalez-Alonso:2016etj} & 0 \\
\hline
\multirow{5}{*}{$e^+e^-\to f\bar{f}$}& $\delta A_{LR}^e$ & 2.0\% & \multirow{5}{*}{SuperKEKB\cite{Banerjee:2022kfy}} & 0.00015 \\
 & $\delta A_{LR}^\mu$ & 1.5\% &  & -0.0006 \\
 & $\delta A_{LR}^\tau$ & 2.4\% &  & -0.0006 \\
 & $\delta A_{LR}^c$    & 0.5\%  &  & -0.005 \\
 & $\delta A_{LR}^b$    & 0.4\%  &  & -0.020 \\
\hline
\end{tabular}
\end{adjustbox}
\end{center}
\caption{Low-energy observables included in the global fit to break possible degeneracies for the $4\ell$ and $2\ell2q$ operators listed in \autoref{tab:4fope}. The last entry for $A_{LR}^f$ is taken from \cite{Banerjee:2022kfy} for SuperKEKB that would operate at 10.58\,GeV with 40\,$\rm ab^{-1}$ integrated luminosity and a beam polarization of $(70\pm0.3)\%$.}
\label{tab:leobs}
\end{table}

On the other hand, well below the $Z$ pole, various precision observables can also be utilized to constrain these 4-fermion operators. These observables, together the corresponding processes and the relevant experiments, are summarized in \autoref{tab:leobs}. As implied in the table, these observables are conventionally parameterized using the low-energy EFT (LEFT). We adopt the notations in \cite{Falkowski:2017pss} and parameterize the LEFT as
\eqalsplit{\mathcal{L}_{\rm {LEFT }} \supset &\, 
% CC neutrino NSIs
-\sum_{I,J=1,2}\frac{2 \widetilde{V}_{u d_I}^{e_J}}{v^{2}}\left[\left(1+{\epsilon}_{L}^{d_I e_{J}}\right)\left(\bar{e}_{J} \bar{\sigma}_{\mu} \nu_{J}\right)\left(\bar{u}\, \bar{\sigma}^{\mu} d_I\right)+\epsilon_{R}^{d_I e_J}\left(\bar{e}_{J} \bar{\sigma}_{\mu} \nu_{J}\right)\left(u^{c} \sigma^{\mu} \bar{d}_I^{c}\right)\right.\\
&\quad\quad\quad\quad\left.\,+\frac{\epsilon_{S}^{d_I e_{J}}+\epsilon_{P}^{d_I e_{J}}}{2}\left(e_{J}^{c} \nu_{J}\right)\left(u^{c} d_I\right)+\frac{\epsilon_{S}^{d_I e_{J}}-\epsilon_{P}^{d_I e_{J}}}{2}\left(e_{J}^{c} \nu_{J}\right)\left(\bar{u} \bar{d}_I^{c}\right)\right.\\
&\quad\quad\quad\quad\left.\,+\epsilon_{T}^{d_I e_{J}}\left(e_{J}^{c} \sigma_{\mu \nu} \nu_{J}\right)\left(u^{c} \sigma_{\mu \nu} d_I\right) +  h.c. \right]\\
% NC neutrino NSIs
&\,-\sum_{q=u,d}\frac{2}{v^{2}}\left(\bar{\nu}_{J} \bar{\sigma}^{\mu} \nu_{J}\right)\left(g_{L L}^{\nu_{J} q} \bar{q} \bar{\sigma}_{\mu} q+g_{L R}^{\nu_{J} q} q^{c} \sigma_{\mu} \bar{q}^{c}\right)\\
% eeqq part
&\,+ \sum_{q=u,d}\frac{1}{2 v^{2}}\left[g_{A V}^{e_{J} q}\left(\bar{e}_{J} \gamma_{\mu} \gamma_{5} e_{J}\right)\left(\bar{q} \gamma_{\mu} q\right)+g_{V A}^{e_{J} q}\left(\bar{e}_{J} \gamma_{\mu} e_{J}\right)\left(\bar{q} \gamma_{\mu} \gamma_{5} q\right)\right.\\
&\quad\quad\quad\left.\,+ g_{V V}^{e_{J} q}\left(\bar{e}_{J} \gamma_{\mu} e_{J}\right)\left(\bar{q} \gamma_{\mu} q\right)+g_{A A}^{e_{J} q}\left(\bar{e}_{J} \gamma_{\mu} \gamma_{5} e_{J}\right)\left(\bar{q} \gamma_{\mu} \gamma_{5} q\right)\right]\\
% 4 charge lepton
&\,-\sum_{I,J=1,2}\frac{1}{v^{2}}\left(\bar{\nu}_{J} \bar{\sigma}_{\mu} \nu_{J}\right)\left[\left(g_{L V}^{\nu_{J} e_{I}}+g_{L A}^{\nu_{J} e_{I}}\right)\left(\bar{e}_{I} \bar{\sigma}_{\mu} e_{I}\right)+\left(g_{L V}^{\nu_{J} e_{I}}-g_{L A}^{\nu_{J} e_{I}}\right)\left(e_{I}^{c} \sigma_{\mu} \bar{e}_{I}^{c}\right)\right],\label{eq:LEFTLag}}
where the first (second) term is usually referred to as the charge- (neutral-) current neutrino non-standard interactions (NSIs), and the third (last) term is the 4-fermion operators of $2\ell2q$ ($4\ell$) types. Note also that the SMEFT framework we adopt in this study forces $\epsilon_R^{de}=\epsilon_R^{d\mu}$ up to $\mathcal{O}(1/\Lambda^4)$ as a result of the $SU(2)_L\times U(1)_Y$ invariance \cite{Bernard:2006gy,Cirigliano:2009wk,Alonso:2015sja}.

Due to the presence of the SMEFT operators, the weak currents will be modified and as a result, the $W$ couplings to quarks will no longer be unitary. An overall $\widetilde{V}_{ud_I}^{e_J}$ is factored out to reflect this fact and its ``11'' element would be related to the actual $V_{ud}^e$ through $V_{ud}^e=\widetilde{V}_{ud}^e(1+\delta V_{ud}^e)$ with $\delta V_{ud}^e$ chosen such that ${\epsilon}_L^{de}=-\epsilon_R^{de}$. As a consequence, $\delta V_{ud}^e$ effectively only depends on $\epsilon_S^{de}$. Note that corrections from $\epsilon_S^{de}$ can not always be absorbed away from a redefinition of the hadronic form factors for a generic process, the nuclear or the neutron beta decay for example. For this reason, one shall take $\widetilde{V}_{ud}$ as a free parameter in the fit to be consistent. In practice, $\widetilde{V}_{ud}^e$ can be precisely measured from the superallowed nuclear beta decay \cite{Hardy:2014qxa}, which in turn allows one to perform a unitarity test of the first row of the CKM matrix, via a combined analysis of the $K_{e3(\gamma)}$ decay rates to extract $\widetilde{V}_{us}^e$. Calculations of the decay rates, on the other hand, rely on a global fit to the kinematic distributions of the decay, as well as extra observables, such as the lepton-universality ratios $R_{\pi,K}$, from the (inclusive) (semi)leptonic decays of pions and kaons, plus nuclear, neutron and hyperon beta decays in order to lift the flat directions in the fit \cite{Gonzalez-Alonso:2016etj}. A global fit on the $\epsilon$ parameters has been done in \cite{Gonzalez-Alonso:2016etj}, with the full correlation matrix provided. One can thus readily reconstruct the $\chi^2$ from their results and add it to the global fit in a larger framework, the SMEFT for example, as long as the matching and running effects are properly included.

We have independently performed the matching between the LEFT in \autoref{eq:LEFTLag} and the SMEFT in the Higgs basis, perfect agreement has been found with those in \cite{Falkowski:2017pss}. Note that the validity of EFTs requires the LEFT and the SMEFT to be defined at very different energy scales, thus this matching procedure is only meaningful at the weak scale. Therefore, to consistently combine the low-energy observables with the high-energy ones, we evolve the LEFT Wilson coefficients, typically defined at 2\,GeV, up to the weak scale through the renormalization group equations. While the running effects are generically small at the percent level and can be neglected for our purpose, it is not quite accurate for the $2\ell2q$ type operators, where higher-loop QCD effects were found to be significant \cite{Gonzalez-Alonso:2017iyc,Falkowski:2017pss}. For this reason, one-loop QED and electroweak\cite{Celis:2017hod,Aebischer:2017gaw,Alonso:2013hga}, as well as three-loop QCD\cite{Gracey:2000am} including $b$ and $t$ threshold effects\cite{Chetyrkin:1997un,Misiak:2010sk} are included in our analysis, but Yukawa suppressed corrections in \cite{Jenkins:2013wua} are generically ignored.

Following the prescription discussed above, the $\chi^2$ for the high- and low-energy processes is then constructed from these observables summarized in \autoref{tab:zwpoleobs}, \autoref{tab:fpairobs}, and \autoref{tab:leobs}, with the exception for the muon neutrino scattering and the $d_I\to u_J\ell\bar{\nu}_\ell(\gamma)$ ($I,J=1,2$) processes.\footnote{As an intermediate step, we derive all the analytical results for all the observables used in our fit, and find agreement with those in \cite{Efrati:2015eaa,Falkowski:2017pss}. Results from our global fit also agree those in \cite{Efrati:2015eaa,Falkowski:2017pss}.} For the former, due to the strong and non-Gaussian correlations between $g_{LL}^{\nu_J q}$ and $g_{LR}^{\nu_J q}$, we use the PDG fit in \cite{ParticleDataGroup:2016lqr} using the $g_{L,R}^{\nu_\mu}$ and $\theta_{L,R}^{\nu_\mu}$ parameterization. These variables are related to our notations in \autoref{eq:LEFTLag} through
\eqal{\left(g_{L / R}^{\nu_{J}}\right)^{2} \equiv \frac{\left(g_{L L / L R}^{\nu_{J} u}\right)^{2}+\left(g_{L L / L R}^{\nu_{J} d}\right)^{2}}{\left(1+{\epsilon}_{L}^{d e_{J}}\right)^{2}}, \quad \theta_{L / R}^{\nu_{J}} \equiv \arctan \left(\frac{g_{L L / L R}^{\nu_{J} u}}{g_{L L / L R}^{\nu_{J} d}}\right),}
and are expanded consistently to the linear order in terms of the dimension-six SMEFT operators in practice. While for the latter, a global fit for the $\epsilon$ parameters in \autoref{eq:LEFTLag} has been performed in \cite{Gonzalez-Alonso:2016etj} taking into account both inclusive and exclusive (semi)leptonic decay of pions and kaons, as well as nuclear, neutron and hyperon decays. We reconstruct the full $\chi^2$ from their results with the full correlations taken into account, and then marginalize over $\widetilde{V}_{ud}^e$ and the effective couplings of the strange quark.\footnote{Inclusion of the strange couplings will introduce the dependence on off-diagonal flavor couplings, which will not be covered in this study.}

%%%%%%%%%%%%%%%%%
\subsection{Flat directions}
%%%%%%%%%%%%%%%%%
The observables discussed in previous subsections, however, are not sufficient enough to separately constrain all the Wilson coefficients involved for the following reasons:
\begin{itemize}
\item Since LEP was run below the top-pair threshold, flat directions will thus show up for the 4-fermion operators of the $2\ell2q$ type, which we denote by ${\mathbf{Flat}}{[\rm top]}$. These flat directions may be lifted at future lepton colliders by running above the top-pair threshold, see \autoref{sec:topfit}.
\item Though the total cross section and the forward-backward asymmetry for $c\bar{c}$ are included using the experimental results from VENUS and TOPAZ, the corresponding information for the strange quark is missing. As a result, similar flat directions as ${\mathbf{Flat}}{[\rm top]}$, i.e., ${\mathbf{Flat}}{[\rm strange]}$, arise. These flat directions could be lifted when $\sigma_s$ and $A_{\rm FB}^s$ becomes available at future lepton colliders. 
\item The parity-conserving $(\bar{e}\gamma_\mu\gamma_5e)(\bar{q}_1\gamma_\mu\gamma_5q_1)$ and the axial vector neutrino-quark $(\bar{\nu}_L\gamma_\mu\nu_L)(\bar{q}_1\gamma_\mu\gamma_5q_1)$ operators remain unconstrained at the low-energy parity-violating scattering experiments or LEP. We denote these flat directions as ${\mathbf{Flat}}{[\rm {parity}]}$. These flat directions could possibly be eliminated by future low-energy parity-violating electron-nucleus scattering experiments, the P2 experiment as MESA \cite{Berger:2015aaa}, for example.
\item The muon scattering off the Carbon target at CERN SPS is insufficient to disentangle the contributions from $\mathcal{O}_{eq,eu,ed}$. We denote the resultant flat directions as ${\mathbf{Flat}}{[\rm SPS]}$. Varying the muon beam polarization $\lambda$ alone will not help in lifting this flat direction, but precision measurements of the neutral-current charged-lepton and quark interactions would, light quark pair production at a future muon collider for example. 
\item The trident process is the only low-energy channel sensitive to the four-muon operators $\mathcal{O}_{\ell\ell,\ell e}$, we denote this flat direction by ${\mathbf{Flat}}{[\rm trident]}$. We note that the $Z\to4\mu$ branching ratio has been measured at the LHC \cite{CMS:2012bw,ATLAS:2014jlg,CMS:2017dzg} and can thus be used to probe these operators.\footnote{We thank Radja Boughezal for pointing this observable to us.} However, this branching ratio is also dependent on $\mathcal{O}_{ee}$. Therefore, to eventually close the fit, one could, for example, measure the cross section and the asymmetry of muon pair production at a future muon collider. Note in particular that the muon beam polarization would be expected to help improve the fit significantly in this respect. 
\item The $\pi_{\mu2}$ decay rate only provides one constraint on $\epsilon_P^{d\mu}$ from the flavor observable $R_\pi$. Interpreting in the SMEFT, $\epsilon_P^{d\mu}$ is sourced from both $\mathcal{O}_{ledq}$ and $\mathcal{O}_{lequ}$. The lack of additional flavor observables sensitive to $\epsilon_P^{d\mu}$ results in an additional flat direction, which we call ${\mathbf{Flat}}{[\rm flavor]}$.
\end{itemize}

To separate these flat directions, we define \cite{Falkowski:2017pss}
\begin{itemize}
\item  ${\mathbf{Flat}}{[\rm top]}$:
\eqal{
{\left[\hat{c}_{\ell q}^{(3)}\right]_{1133} } &=\left[c_{\ell q}^{(3)}\right]_{1133}+\left[c_{\ell q}\right]_{1133}.
}
\item  ${\mathbf{Flat}}{[\rm strange]}$:
\eqalsplit{
{\left[\hat{c}_{\ell q}^{(3)}\right]_{1122} } &=\left[c_{\ell q}^{(3)}\right]_{1122}-\left[c_{\ell q}\right]_{1122}, \\
{\left[\hat{c}_{\ell d}\right]_{1122} } &=\left[c_{\ell d}\right]_{1122}+\left(5-\frac{3 g^{2}}{g'^{2}}\right)\left[c_{\ell q}\right]_{1122}-\left[\hat{c}_{e q}\right]_{1111},\\
{\left[\hat{c}_{e d}\right]_{1122} } &=\left[c_{e d}\right]_{1122}-\left(3-\frac{3 g^{2}}{g'^{2}}\right)\left[c_{\ell q}\right]_{1122}-\left[\hat{c}_{e q}\right]_{1111}.}
\item ${\mathbf{Flat}}{[\rm parity]}$:
\eqalsplit{
{\left[\hat{c}_{e q}\right]_{1111} } &=\left[c_{e q}\right]_{1111}+\left[c_{\ell q}\right]_{1111}, \\
{\left[\hat{c}_{\ell u}\right]_{1111} } &=\left[c_{\ell u}\right]_{1111}+\left[c_{\ell q}\right]_{1111}-\left[\hat{c}_{e q}\right]_{1111}, \\
{\left[\hat{c}_{\ell d}\right]_{1111} } &=\left[c_{\ell d}\right]_{1111}+\left[c_{\ell q}\right]_{1111}-\left[\hat{c}_{e q}\right]_{1111}, \\
{\left[\hat{c}_{e u}\right]_{1111} } &=\left[c_{e u}\right]_{1111}-\left[c_{\ell q}\right]_{1111},\\
{\left[\hat{c}_{e d}\right]_{1111} } &=\left[c_{e d}\right]_{1111}-\left[c_{\ell q}\right]_{1111},}
\item ${\mathbf{Flat}}{[\rm SPS]}$:
\eqal{
{\left[\hat{c}_{e q}\right]_{2211} } &=\left[c_{e q}\right]_{2211}+\left[c_{e d}\right]_{2211}-2\left[c_{e u}\right]_{2211},}
\item ${\mathbf{Flat}}{[\rm trident]}$:
\eqal{
{\left[\hat{c}_{\ell \ell}\right]_{2222} } &=\left[c_{\ell \ell}\right]_{2222}+\frac{2g'^{2}}{g^{2}+3 g'^{2}}\left[c_{\ell e}\right]_{2222},
}
\item ${\mathbf{Flat}}{[\rm flavor]}$:
\eqal{
\epsilon_{P}^{d \mu}[2\, \mathrm{GeV}] &=0.86\left[c_{l e d q}\right]_{2211}-0.86\left[c_{l e q u}\right]_{2211}+0.012\left[c_{l e d q}^{(3)}\right]_{2211}.
}
It is worth pointing out that the numbers on the right hand side are directly from the renormalization group evolution, based on \cite{Gonzalez-Alonso:2017iyc}, from $\mu=2\,\rm GeV$ to $\mu=m_Z$, where the matching between the LEFT and the SMEFT is performed. The coefficient of $\left[c_{l e d q}^{(3)}\right]_{2211}$ is much smaller than the other two since it only indirectly matches onto $\epsilon_{P}^{d \mu}[2\, \mathrm{GeV}]$ through the renormalization group evolution.
\end{itemize}

%%%%%%%%%%%%%%%%%
\subsection{SMEFT global fit results}
%%%%%%%%%%%%%%%%%
%

Results for the vertex and the 4-fermion operators are reported in this subsection based on inputs summarized in \autoref{sec:input}. The numerical $1\sigma$ bounds from the global fit are summarized in \autoref{tab:fit2Part1}, and also pictorially in \autoref{fig:fit2-4f1}, \autoref{fig:fit2-4f2} and \autoref{fig:fit2-4f3} at various lepton colliders. As for the Higgs and electroweak fit presented in last section, the relative (absolute) $1\sigma$ errors are always reported whenever their corresponding SM predictions are non-vanishing (vanishing). Several comments are in order:
\begin{itemize}
\item While one can implement special flavor structures, such as the ${\rm U(3)^5}$ global symmetry \cite{Gerard:1982mm,Chivukula:1987py} and the Minimal Flavor Violation symmetry \cite{DAmbrosio:2002vsn}, to reduce the number of parameters in the fit, we do not make any flavor assumptions except for focusing on flavor diagonal operators only in this study.
\item For ILC running at different energies and luminosities, a horizontal white line is drawn in \autoref{fig:fit2-4f1}, \autoref{fig:fit2-4f2} and \autoref{fig:fit2-4f3} to indicate the alternative results if the pole observables from the GigaZ option in \autoref{tab:EWPO} is adopted for the fit.
\item At future circular lepton colliders, we find $\delta g_Z^{\ell\ell}$ would be constrained about one order of magnitude better than $\delta g_Z^{qq}$. Projections on the $R_{uc}$ parameter are expected to improve the sensitivity to $\delta g_Z^{qq}$.
\item The right-handed coupling between $W$ and the first-generation quark only contributes at the quadratic order to high-energy observables, and their corrections to the SM predictions are thus suppressed and ignored. In contrast, the low-energy flavor observables, neutral-current electron-neutrino scattering off nuclei and the CKM unitarity test specifically, have a linear dependence on it. These low-energy flavor observables dominate the constraint on $\hat{\delta}g_{R}^{Wq_1}$.
\item Measurements of the average $\tau$ polarization and its forward-backward asymmetry at VENUS help eliminate the flat direction existed in $\tau$ pair production at LEP. However, results from the VENUS collaboration are not yet very precise and are expected to be surpassed by SuperKEKB \cite{Banerjee:2022kfy} and future $e^+e^-$ \cite{Yumino:2022vqt} results.
\item With polarized beams probing different combinations of the Wilson coefficients at future linear lepton colliders, constraints on some of the 4-fermion operators turn out to be several orders of magnitude better than those at the future circular colliders, $[c_{eq}]_{1122}$ and $[c_{eu}]_{1122}$ for example. This largely seeds in the fact that beam polarizations help reduce the correlations among multiple Wilson coefficients significantly.
\end{itemize}

\begin{table}[]
  \centering
  \begin{adjustbox}{max width=1\textwidth, max height=0.49\textheight}
\begin{tabular}{|c|c|c|cc|cc|cc|ccc|cc|cc|c|c|c|}\hline
in & LEP + SLC &  HL-LHC & \multicolumn{6}{c|}{ILC} & \multicolumn{3}{c|}{CLIC} &  \multicolumn{2}{c|}{FCC-ee}  & \multicolumn{2}{c|}{CEPC} \\ 
\cline{3-16} %\hline
\% & + SLD + D0 & 14\,TeV & \multicolumn{2}{c|}{250} & \multicolumn{2}{c|}{+500} & \multicolumn{2}{c|}{+1\,TeV} & 380 & +1.5\,TeV & +3\,TeV  &  240 & +365 & 240 & +360 \\ 
 & + LHC & &  & Giga-Z &  & Giga-Z  &  & Giga-Z &  &  &  & &  &  &  \\ 
 
\hline
\hline
$\delta g_ {W}^{e\nu}$ & 0.64 & 0.233 & 0.046 & 
  0.043 & 0.042 & 0.0465 & 0.042 & 0.0404 & 0.091
   & 0.085 & 0.083 & 0.0185 & 0.018 & 0.012 & 
  0.0119\\
$\delta g_ {W}^{\mu\nu}$ & 0.59 & 0.58 & 0.047 
  & 0.043 & 0.042 & 0.047 & 0.043 & 0.041 & 
  0.1 & 0.091 & 0.089 & 0.0184 & 0.018 & 0.0119 
  & 0.0118\\
$\delta g_ {W}^{\tau\nu}$ & 0.79 & 0.62 & 0.047 
  & 0.043 & 0.042 & 0.047 & 0.043 & 0.041 & 
  0.1 & 0.092 & 0.089 & 0.0184 & 0.018 & 0.0119 
  & 0.0118\\
$\delta g_ {Z,L}^{ee}$ & 0.102 & 0.098 & 0.0286 &
   0.0153 & 0.0121 & 0.0212 & 0.013 & 0.01 & 
  0.053 & 0.046 & 0.044 & 0.0077 & 0.0076 & 
  0.0053 & 0.0053\\
$\delta g_ {Z,L}^{\mu\mu}$ & 0.45 & 0.13 & 0.064 
  & 0.058 & 0.054 & 0.028 & 0.026 & 0.025 & 
  0.094 & 0.09 & 0.09 & 0.008 & 0.008 & 0.0075 
  & 0.0075\\
$\delta g_ {Z,L}^{\tau\tau}$ & 0.22 & 0.22 & 0.073
   & 0.066 & 0.062 & 0.033 & 0.031 & 0.03 & 
  0.15 & 0.14 & 0.137 & 0.0127 & 0.0127 & 0.008 
  & 0.008\\
$\delta g_ {Z,R}^{ee}$ & 0.116 & 0.113 & 0.029 & 
  0.015 & 0.0118 & 0.0213 & 0.0128 & 0.01 & 0.058
   & 0.051 & 0.049 & 0.0077 & 0.0076 & 0.0053 &
   0.0053\\
$\delta g_ {Z,R}^{\mu\mu}$ & 0.58 & 0.157 & 0.07 
  & 0.064 & 0.06 & 0.028 & 0.026 & 0.0253 & 
  0.11 & 0.106 & 0.105 & 0.008 & 0.008 & 0.0076 
  & 0.0076\\
$\delta g_ {Z,R}^{\tau\tau}$ & 0.266 & 0.265 & 
  0.078 & 0.072 & 0.068 & 0.033 & 0.031 & 0.03 
  & 0.174 & 0.163 & 0.16 & 0.0146 & 0.0145 & 
  0.0084 & 0.0084\\
$\delta g_ {Z,L}^{uu}$ & 6.65 & 1.29 & 0.67 & 
  0.66 & 0.65 & 0.56 & 0.56 & 0.56 & 0.86 & 
  0.84 & 0.84 & 0.55 & 0.55 & 0.55 & 0.55\\
$\delta g_ {Z,L}^{cc}$ & 1.06 & 1.05 & 0.22 & 
  0.198 & 0.183 & 0.144 & 0.138 & 0.132 & 0.24 
  & 0.233 & 0.23 & 0.076 & 0.076 & 0.052 & 
  0.051\\
$\delta g_ {Z,L}^{tt}$ & 11.92 & 11.92 & 11.91 & 
  11.91 & 11.91 & 11.91 & 11.91 & 11.91 & 11.91 
  & 11.91 & 11.91 & 11.91 & 11.91 & 11.91 & 
  11.91\\
$\delta g_ {Z,R}^{uu}$ & 18.1 & 7.36 & 7.05 & 
  7.05 & 7.05 & 7.04 & 7.03 & 7.03 & 7.1 & 
  7.09 & 7.09 & 7.03 & 7.03 & 7.03 & 7.03\\
$\delta g_ {Z,R}^{cc}$ & 3.42 & 3.4 & 0.32 & 0.3
   & 0.28 & 0.16 & 0.153 & 0.148 & 0.4 & 0.39 
  & 0.39 & 0.084 & 0.084 & 0.068 & 0.068\\
$\delta g_ {Z,L}^{dd}$ & 8.05 & 1.61 & 1.46 & 
  1.46 & 1.46 & 1.44 & 1.44 & 1.44 & 1.5 & 
  1.49 & 1.49 & 1.44 & 1.44 & 1.44 & 1.44\\
$\delta g_ {Z,L}^{ss}$ & 6.03 & 2.29 & 1.35 & 
  1.35 & 1.35 & 1.35 & 1.34 & 1.34 & 1.42 & 
  1.41 & 1.41 & 1.33 & 1.33 & 1.33 & 1.33\\
$\delta g_ {Z,L}^{bb}$ & 0.4 & 0.395 & 0.061 & 
  0.055 & 0.051 & 0.042 & 0.04 & 0.038 & 0.094 
  & 0.09 & 0.089 & 0.0173 & 0.0173 & 0.0124 & 
  0.0124\\
$\delta g_ {Z,R}^{dd}$ & 153.85 & 45.75 & 44.02 &
   44. & 43.99 & 43.98 & 43.97 & 43.96 & 44.19 
  & 44.18 & 44.17 & 43.91 & 43.91 & 43.91 & 
  43.91\\
$\delta g_ {Z,R}^{ss}$ & 63.14 & 53.5 & 18.48 & 
  18.34 & 18.24 & 18.06 & 17.79 & 17.69 & 24.02 
  & 23.37 & 23.16 & 16.46 & 16.45 & 16.35 & 
  16.35\\
$\delta g_ {Z,R}^{bb}$ & 11.19 & 11.07 & 0.46 & 
  0.43 & 0.4 & 0.374 & 0.356 & 0.34 & 1.32 & 
  1.18 & 1.16 & 0.16 & 0.16 & 0.16 & 0.16\\
$\hat{\delta} g_ {R}^{Wq_1}$ & 1.69 & 1.69 & 1.69 
  & 1.69 & 1.69 & 1.69 & 1.69 & 1.69 & 1.69 
  & 1.69 & 1.69 & 1.69 & 1.69 & 1.69 & 1.69\\
$[c_{\ell\ell}]_{1111}$ & 0.4 & 0.28 & 0.0051 & 
  0.0017 & 0.00061 & 0.005 & 0.00167 & 0.00059 & 
  0.0059 & 0.0007 & 0.00023 & 0.064 & 0.063 & 
  0.063 & 0.063\\
$[c_{\ell e}]_{1111}$ & 0.22 & 0.074 & 0.0062 & 
  0.00216 & 0.00077 & 0.0048 & 0.00207 & 0.00076 &
   0.005 & 0.0013 & 0.00044 & 0.00385 & 0.0031 & 
  0.00203 & 0.00195\\
$[c_{ee}]_{1111}$ & 0.404 & 0.39 & 0.0065 & 
  0.00217 & 0.00083 & 0.006 & 0.0021 & 0.00082 & 
  0.0086 & 0.00195 & 0.00066 & 0.067 & 0.066 & 
  0.066 & 0.065\\
$[c_{\ell\ell}]_{1221}$ & 1.63 & 1.23 & 0.162 & 
  0.155 & 0.152 & 0.177 & 0.166 & 0.16 & 0.32 
  & 0.3 & 0.29 & 0.057 & 0.056 & 0.0385 & 
  0.038\\
$[c_{\ell\ell}]_{1122}$ & 1.82 & 1.81 & 0.162 & 
  0.154 & 0.152 & 0.176 & 0.165 & 0.16 & 0.32 
  & 0.3 & 0.29 & 1.66 & 1.2 & 1.66 & 1.21\\
$[c_{\ell e}]_{1122}$ & 1.87 & 1.86 & 0.0089 & 
  0.0029 & 0.00101 & 0.0085 & 0.0028 & 0.001 & 
  0.0088 & 0.0011 & 0.00037 & 1.71 & 1.23 & 1.71 
  & 1.24\\
$[c_{\ell e}]_{2211}$ & 1.8 & 1.8 & 0.0089 & 
  0.0029 & 0.00102 & 0.0086 & 0.0028 & 0.001 & 
  0.0089 & 0.0021 & 0.00072 & 1.71 & 1.23 & 1.71 
  & 1.24\\
$[c_{ee}]_{1122}$ & 2.38 & 1.91 & 0.0067 & 
  0.00224 & 0.0008 & 0.0063 & 0.00215 & 0.00078 & 
  0.0069 & 0.00164 & 0.00056 & 1.79 & 1.29 & 1.79
   & 1.3\\
$[c_{\ell\ell}]_{1331}$ & 1.34 & 1.32 & 0.49 & 
  0.49 & 0.48 & 0.49 & 0.49 & 0.49 & 0.56 & 
  0.55 & 0.55 & 0.46 & 0.46 & 0.46 & 0.46\\
$[c_{\ell\ell}]_{1133}$ & 11.29 & 5.32 & 0.49 & 
  0.49 & 0.48 & 0.49 & 0.49 & 0.49 & 0.56 & 
  0.55 & 0.55 & 5.17 & 1.82 & 5.17 & 1.86\\
$[c_{\ell e}]_{1133}$ & 7.22 & 5.32 & 0.01 & 
  0.00324 & 0.00114 & 0.0096 & 0.00315 & 0.00112 &
   0.0101 & 0.00124 & 0.000416 & 5.3 & 1.81 & 5.3
   & 1.86\\
$[c_{\ell e}]_{3311}$ & 7.22 & 5.32 & 0.01 & 
  0.00324 & 0.00115 & 0.0097 & 0.0032 & 0.00113 & 
  0.01 & 0.0024 & 0.00081 & 5.3 & 1.81 & 5.3 &
   1.86\\
$[c_{e e}]_{1133}$ & 12.16 & 5.59 & 0.0074 & 
  0.00247 & 0.00088 & 0.007 & 0.00237 & 0.00086 & 
  0.0077 & 0.0018 & 0.00062 & 5.56 & 1.89 & 5.56 
  & 1.94\\
$[\hat{c}_{\ell \ell}]_{2222}$ & 22.92 & 22.92 & 
  22.84 & 22.84 & 22.84 & 22.84 & 22.84 & 22.84 
  & 22.84 & 22.84 & 22.84 & 22.84 & 22.84 & 
  22.84 & 22.84\\
$[c_{\ell \ell}]_{2332}$ & 2.25 & 1.96 & 1.81 & 
  1.81 & 1.81 & 1.81 & 1.81 & 1.81 & 1.83 & 
  1.83 & 1.83 & 1.8 & 1.8 & 1.8 & 1.8\\
$[c_{\ell q}^{(3)}]_{1111}$ & 2.9 & 1.8 & 1.78 & 
  1.78 & 1.78 & 1.78 & 1.78 & 1.78 & 1.79 & 
  1.79 & 1.79 & 1.78 & 1.78 & 1.78 & 1.78\\
$[\hat{c}_{eq}]_{1111}$ & 179.61 & 178.79 & 178.74 
  & 178.74 & 178.74 & 178.74 & 178.74 & 178.74 
  & 178.75 & 178.75 & 178.75 & 178.74 & 178.74 
  & 178.74 & 178.74\\
$[\hat{c}_{\ell u}]_{1111}$ & 8.75 & 7.68 & 7.65 
  & 7.65 & 7.65 & 7.65 & 7.65 & 7.65 & 7.66 
  & 7.66 & 7.66 & 7.65 & 7.65 & 7.65 & 7.65\\
$[\hat{c}_{\ell d}]_{1111}$ & 17.92 & 11.61 & 11.56
   & 11.56 & 11.56 & 11.56 & 11.56 & 11.56 & 
  11.57 & 11.57 & 11.57 & 11.56 & 11.56 & 11.56 
  & 11.56\\
$[\hat{c}_{e u}]_{1111}$ & 9.38 & 7.86 & 7.84 & 
  7.84 & 7.84 & 7.84 & 7.84 & 7.84 & 7.84 & 
  7.84 & 7.84 & 7.83 & 7.83 & 7.83 & 7.83\\
$[\hat{c}_{e d}]_{1111}$ & 17.11 & 11.55 & 11.51 
  & 11.51 & 11.51 & 11.51 & 11.51 & 11.51 & 
  11.52 & 11.52 & 11.52 & 11.51 & 11.51 & 11.51 
  & 11.51\\
$[\hat{c}_{\ell q}^{(3)}]_{1122}$ & 28.94 & 13.43 & 
  0.0245 & 0.0081 & 0.0028 & 0.023 & 0.0077 & 
  0.0027 & 0.023 & 0.0029 & 0.00097 & 10.83 & 
  0.29 & 10.83 & 0.283\\
$[{c}_{\ell u}]_{1122}$ & 7.76 & 7.59 & 0.034 & 
  0.011 & 0.0039 & 0.0335 & 0.0108 & 0.00385 & 
  0.0335 & 0.00415 & 0.0014 & 6.76 & 0.185 & 6.76
   & 0.172\\
$[\hat{c}_{\ell d}]_{1122}$ & 121.45 & 86.39 & 
  44.68 & 44.68 & 44.68 & 44.68 & 44.68 & 44.68 
  & 44.7 & 44.69 & 44.69 & 75.47 & 44.7 & 
  75.47 & 44.7\\
$[{c}_{eq}]_{1122}$ & 26.26 & 25.75 & 0.061 & 
  0.017 & 0.0058 & 0.061 & 0.0168 & 0.0058 & 
  0.054 & 0.012 & 0.0041 & 18.89 & 0.46 & 18.89 
  & 0.47\\
$[{c}_{e u}]_{1122}$ & 41.3 & 19.08 & 0.023 & 
  0.0078 & 0.0028 & 0.022 & 0.0076 & 0.00276 & 
  0.024 & 0.0058 & 0.002 & 15.16 & 0.4 & 15.16 
  & 0.395\\
$[\hat{c}_{ed}]_{1122}$ & 128.73 & 86.56 & 25.43 
  & 25.43 & 25.43 & 25.43 & 25.43 & 25.43 & 
  25.45 & 25.44 & 25.44 & 69.8 & 25.48 & 69.8 
  & 25.47\\
$[\hat{c}_{\ell q}^{(3)}]_{1133}$ & 7.57 & 1.44 & 
  0.0107 & 0.0035 & 0.00122 & 0.0102 & 0.0034 & 
  0.0012 & 0.0104 & 0.00128 & 0.00043 & 0.42 & 
  0.082 & 0.42 & 0.082\\
$[{c}_{\ell d}]_{1133}$ & 9.54 & 6.65 & 0.0228 & 
  0.007 & 0.0024 & 0.0225 & 0.0069 & 0.0024 & 
  0.022 & 0.0026 & 0.00087 & 3.55 & 0.103 & 3.55 
  & 0.11\\
$[{c}_{e q}]_{1133}$ & 4.74 & 3.37 & 0.0122 & 
  0.0044 & 0.00165 & 0.012 & 0.0044 & 0.00164 & 
  0.0136 & 0.0035 & 0.0012 & 2.26 & 0.073 & 2.26 
  & 0.071\\
$[{c}_{e d}]_{1133}$ & 18.57 & 3.05 & 0.0122 & 
  0.004 & 0.00144 & 0.0118 & 0.00395 & 0.00142 & 
  0.0135 & 0.00305 & 0.00103 & 1. & 0.19 & 1. 
  & 0.19\\
$[{c}_{\ell q}^{(3)}]_{2211}$ & 3.57 & 2.76 & 2.02 
  & 2.02 & 2.01 & 2.02 & 2.02 & 2.02 & 2.05 
  & 2.05 & 2.04 & 2.01 & 2.01 & 2. & 2.\\
$[{c}_{\ell q}]_{2211}$ & 5.93 & 2.77 & 2.68 & 
  2.68 & 2.68 & 2.67 & 2.67 & 2.67 & 2.7 & 
  2.7 & 2.7 & 2.67 & 2.67 & 2.67 & 2.67\\
$[{c}_{\ell u}]_{2211}$ & 8.11 & 6.22 & 6.13 & 
  6.13 & 6.13 & 6.13 & 6.13 & 6.13 & 6.14 & 
  6.14 & 6.14 & 6.13 & 6.13 & 6.13 & 6.13\\
$[{c}_{\ell d}]_{2211}$ & 26.34 & 13.11 & 12.96 &
   12.96 & 12.96 & 12.96 & 12.96 & 12.96 & 12.98 
  & 12.97 & 12.97 & 12.96 & 12.96 & 12.96 & 
  12.96\\
$[\hat{c}_{e q}]_{2211}$ & 40.71 & 38.89 & 38.88 
  & 38.88 & 38.88 & 38.88 & 38.88 & 38.88 & 
  38.88 & 38.88 & 38.88 & 38.88 & 38.88 & 38.88 
  & 38.88\\
$[{c}_{\ell e qu}]_{1111}$ & 0.076 & 0.076 & 0.076 
  & 0.076 & 0.076 & 0.076 & 0.076 & 0.076 & 
  0.076 & 0.076 & 0.076 & 0.076 & 0.076 & 0.076 
  & 0.076\\
$[{c}_{\ell e qd}]_{1111}$ & 0.076 & 0.076 & 0.076 
  & 0.076 & 0.076 & 0.076 & 0.076 & 0.076 & 
  0.076 & 0.076 & 0.076 & 0.076 & 0.076 & 0.076 
  & 0.076\\
$[{c}_{\ell e qu}^{(3)}]_{1111}$ & 0.194 & 0.194 & 
  0.194 & 0.194 & 0.194 & 0.194 & 0.194 & 0.194 
  & 0.194 & 0.194 & 0.194 & 0.194 & 0.194 & 
  0.194 & 0.194\\
$\epsilon_P^{d\mu}[{\rm 2\,GeV}]$ & 0.144 & 0.141 
  & 0.13 & 0.13 & 0.13 & 0.13 & 0.13 & 0.13 
  & 0.13 & 0.13 & 0.13 & 0.13 & 0.13 & 0.13 
  & 0.13\\
  \hline
  \end{tabular}
  \end{adjustbox}
\caption{Precision reach (in percentage) on the effective couplings from a SMEFT global analysis of the 4-fermion operators at various future lepton colliders.}
\label{tab:fit2Part1}
\end{table}

\iffalse
\begin{figure}%[t]
    \includegraphics[width=\textwidth]{plots/4f/4f1}\\
    \includegraphics[width=\textwidth]{plots/4f/4f2}\\
    \includegraphics[width=\textwidth]{plots/4f/4f3}\\
    \includegraphics[width=\textwidth]{plots/4f/4f4}
    \caption{Precision reach on the effective couplings from a SMEFT global analysis of the 4-fermion operators at various future lepton colliders. The horizontal white line for ILC suggests the global fit results when applying the pole observables from its GigaZ option.}
    \label{fig:fit2-4f1}
\end{figure}

\begin{figure}%[t]
    
    \includegraphics[width=\textwidth]{plots/4f/4f5}\\
    \includegraphics[width=\textwidth]{plots/4f/4f6}\\
    \includegraphics[width=\textwidth]{plots/4f/4f7}\\
    \includegraphics[width=\textwidth]{plots/4f/4f8}
    \caption{\autoref{fig:fit2-4f1} continued.}
    \label{fig:fit2-4f2}
\end{figure}

\begin{figure}%[t]
    \includegraphics[width=\textwidth]{plots/4f/4f9}\\
    \includegraphics[width=\textwidth]{plots/4f/4f10}
    \caption{\autoref{fig:fit2-4f2} continued.}
    \label{fig:fit2-4f3}
\end{figure}
\fi

\begin{figure}%[t]
    \includegraphics[width=\textwidth]{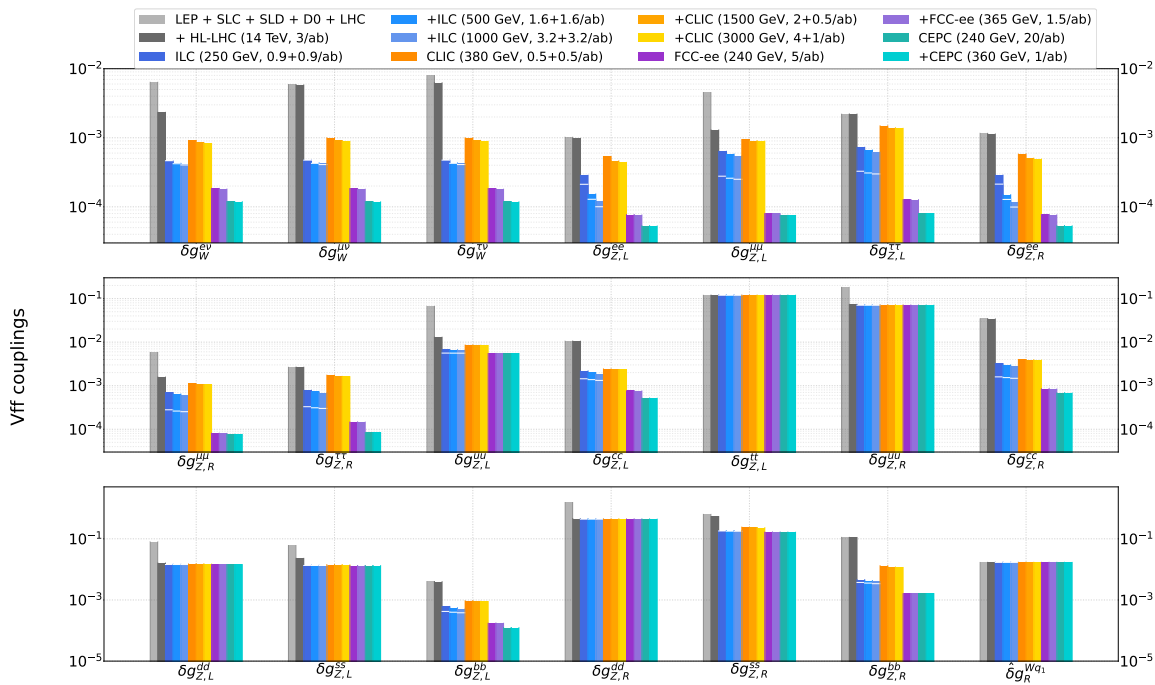}
    \caption{Precision reach on the effective couplings from a SMEFT global analysis of the 4-fermion operators at various future lepton colliders. The horizontal white line for ILC suggests the global fit results when applying the pole observables from its GigaZ option.}
    \label{fig:fit2-4f1}
\end{figure}

\begin{figure}%[t]
    
    \includegraphics[width=\textwidth]{plots/4f/4f-4l}
    \caption{\autoref{fig:fit2-4f1} continued.}
    \label{fig:fit2-4f2}
\end{figure}

\begin{figure}%[t]
    \includegraphics[width=\textwidth]{plots/4f/4f-2l2q}
    \caption{\autoref{fig:fit2-4f2} continued.}
    \label{fig:fit2-4f3}
\end{figure}

%%%%%%%%%%%%%%%%%
\subsection{Implication on some benchmark UV models}
%%%%%%%%%%%%%%%%%
We discuss the implication of the global 4-fermion fit on specific UV models in this section. To that end, we focus on two specific models in the following: (1) The Y-Universal $Z'$ model, and (2) the leptoquark model with two colored scalar leptoquarks, $(\bar{\bf{3}},\bf{1})_{\frac{1}{3}}$ and $(\bar{\bf{3}},\bf{3})_{\frac{1}{3}}$.

\subsubsection{The Y-Universal $Z'$ model}
The Y-Universal $Z'$ model is interesting since its couplings to the SM are flavor-diagonal, avoiding stringent flavor constraints. Our discussion on this model \cite{Appelquist:2002mw} is similar to that in \cite{EuropeanStrategyforParticlePhysicsPreparatoryGroup:2019qin}: We first translate the results for the 4-fermion fit into a global fit for the oblique parameters by marginalizing over all the other Wilson coefficients. We then perform an individual fit for $\mathcal{O}_{2W,2B}$ operators as defined in \cite{Giudice:2007fh}. The results are shown in \autoref{fig:fit2O2W2B}. Then, given that the Y-Universal $Z'$ model only matches onto the $\mathcal{O}_{2B}$ operator, the bound on this operator can thus be straightforwardly transferred onto the parameter space of this model for various colliders. This is shown in \autoref{fig:fit2zprime}. We comment on that, for the hadron colliders, the difference between our result and that in \cite{EuropeanStrategyforParticlePhysicsPreparatoryGroup:2019qin} comes from the fact that we only consider the neutral Drell-Yan processes for the 4-fermion fit in this work. For future lepton colliders, our results are generically improved due to the updated inputs used in this study.

\begin{figure}
\centering
    \includegraphics[width=0.9\textwidth]{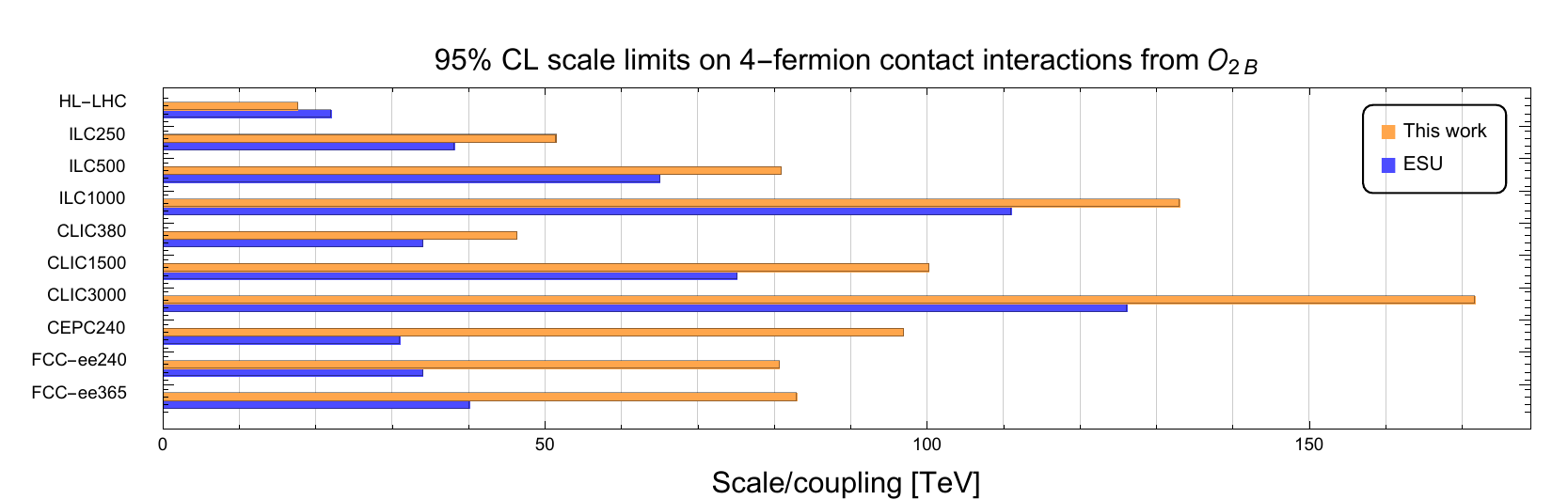}\\
    \includegraphics[width=0.9\textwidth]{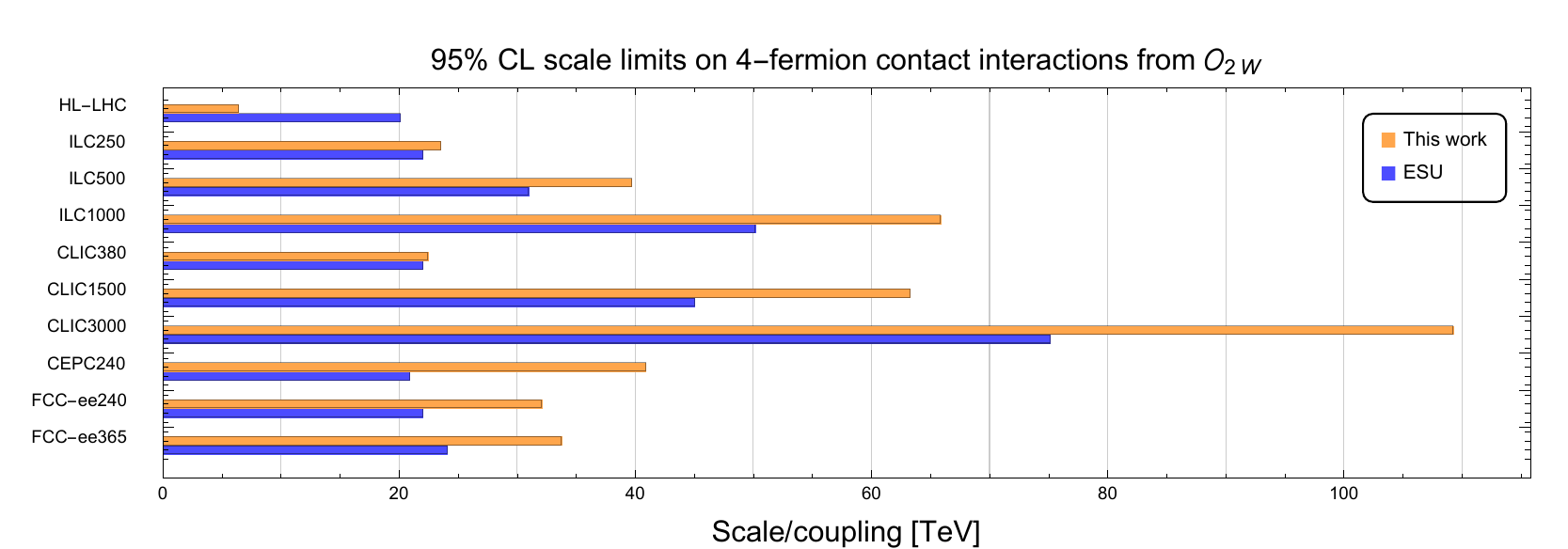}
    \caption{Constraints on the $\mathcal{O}_{2W,2B}$ from the global 4-fermion fit and the comparison with ESU.}
    \label{fig:fit2O2W2B}
\end{figure}

\begin{figure}
\centering
    \includegraphics[width=0.55\textwidth]{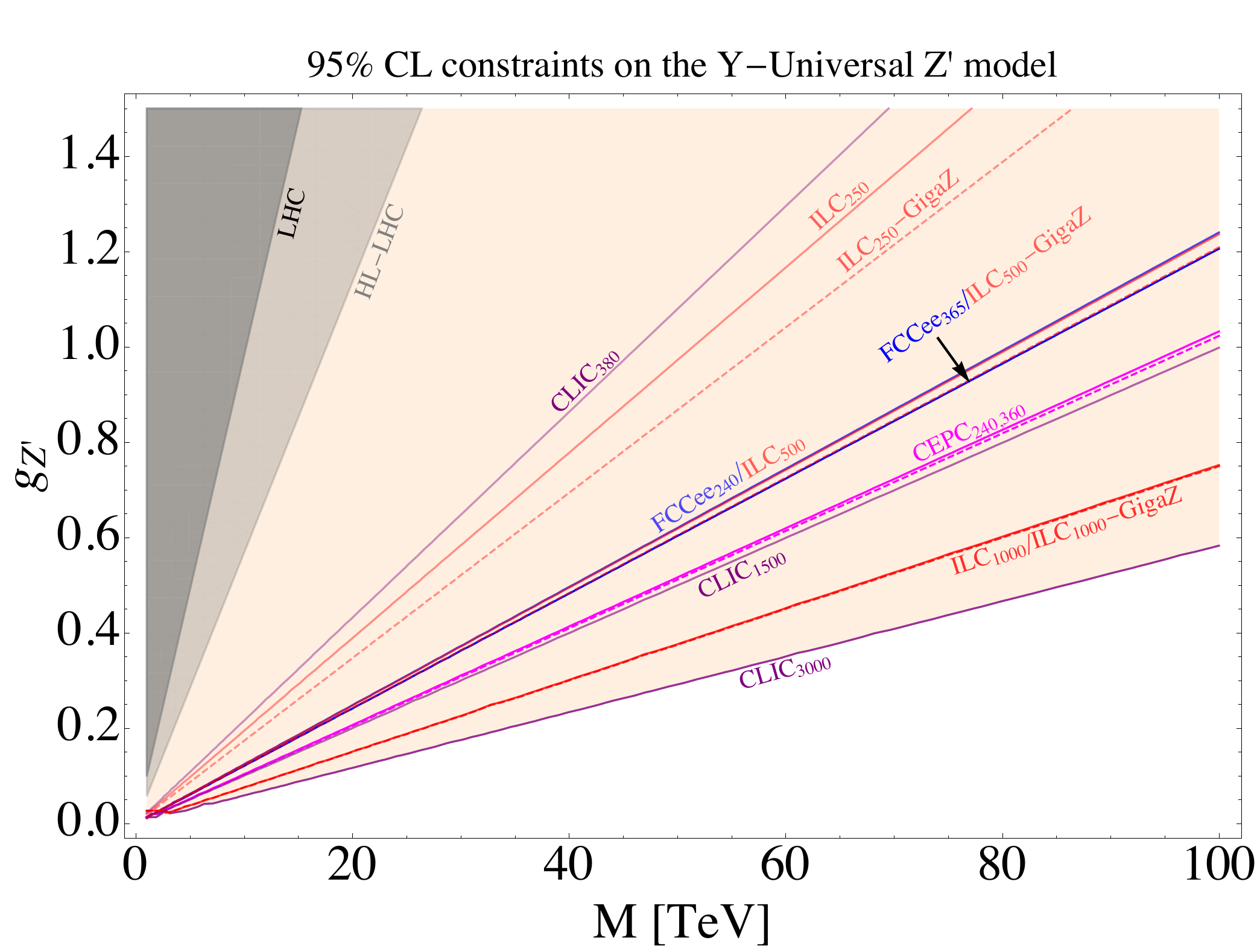}
    \caption{Constraints on the Y-Universal $Z'$ model from the global 4-fermion fit.}
    \label{fig:fit2zprime}
\end{figure}

\subsubsection{The scalar leptoquark model}
This model is obtained by extending the SM with two colored scalar leptoquarks, $(\bar{\bf{3}},\bf{1})_{\frac{1}{3}}$ and $(\bar{\bf{3}},\bf{3})_{\frac{1}{3}}$, and the relevant part in the Lagrangian for our discussion below can be written as 
\eqal{
\mathcal{L}_{\mathrm{LQ}}^{\rm Yukawa} &\supset \left(\lambda^{1 L}_{i \alpha} \bar{q}_{i}^{c} \epsilon \ell_{\alpha}+\lambda^{1 R}_{i \alpha} \bar{u}_{i}^{c} e_{\alpha}\right) S_{1}+\lambda^{3 L}_{i \alpha} \bar{q}_{i}^{c} \epsilon \sigma^{I} \ell_{\alpha} S_{3}^{I}+\text { h.c. }\,.
}
At tree level, the full model only matches onto the following operators with the matching relations given by \cite{Gherardi:2020det,Aebischer:2021uvt}
\eqal{
&{\left[c_{l q}^{(1)}\right]_{\alpha \beta i j}=\frac{\lambda_{i \alpha}^{1 L *} \lambda_{j \beta}^{1 L}v^2}{4 M_1^2}+\frac{3 \lambda_{i \alpha}^{3 L *} \lambda_{j \beta}^{3 L}v^2}{4 M_3^2},} \quad {\left[c_{l q}^{(3)}\right]_{\alpha \beta i j}=-\frac{\lambda_{i \alpha}^{1 L *} \lambda_{j \beta}^{1 L}v^2}{4 M_1^2}+\frac{\lambda_{i \alpha}^{3 L *} \lambda_{j \beta}^{3 L}v^2}{4 M_3^2},} \\
&{\left[c_{l e q u}^{(1)}\right]_{\alpha \beta i j}=\frac{\lambda_{j \beta}^{1 R} \lambda_{i \alpha}^{1 L *}v^2}{2 M_1^2},} \quad {\left[c_{l e q u}^{(3)}\right]_{\alpha \beta i j}=-\frac{\lambda_{j \beta}^{1 R} \lambda_{i \alpha}^{1 L *}v^2}{8 M_1^2},} \quad {\left[c_{e u}\right]_{\alpha \beta i j}=\frac{\lambda_{i \alpha}^{1 R *} \lambda_{j \beta}^{1 R}v^2}{2 M_1^2} .}
}
For simplicity, we will work in the universal Yukawa scenario for the following discussion. As a result, these five Wilson coefficients will only depend on two ratios: $\lambda_1/M_1$ and $\lambda_3/M_3$. Constraints on this model from our global fit are then shown in \autoref{fig:fit2leptoquark}, marginalizing over the other Wilson coefficients that cannot be matched from this model at tree level. Since the 4-fermion global fit presented in this section does not involve any top operators, in \autoref{fig:fit2leptoquark}, we only show these collider options running below the top pair production threshold. We conclude that future lepton colliders will surpass the LHC or its high-luminosity era significantly in exploring the parameter space of this model. In particular, due to the large luminosity of CEPC, it will be more competitive than FCC-ee in probing both ratios. In contrast, the linear colliders will be more powerful than the circular colliders in constraining $\lambda_3/M_3$, or equivalently orders of magnitude better in constraining $\hat{c}_{\ell q}^{(3)}$, due to beam polarization.

\begin{figure}
\centering
    \includegraphics[width=0.55\textwidth]{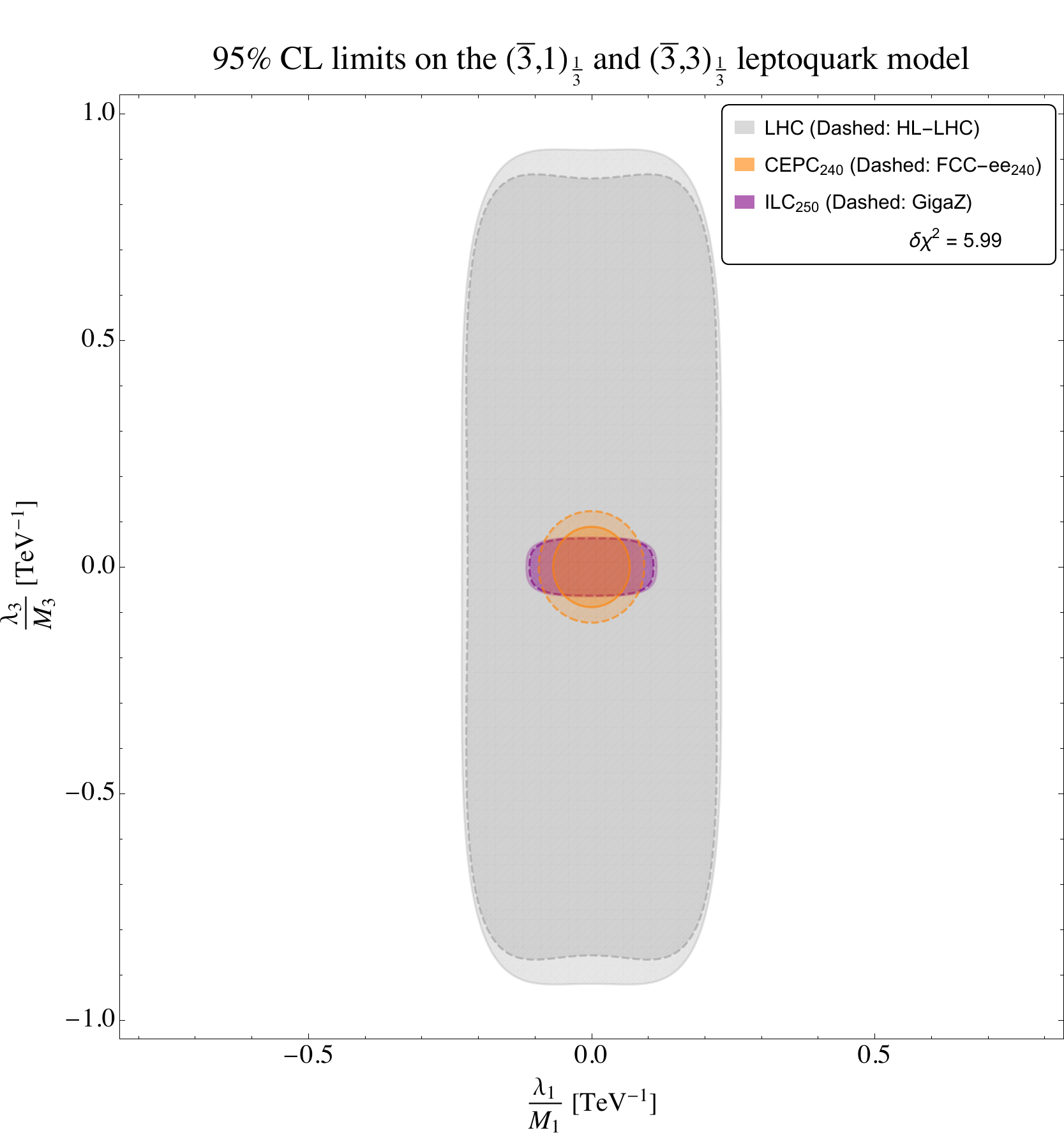}
    \caption{Constraints on the scalar leptoquark model from the global 4-fermion fit.}
    \label{fig:fit2leptoquark}
\end{figure}

%%%%%%%%%%%%%%%%%%%%%%%%%%%%%%%%%%
\section{CP-odd operators \label{sec:CPodd}}
%%%%%%%%%%%%%%%%%%%%%%%%%%%%%%%%%%
%[editor: Christophe, Jorge, Yong]

All the results presented thus far pertain only to CP-even SMEFT interactions.
Here we explore the constraints that CP observables can be set on the bosonic sector of the dimension-six SMEFT.

There are in total six pure bosonic CPV SMEFT operators in the Warsaw basis, which are summarized in \autoref{eq:LSMEFT-CPodd-bos}. The two operators involving gluons can be very stringently constrained by neutron and chromo electronic dipole moments\cite{Cirigliano:2016nyn}, we thus do not include them in our fit. These remaining operators will affect the triple gauge couplings (TGCs) that are phenomenologically parameterized as\cite{Hagiwara:1986vm}, in the broken phase,
\eqal{\left(\mathcal{L}_{\rm CPV}^{\rm bosonic}\right)_{\rm broken}^{V_1V_2V_3} = & \, i e\left(\widetilde{\kappa}_{\gamma} \widetilde{F}_{\mu \nu} W^{+\mu} W^{-\nu}+\frac{\widetilde{\lambda}_{\gamma}}{M_{{W}}^{2}} \widetilde{F}^{\nu \lambda} W_{\lambda \mu}^{+} W^{-\mu}{}_{\nu}\right.\nonumber\\
&\quad\quad\left.\,+\cot\theta_{{w}} \widetilde{\kappa}_{{Z}} \tilde{Z}_{\mu \nu} W^{+\mu} W^{-\nu} + \cot\theta_{{w}} \frac{\widetilde{\lambda}_{{Z}}}{M_{{W}}^{2}} \widetilde{Z}^{\nu \lambda} W_{\lambda \mu}^{+} W^{-\mu}_{\nu}\right)\nonumber\\
&\, + e \cot\theta_{{w}} \widehat{\kappa}_{{Z}}\left(\partial^{\mu} Z^{\nu}+\partial^{\nu} Z^{\mu}\right) W_{\mu}^{+} W_{\nu}^{-}.\label{eq:cpvbosonic}}
Clearly, all the terms in the first bracket violate both $\widehat{P}$ and $\widehat{CP}$ while conserve $\widehat{C}$, and conversely, the last term in \autoref{eq:cpvbosonic} violates both $\widehat{C}$ and $\widehat{CP}$ while conserves $\widehat{P}$. In contrast, all the operators in \autoref{eq:LSMEFT-CPodd-bos} conserve $\widehat{C}$ but violate $\widehat{P}$ and $\widehat{CP}$. Therefore, fixing our notations as in\cite{Barklow:2017awn} in the unbroken phase,
\eqal{\left(\mathcal{L}_{\rm CPV}^{\rm bosonic}\right)_{\rm unbroken} =&\,+\frac{g^{2} \widetilde{c}_{W W}}{m_{W}^{2}} \phi^{\dagger} \phi W_{\mu \nu}^{a} \widetilde{W}^{a \mu \nu}+\frac{4 g g^{\prime} \widetilde{c}_{W B}}{m_{W}^{2}} \phi^{\dagger} \frac{\sigma_{a}}{2} \phi W_{\mu \nu}^{a} \widetilde{B}^{\mu \nu}\nonumber\\
&\,+\frac{g^{2} \widetilde{c}_{B B}}{m_{W}^{2}} \phi^{\dagger} \phi B_{\mu \nu} \widetilde{B}^{\mu \nu}+\frac{g^{3} \widetilde{c}_{3 W}}{m_{W}^{2}} \varepsilon_{abc} W_{\mu \nu}^{a} W^{b \nu}{ }_{\rho} \widetilde{W}^{c \rho \mu} ,\label{eq:cpvunbroken}}
one can readily obtain the matching between these two formalisms:
\eqal{& \widetilde{\kappa}_{\gamma} = -8\widetilde{c}_{WB} ,\quad\quad\quad\quad \widetilde{\kappa}_{Z} = \frac{8s_w^2}{c_w^2}\widetilde{c}_{WB}=-\frac{s_w^2}{c_w^2}\widetilde{\kappa}_{\gamma},\\
& \widetilde{\lambda}_\gamma = \widetilde{\lambda}_Z = 6 g^2 \widetilde{c}_{3W},\quad\,\,\,\, \widehat{\kappa}_Z = 0.\label{eq:cpvmatching}
}
Note that all results are in perfect agreement with those in \cite{LHCHiggsCrossSectionWorkingGroup:2016ypw,Azatov:2022kbs} after a notation transformation.

The OPAL collaboration has reported their measurements of $\widetilde{\lambda}_Z=-0.18_{-0.16}^{+0.24}$ and $\widetilde{\kappa}_{Z}=-0.20_{-0.07}^{+0.10}$ in \cite{OPAL:2000wbs}, which can thus be used to constrain $\widetilde{c}_{WB}$ and $\widetilde{c}_{3W}$ in \autoref{eq:cpvunbroken}. While these bounds are weak, it is essential to include them to lift the flat directions. On the other hand, these operators also modify the production and the decay of the Higgs at the LHC \cite{Gritsan:2020pib,Davis:2021tiv,Gritsan:2022php}. As a result, stringent bounds on these operators have been obtained from the $h\to4\ell$ channel, which are included in our fit. In addition, for future lepton colliders, we also utilize the angular asymmetries $\mathcal{A}_\phi^{(1)}$ and $\mathcal{A}_\phi^{(2)}$ from $e^+e^-\to ZH$ production\cite{Beneke:2014sba} and their projections at both future circular lepton colliders\cite{Craig:2015wwr} and linear ones\cite{Ogawa:2017bmg}. Recall that these angular asymmetric observables in \cite{Craig:2015wwr} are parameterized using the mass eigenstates
\eqal{\left(\mathcal{L}_{\rm CPV}^{\rm bosonic}\right)_{\rm broken}^{hV_1V_2} = \frac{\widehat{\alpha}_{Z\widetilde{Z}}}{v}hZ_{\mu\nu}\widetilde{Z}^{\mu\nu} +  \frac{\widehat{\alpha}_{A\widetilde{Z}}}{v} hA_{\mu\nu}\widetilde{Z}^{\mu\nu},}
one can readily match these $\widehat{\alpha}$'s onto those in our notations in \autoref{eq:cpvbosonic} and find
\begin{itemize}
\item For circular colliders:
\eqal{\widehat{\alpha}_{Z\widetilde{Z}} =&\, \frac{4}{c_w^2}\widehat{\widetilde{c}}_{BB},\label{eq:hvvmatching1}\\
\widehat{\alpha}_{A\widetilde{Z}} = &\,\frac{8s_w}{c_w}\widehat{\widetilde{c}}_{WW}.\label{eq:hvvmatching2}}
\item For polarized linear colliders with the subscripts indicating the beam polarization: 
\eqal{(\widehat{\alpha}_{Z\widetilde{Z}})_{e^-_Le^+_R} =&\, \frac{8\widehat{\widetilde{c}}_{BB}}{c_w^2} + \frac{s_w^2}{1/2-s_w^2}\frac{s-m_Z^2}{s}8\widehat{\widetilde{c}}_{WW},\label{eq:hvvmatching3}\\
(\widehat{\alpha}_{Z\widetilde{Z}})_{e^-_Re^+_L} = &\, \frac{8\widehat{\widetilde{c}}_{BB}}{c_w^2} - \frac{s-m_Z^2}{s}8\widehat{\widetilde{c}}_{WW},\label{eq:hvvmatching4}}
where $\sqrt{s}$ is the center of mass energy, and we define the following variables for simplicity:
\eqal{\widehat{\widetilde{c}}_{BB}\equiv&\, c_w^4\widetilde{c}_{WW} + s_w^4\widetilde{c}_{BB} + 2c_w^2s_w^2 \widetilde{c}_{WB},\\
\widehat{\widetilde{c}}_{WW}\equiv&\, c_w^2(\widetilde{c}_{WW} - \widetilde{c}_{WB}) + s_w^2(\widetilde{c}_{WB} - \widetilde{c}_{BB}).}
\end{itemize}
Utilizing $\widetilde{\lambda}_Z$ and $\widetilde{\kappa}_{Z}$ from OPAL, the $h\to4\ell$ decay channel at the HL-LHC, and the two angular asymmetries $\mathcal{A}_\phi^{(1)}$ and $\mathcal{A}_\phi^{(2)}$ from $ZH$ production as just discussed above, the fit for the four CPV operators can then be closed. The results are shown in \autoref{fig:fit4cpv1} using the notations in \cite{Gritsan:2020pib,Davis:2021tiv,Gritsan:2022php} to compare different colliders, and we do not combine future colliders with the HL-LHC for this purpose. The upper row of \autoref{fig:fit4cpv1} is obtained by using the aTGC results from OPAL in \cite{OPAL:2000wbs}, and these results are estimated to, to be conservative, get improved by a factor of 10 (100) for the HL-LHC (future lepton colliders). We then obtain the results in the second row of \autoref{fig:fit4cpv1} based on this estimation. We find the HL-LHC could better constrain $g_4^{\gamma\gamma}$, while it will not be as competitive as future colliders in terms of constraining $g_4^{ZZ,Z\gamma}$. Furthermore, the linear colliders will in general surpass the circular ones due to beam polarization. Interestingly, we also find the hadron and the lepton colliders are sensitive to very different combinations of these $g_4$ couplings, or equivalently, $\widetilde{c}_{WW,WB,BB}$ in \autoref{eq:cpvunbroken} in the Warsaw basis. This is explicitly shown in \autoref{fig:fit4cpv2}, where the blue region is for the HL-LHC, and the orange for ILC250. We present in the left panel the region plot with $\delta\chi^2=1$ from the current fit, and the right one from the future fit based on the aforementioned estimation. This complementarity between the HL-LHC and future lepton colliders helps improve the global fit for the CPV operators when they are combined together, as is shown in \autoref{fig:fit4cpv3} and numerically summarized in \autoref{tab:fit4cpv}.

\begin{figure}
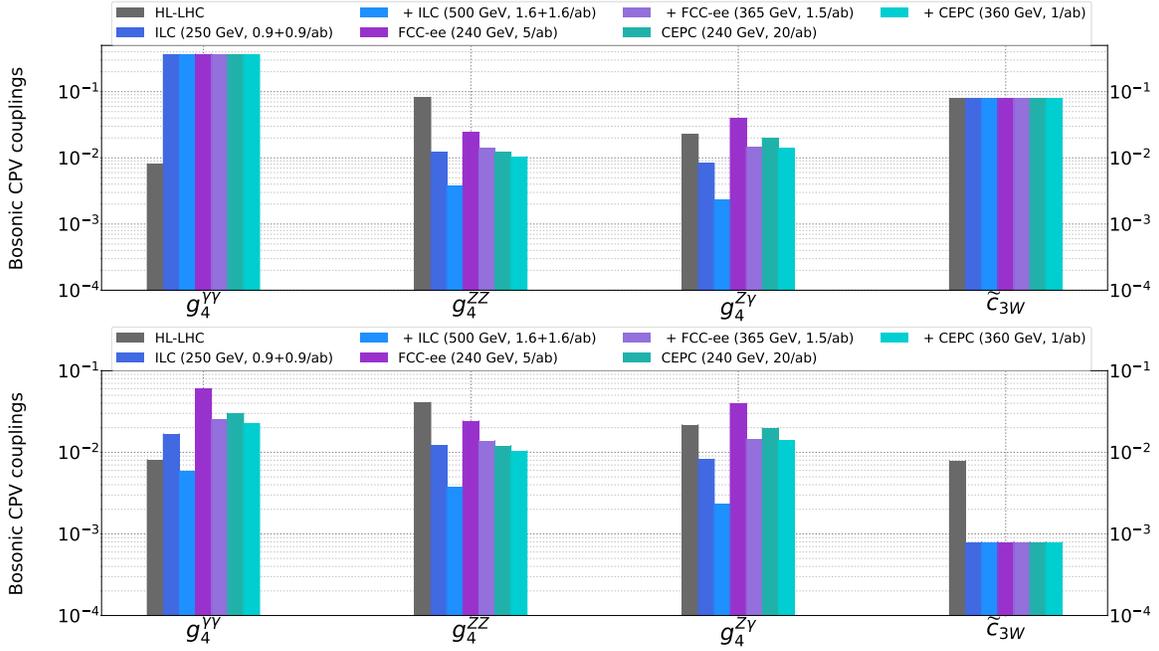

\centering
    \includegraphics[width=\textwidth]{plots/cpv/CPVg4/CPV-comparison}\\
    \includegraphics[width=\textwidth]{plots/cpv/CPVg4/CPV-comparison-scaled}
    \caption{Global fit results for the CPV $g_4$ couplings. Above: Results from the current fit. Bottom: Assuming OPAL precision on aTGCs is improved by a factor of 10 (100) for HL-LHC (future colliders).}
    \label{fig:fit4cpv1}
\end{figure}

\begin{figure}
\centering
    \includegraphics[width=0.45\textwidth]{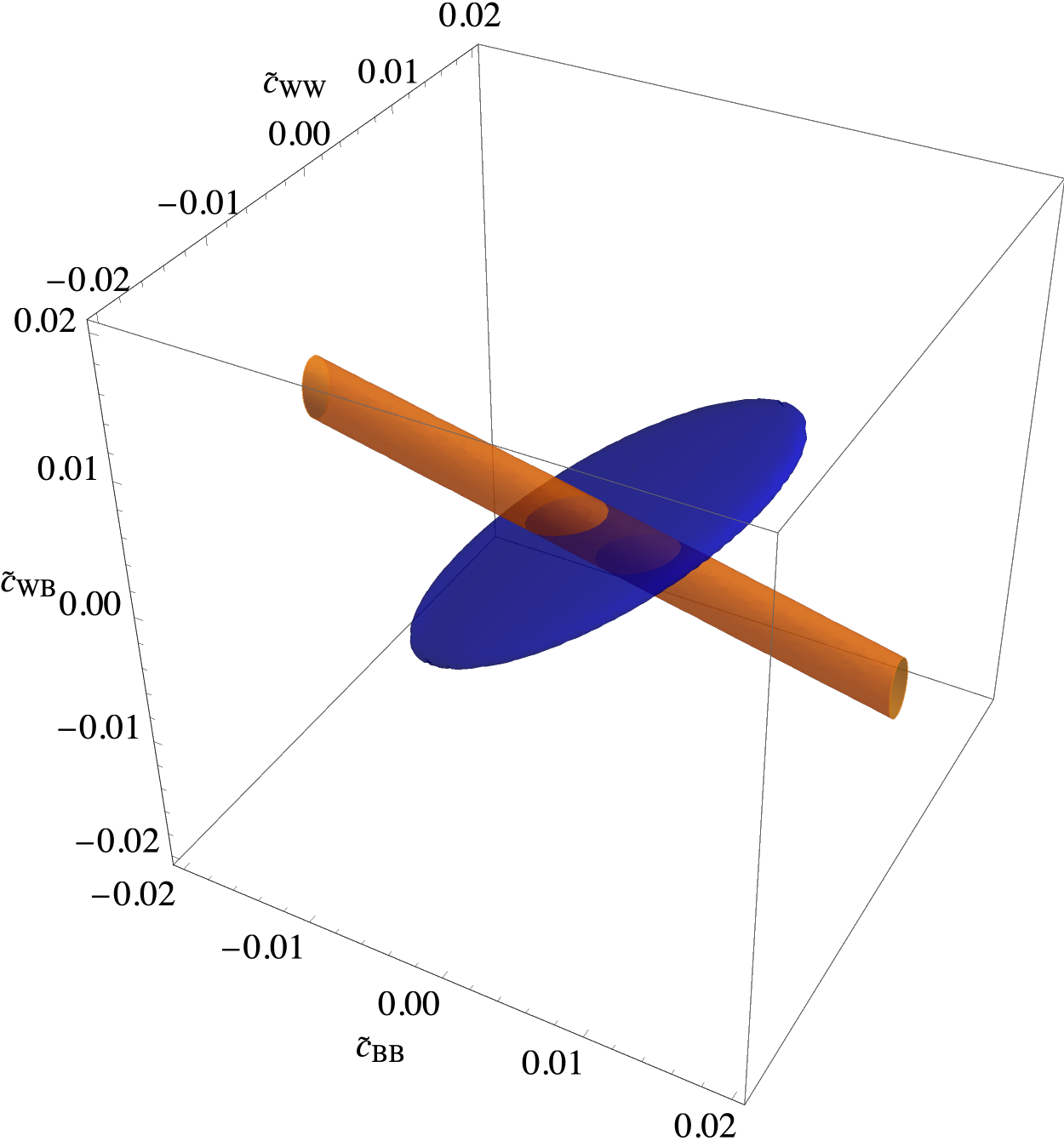}  \hspace{0.2cm}
    \includegraphics[width=0.45\textwidth]{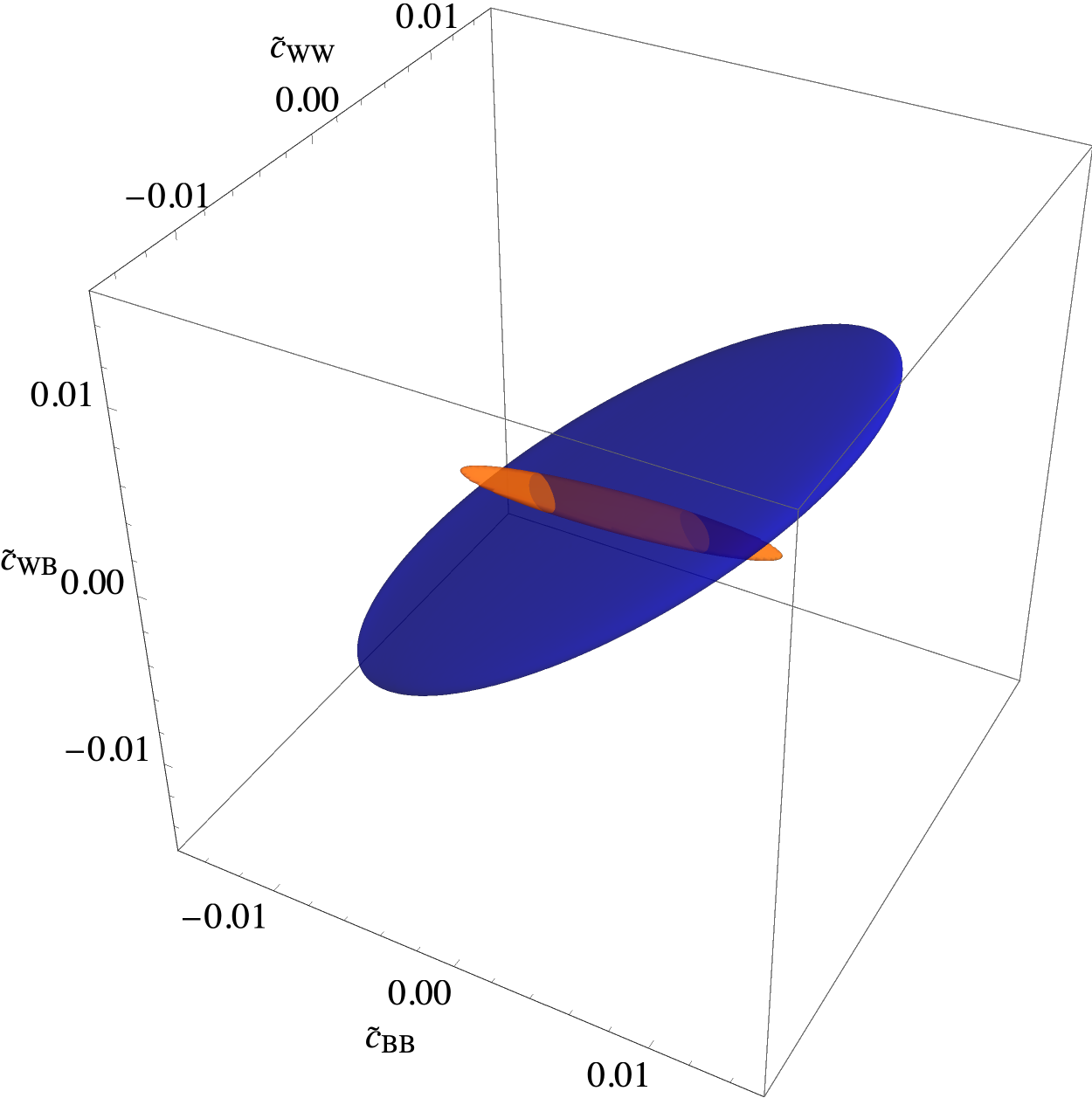}\\
    \caption{The 3D region plot with $\delta\chi^2=1$, where blue is for the HL-LHC and orange for ILC250. Left: Results from the current fit. Right: Assuming OPAL precision on aTGCs is improved by a factor of 10 (100) for HL-LHC (future colliders).}
    \label{fig:fit4cpv2}
\end{figure}

\begin{figure}
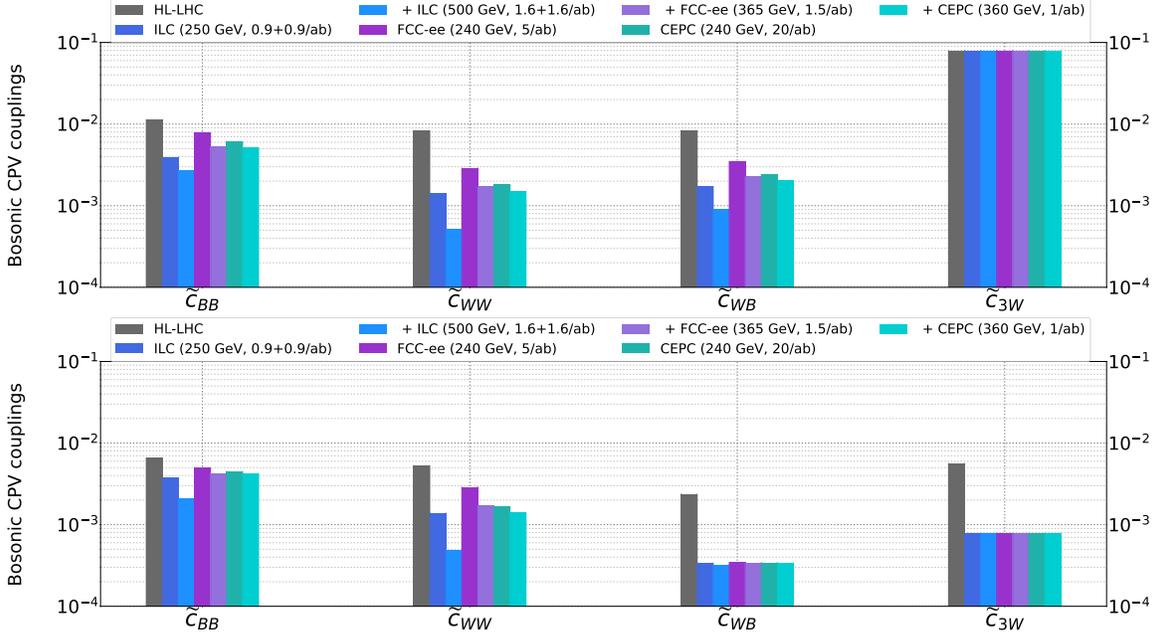

\centering
    \includegraphics[width=\textwidth]{plots/cpv/CPVwarsaw/CPV}\\
    \includegraphics[width=\textwidth]{plots/cpv/CPVwarsaw/CPV-scaled}
    \caption{Global fit results for the CPV operators in the Warsaw basis. Above: Results from the current fit. Bottom: Assuming OPAL precision on aTGCs is improved by a factor of 10 (100) for HL-LHC (future colliders).}
    \label{fig:fit4cpv3}
\end{figure}

\begin{table}[]
  \centering
  \begin{adjustbox}{max width=1\textwidth, max height=0.49\textheight}
\begin{tabular}{|c|cc|cc|cc|cc|cc|cc|cc|}\hline
in & \multicolumn{2}{c|}{HL-LHC} &  \multicolumn{4}{c|}{ILC} & \multicolumn{4}{c|}{FCC-ee}  & \multicolumn{4}{c|}{CEPC} \\ 
\cline{2-15} %\hline
\% & \multicolumn{2}{c|}{14\,TeV} & \multicolumn{2}{c|}{250} & \multicolumn{2}{c|}{+500} &  \multicolumn{2}{c|}{240} & \multicolumn{2}{c|}{+365} & \multicolumn{2}{c|}{240} & \multicolumn{2}{c|}{+360} \\ 
 & & est. &  & est. &  & est. &  & est. &  & est. &  & est. & & est.\\
\hline
\hline
$\widetilde{c}_{BB}$ & 1.13   & 0.66    & 0.39 & 0.376 & 0.27   & 0.21   & 0.78 & 0.50     & 0.53   & 0.43   & 0.61 & 0.45   & 0.51   & 0.425\\
$\widetilde{c}_{WW}$ & 0.83 & 0.53    & 0.14 & 0.14   & 0.051 & 0.049 & 0.29 & 0.29     & 0.17   & 0.17   & 0.18 & 0.167 & 0.15   & 0.14\\
$\widetilde{c}_{WB}$ & 0.83  & 0.237  & 0.17 & 0.034 & 0.09   & 0.032 & 0.35 & 0.0344 & 0.225 & 0.034 & 0.24 & 0.034 & 0.203 & 0.034 \\
$\widetilde{c}_{3W}$ & 7.93  & 0.56    & 7.93  & 0.078 & 7.93  & 0.078 & 7.93 & 0.078   & 7.93   & 0.078 & 7.93 & 0.078 & 7.93    & 0.078\\
  \hline
  \end{tabular}
  \end{adjustbox}
\caption{Precision reach (in percentage) on the effective couplings from a SMEFT global analysis of the CPV operators at various future lepton colliders. Here ``est.'' means the results from the future global fit based on our estimation for the aTGC precision at the HL-LHC and future colliders. See text for details.}
\label{tab:fit4cpv}
\end{table}

Could these constraints be further improved by including others observables that are sensitive to these CPV operators? Realizing that these operators could also modify the $h\to\gamma\gamma$ decay rate, which would be measured at the percent level at the high-luminosity era of the LHC or future lepton colliders \cite{deBlas:2019rxi}, it is natural to investigate the impact of this decay rate on possibly improving the fit. To that end, we note that, generically, the $h\to\gamma\gamma$ decay rate will receive corrections from both the CP-even and the CP-odd operators. The relevant Lagrangian in the broken phase can be expressed as
\eqal{\mathcal{L}_{h\gamma\gamma} = \frac{1}{2v}h\zeta_A A_{\mu\nu}A^{\mu\nu} + \frac{1}{2v}h\widetilde{\zeta}_A A_{\mu\nu}\widetilde{A}^{\mu\nu}.}
The $\zeta_A$ and $\widetilde{\zeta}_A$ parameters in the broken phase can be straightforwardly matched onto those Wilson coefficients in the unbroken phase, and the former case has been investigated in \cite{Barklow:2017awn} and independently checked to give\footnote{Note the typo of an extra factor of ``8'' in \cite{Barklow:2017awn}.}
\eqal{\zeta_A = s_w^2\, [8c_{WW} + 8 c_{BB} - 2(8c_{WB})],}
while for the latter, we find
\eqal{\widetilde{\zeta}_A = s_w^2\, [8\widetilde{c}_{WW} + 8 \widetilde{c}_{BB} - 2(8\widetilde{c}_{WB})],}
and both results are in perfect agreement with those in \cite{LHCHiggsCrossSectionWorkingGroup:2016ypw,Azatov:2022kbs} after a notation transformation.

Due to the different transformation properties under $\widehat{CP}$, the CP-even and the CP-odd sectors do not interfere with one another and the CP-odd operators would only contribute at the quadratic order. However, since the SM contribution to $h\to\gamma\gamma$ is loop suppressed, one could thus expect the CP-odd operators to contribute at the same order compared with the SM, or the leading-order interference between the SM and the CP-even operators, or the quadratic contributions from the CP-even operators. For estimation, one can ignore these contributions from the CP-even operators since their corresponding Wilson coefficients are constrained at $\mathcal{O}(10^{-5})$ from a global study on the Higgs couplings at ILC250+500 in \cite{Barklow:2017awn}. We then find adding this rate to the global fit could further improve the $1\sigma$ bounds on these CPV operators by a factor of a few except for ${W}^a_{\mu\nu}{W}^{b\nu}_{\rho}\widetilde{W}^c_{\rho\mu}$ due to its vanishing contribution to this rate.

%\begin{figure}%[t]
%    \includegraphics[width=\textwidth]{plots/cpv/CPV}\caption{Precision reach on the effective couplings from a SMEFT global analysis of the bosonic and gluon-free CPV operators at various future lepton colliders. The light (solid) color is obtained without (with) the inclusion of the $h\to\gamma\gamma$ decay rate.}
%    \label{fig:fit4-cpvbosonic}
%\end{figure}

%%%%%%%%%%%%%%%%%%%%%%%%%%%%%%%%%%
\section{The top-quark sector in the global EFT fit}\label{sec:topfit}
%%%%%%%%%%%%%%%%%%%%%%%%%%%%%%%%%%
%[editor: Victor, Marcel, Eleni]

The Tevatron and LHC have characterized top-quark interactions to excellent precision. Differential measurements of top quark pair production well into the boosted regime provide a strong constraint on the $t\bar{t}-$gluon vertex and $q\bar{q}t\bar{t}$ operators~\cite{Rosello:2015sck}. Top quark decay, elecro-weak single top-quark production and associated production with a $Z-$ or $W-$boson, a photon or a Higgs boson constrain the electro-weak interactions and Yukawa coupling directly~\cite{Miralles:2021dyw,Durieux:2019rbz,Maltoni:2019aot}. Four-top production and $t\bar{t}b\bar{b}$ production, finally, constrain the four-heavy-quark operators~\cite{Banelli:2020iau}. Several groups have performed fits of the top sector of the SMEFT to these data~\cite{Hartland:2019bjb, Brivio:2019ius, Buckley:2015nca, Buckley:2015lku} and even explored the subtle interplay between the top sector and the Higgs/EW sectors, combining Higgs, electro-weak and top data in comprehensive SMEFT fits with several tens of parameters~\cite{Ethier:2021bye,Ellis:2020unq}. 

The prospects for the complete LHC programme, including the high-luminosity phase that collects an integrated luminosity of 3~ab$^{-1}$, are based on an extrapolation of current run 2 results. The measurements that form the basis of our projection are listed in Table~\ref{tab:measurements_top_sector}. For rare associated production processes the S2 scenario, also used for Higgs physics projections~\cite{Cepeda:2019klc}, is adopted. In this scenario, experimental systematic uncertainties, as well as statistical uncertainties, are assumed to scale with the inverse of the integrated luminosity. Theoretical and modelling uncertainties are reduced by a factor two.
For top quark production, where measurements already reach a precision of a few \%, the systematic uncertainty is divided by two. Further details are provided in Ref.~\cite{Durieux:2022cvf}.

\begin{table*}[!ht]
\centering
\resizebox{\textwidth}{!}{  
\begin{tabular}{|l|c|c|c|c|c|c|}
\hline
Process & Observable & $\sqrt{s}$  & $\int \cal{L}$  & Experiment & SM  & Ref.\\ \hline
%$p p \rightarrow t H q$ & x-sec upper limit & 13 TeV & 140~\ifb & CMS & 
%- & - \\ 
$pp \rightarrow \ttbar $ & $d\sigma/dm_\ttbar$ (15+3 bins) & 13 TeV & 140~fb$^{-1}$ & CMS & \cite{Czakon:2013goa} & \cite{CMS:2021vhb} \\
$pp \rightarrow \ttbar $ & $dA_C/dm_\ttbar$ (4+2 bins) & 13 TeV & 140~fb$^{-1}$ & ATLAS & \cite{Czakon:2013goa} & \cite{ATLAS-CONF-2019-026} \\
$p p \rightarrow t \bar{t} H+ tHq$ & $\sigma$ & 13 TeV & 140~fb$^{-1}$ & ATLAS & 
\cite{deFlorian:2016spz} &  \cite{ATLAS:2020qdt} \\ 
%$p p \rightarrow t \bar{t} Z$ &  inclusive x-sec. & 13 TeV &  140~\ifb & ATLAS  & 
%\cite{Bylund:2016phk} & \cite{Aaboud:2019njj} \\ 
$p p \rightarrow t \bar{t} Z$ &  $d\sigma/dp_T^Z$ (7 bins) & 13 TeV &  140~fb$^{-1}$  & ATLAS &
\cite{Broggio:2019ewu} & \cite{ATLAS:2020cxf} \\ 
%$p p \rightarrow t \bar{t} \gamma$ & inclusive x-sec. & 13 TeV & 140~\ifb & ATLAS &
%\cite{Bylund:2016phk} & \cite{Aaboud:2018hip}  \\ 
$p p \rightarrow t \bar{t} \gamma$ & $d\sigma/dp_T^\gamma$ (11 bins) & 13 TeV & 140~fb$^{-1}$ & ATLAS &
\cite{Bevilacqua:2018woc,Bevilacqua:2018dny} & \cite{Aad:2020axn}  \\ 
$p p \rightarrow tZq$ & $\sigma$ & 13 TeV & 77.4~fb$^{-1}$  & CMS &
 \cite{Sirunyan:2017nbr} & \cite{Sirunyan:2018zgs} \\ 
$p p \rightarrow t\gamma q$ & $\sigma$ & 13 TeV & 36~fb$^{-1}$  & CMS &
 \cite{Sirunyan:2018bsr} & \cite{Sirunyan:2018bsr} \\
 $p p \rightarrow t \bar{t} W$ & $\sigma$ & 13 TeV & 36~fb$^{-1}$  & CMS &
\cite{deFlorian:2016spz,Frederix:2017wme} &  \cite{Sirunyan:2017uzs} \\
 $p p \rightarrow t\bar{b}$ (s-ch) & $\sigma$ & 8~TeV & 20~fb$^{-1}$  & LHC &
\cite{Aliev:2010zk,Kant:2014oha} & \cite{Aaboud:2019pkc} \\ 
$p p \rightarrow tW$ & $\sigma$ & 8~TeV & 20~fb$^{-1}$   & LHC &
 \cite{Kidonakis:2010ux} &  \cite{Aaboud:2019pkc}   \\ 
$p p \rightarrow tq$ (t-ch) & $\sigma$ & 8~TeV & 20~fb$^{-1}$  & LHC &
\cite{Aliev:2010zk,Kant:2014oha} & \cite{Aaboud:2019pkc} \\ 
$t \rightarrow Wb $ & $F_0$, $F_L$  & 8~TeV & 20~fb$^{-1}$  & LHC &
\cite{Czarnecki:2010gb}  & \cite{Aad:2020jvx} \\
$p\bar{p} \rightarrow t\bar{b}$ (s-ch) & $\sigma$ & 1.96~TeV & 9.7~\ifb & Tevatron & \cite{Kidonakis:2010tc} & \cite{CDF:2014uma} \\
$e^{-} e^{+} \rightarrow b \bar{b} $ & $R_{b}$ ,  $A_{FBLR}^{bb}$ & $\sim$ 91~GeV & 202.1~pb$^{-1}$  & LEP/SLD &
 $-$ & \cite{ALEPH:2005ab}  \\ \hline
\end{tabular}
}
\caption{Measurements included in the EFT fit of the top-quark electroweak sector. For each measurement, the process, the observable, the center-of-mass energy, the integrated luminosity and the experiment/collider are given. The last two columns list the references for the predictions and measurements that are included in the fit. LHC refers to the combination of ATLAS and CMS measurements. In a similar way, Tevatron refers to the combination of CDF and D0 results, and LEP/SLD to different experiments from those two accelerators.}
\label{tab:measurements_top_sector}
\end{table*}

A future electron-positron collider is expected to improve the measurements of the bottom EW couplings, and, when operated above the $t\bar{t}$ threshold, the top EW couplings~\cite{Janot:2015yza,Amjad:2015mma, Amjad:2013tlv} and provide strong bounds on $e^+e^-t\bar{t}$ operators~\cite{Durieux:2018tev}.

Prospects for the $e^+e^- \rightarrow b\bar{b}$ process are included that are based on the full-simulation studies of the ILD concept~\cite{Okugawa:2019ycm} at $\sqrt{s} = $ 250~GeV. The prospects are based on realistic estimates of efficiency and acceptance, including the signal losses required to ensure a robust calibration of the flavour tagging efficiency. The statistical uncertainties on the measurements of the cross section and forward-backward asymmetry are complemented by polarisation and flavour-tagging systematics. For the $Z$-pole runs we use the projections for $R_b$ and $A_{FB}$ provided by the FCCee and CEPC projects for their ``TeraZ'' runs at the $Z$-pole, shown in Table~\ref{tab:EWPO}.

The $e^+e^- \rightarrow t\bar{t}$ process opens up for centre-of-mass energies that exceed twice the top mass (i.e. $\sqrt{s} \gtrsim$ 350~GeV) and probes the electroweak couplings of the top quark at tree-level.
Data taken with different beam polarisations at linear colliders can be used to distinguish the photon and $Z$-boson couplings~\cite{Amjad:2013tlv,Amjad:2015mma,Durieux:2018tev,CLICdp:2018esa}.
At circular colliders, a measurement of the final state polarisation using the semi-leptonically decaying top quarks can also be used to separate the two contributions~\cite{Janot:2015yza}.
We base our prospects on the study of statistically optimal observables defined at leading order on the $e^+e^- \to t\bar{t} \rightarrow WbWb$ differential distribution~\cite{Durieux:2018tev}.
This $WbWb$ final state also receives contribution from single top production which become sizeable at high centre-of-mass energies.
Realistic acceptance, identification and reconstruction efficiencies are estimated from full-simulation studies for the ILC and CLIC in Ref.~\cite{Amjad:2013tlv,Abramowicz:2016zbo}.
Since they were performed only for sub-set of centre-of-mass energies and beam polarisations, overall efficiency factors are extrapolated as a functions of the centre-of-mass energy.
They drop significantly for the TeV centre-of-mass energies of ILC and CLIC since a degradation of top-selection and flavour-tagging capabilities is expected in this regime.

The top-quark Yukawa coupling can be determined in a robust manner through the tree-level dependence of the associated $e^+e^- \rightarrow t\bar{t}H$ production process. This process is accessible at centre-of-mass energy above the $t\bar{t}H$ production threshold at $\sqrt{s}=$ 500--550~GeV. At linear colliders, where the luminosity grows with energy, there is a broad plateau up to about 1.5~TeV where $e^+e^- \rightarrow t\bar{t}H$ is accessible. We base our projections on full-simulation studies by ILC and CLIC~\cite{Abramowicz:2016zbo,Price:2014oca,Yonamine:2011jg}.

Several studies have been published for an energy-frontier hadron collider~\cite{FCC:2018vvp,Mangano:2016jyj,Mangano:2015aow,Aguilar-Saavedra:2014iga}, but no systematic projections have been performed of the broad top physics program and no quantitative results are presented here. For a qualitative discussion on this topic we refer to Ref.~\cite{Durieux:2022cvf}.

\begin{figure}[h!]
\hspace*{-1.5 cm}%
\includegraphics[width=1.25\columnwidth]{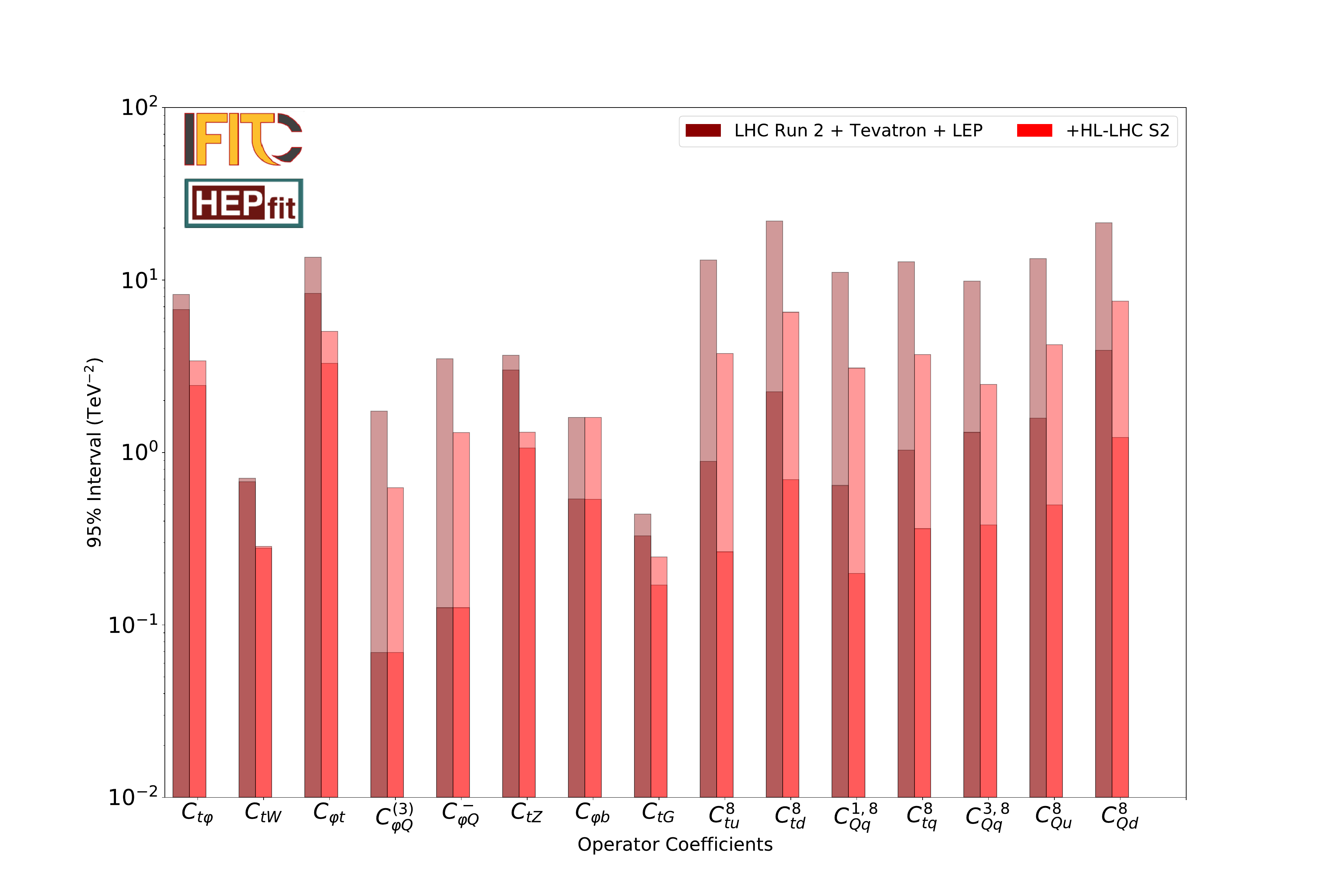}
\caption{\label{fig:hllhc_projection} The 95\% probability bounds on the Wilson coefficients for dimension-six operators that affect the top-quark production and decay measurements listed in Table~\ref{tab:measurements_top_sector} after run 2 of the LHC (in dark red) and prospects for the bounds expected after completion of the complete LHC program, including the high-luminosity stage (in light red). Only linear terms proportional to $\Lambda^{-2}$ are taken into account in the dependence of the observables on the Wilson coefficients. The individual bounds obtained from a single-parameter fit are shown as solid bars, while the global or marginalised bounds obtained fitting all Wilson coefficients at once are indicated by the full bars (shaded region in each bar). }
\end{figure}

In Fig.~\ref{fig:hllhc_projection} the 95\% probability bounds from a fit to the current data are shown in the dark red bars, as well as the limits obtained from the extrapolations of the complete HL-LHC program, with an integrated luminosity of 3~\iab{}, in light red. These fits, and the others of this section, have been performed using the \texttt{HEPfit} package \cite{DeBlas:2019ehy}.

Across the board, the HL-LHC program is expected to improve the bounds by a factor of two to four with respect to the current run 2 limits, both for individual bounds and global fit results. Exceptions are the individual bounds on $C_{\phi Q}^-$ and $C_{\phi Q}^3$, that continue to depend on the $Zb\bar{b}$ measurements at the $Z$-pole. 

The marginalised bounds on the four-fermion operators remain an order of magnitude worse than the individual bounds after the HL-LHC, even if both individual and global bounds improve considerably. This is due to unresolved correlations between the coefficients. The same feature is observed in recent fits to the top sector of the SMEFT~\cite{Brivio:2019ius,Hartland:2019bjb} and in global Higgs/EW/top fits~\cite{Ethier:2021bye,Ellis:2020unq}. Stricter limits can be obtained if the dimension-six-squared terms proportional to $\Lambda^{-4}$ are included in the fit~\cite{Ethier:2021bye}.

\begin{figure}[h!]
\hspace*{-1.5 cm}%
\includegraphics[width=1.25\columnwidth]{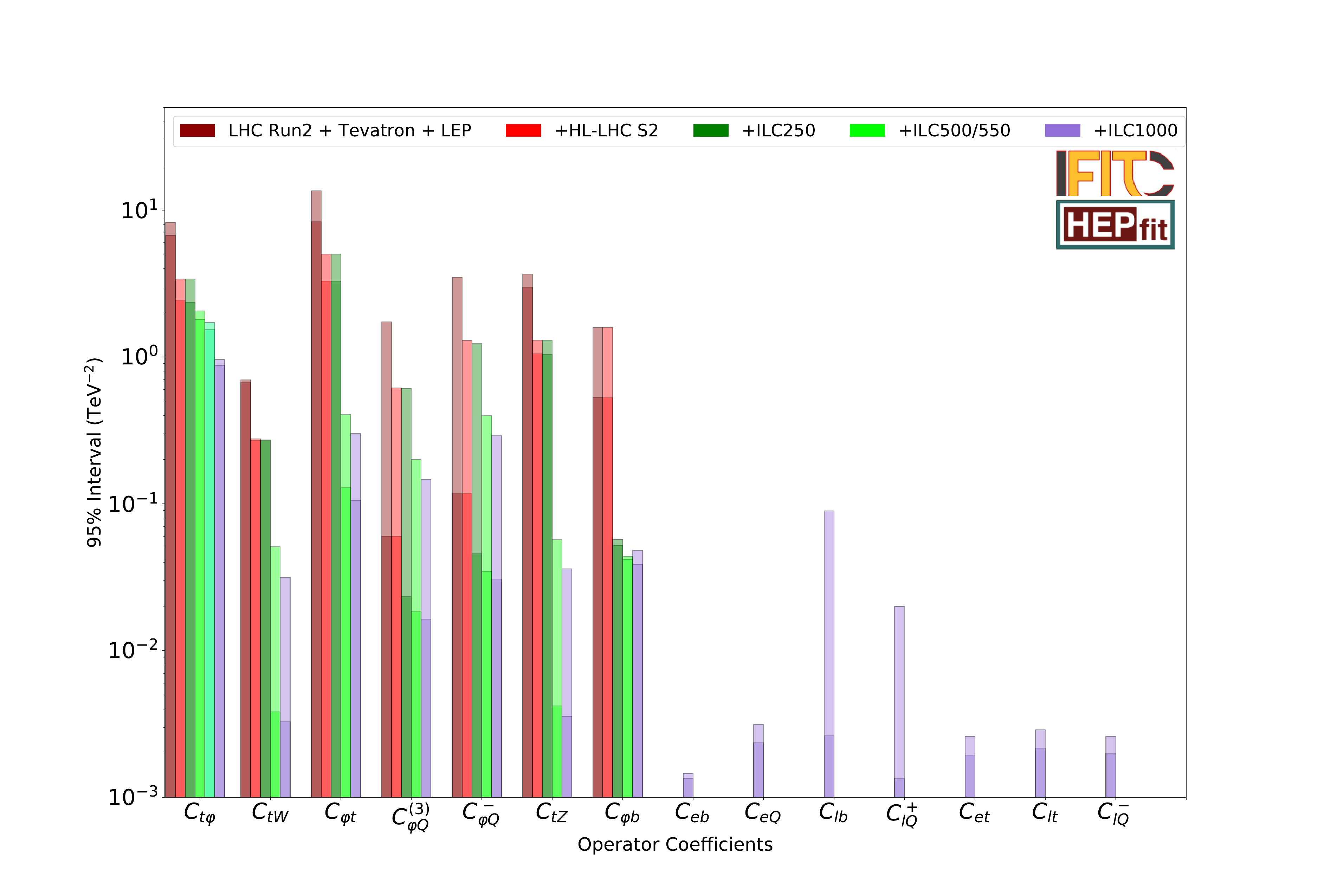}
\caption{\label{fig:hllhc_ILC_projection} Comparison of current LHC constraints with HL-LHC ones, and those deriving from ILC runs at 250, 500 and 1000~GeV.
The limits on the $q\bar{q}t\bar{t}$ and $C_{tG}$ coefficients are not shown, since the $e^+e^-$ collider measurements considered are not sensitive to them, but all operators are included in the global fit.
The improvement expected from the HL-LHC on these coefficients is shown in Fig.~\ref{fig:hllhc_projection}. The additional bar included for $C_{t\phi}$ in light green shows the effect on this operator of ILC working at 550 GeV. The solid bars provide the individual limits of the single-parameter fit and the shaded ones the marginalised limits of the global fit.}
\end{figure}

In Fig.~\ref{fig:hllhc_ILC_projection}, the impact of runs of electron-positron machines at different centre-of-mass energies is illustrated.  The current bounds in brown are compared to HL-LHC ones in red.
The subsequent bars add data at $\sqrt{s}= 250~\rm{GeV}$, 500~GeV and 1~TeV.
The beam polarizations and integrated luminosities of the different ILC stages are summarised in Table~\ref{tab:epem_setup}.
Only the electroweak operators are presented, as the $e^+e^-$ data have the strongest impact there, but results corresponds to a global analysis, including also the $q\bar{q}t\bar{t}$ operators and $C_{t G}$.

The dark green bar shows that the ``Higgs factory" run improves the bounds on bottom-quark operators, including $C_{\phi Q}^3$ and $C_{\phi Q}^-$ and $C_{\phi b}$. The improvement is especially pronounced for the individual bounds. As expected, data above the top-quark pair production threshold is required to improve the bounds on the top-quark operators. 

Runs at two different centre-of-mass energies above the top-quark pair production threshold are required to disentangle the $e^+e^-t\bar{t}$ operator coefficients from the two-fermion operator coefficients~\cite{Durieux:2018tev}.
The two sets of operators have very different scaling with energy: the sensitivity to four-fermion operators grows quadratically, while it is constant or grows only linearly for two-fermion operators.
In a fit to data taken at a single centre of mass, linear combinations of their coefficients remain degenerate and form blind directions.
The combination of runs at two different centre-of-mass energies effectively disentangles them and provides global fit constraints close to the individual bounds.
Note that the two-quark two-lepton operators could also be probed at the LHC, although we have ignored them in our LHC and HL-LHC analysis. 
Dedicated signal regions, for instance with off-$Z$-peak dilepton invariant masses in $pp\to t\bar{t}\ell^+\ell^-$~\cite{Durieux:2014xla, Chala:2018agk, CMS:2020lrr}, would increase their sensitivity.

\begin{figure}[t!]
\hspace*{-1.5 cm}%
\includegraphics[width=1.25\columnwidth]{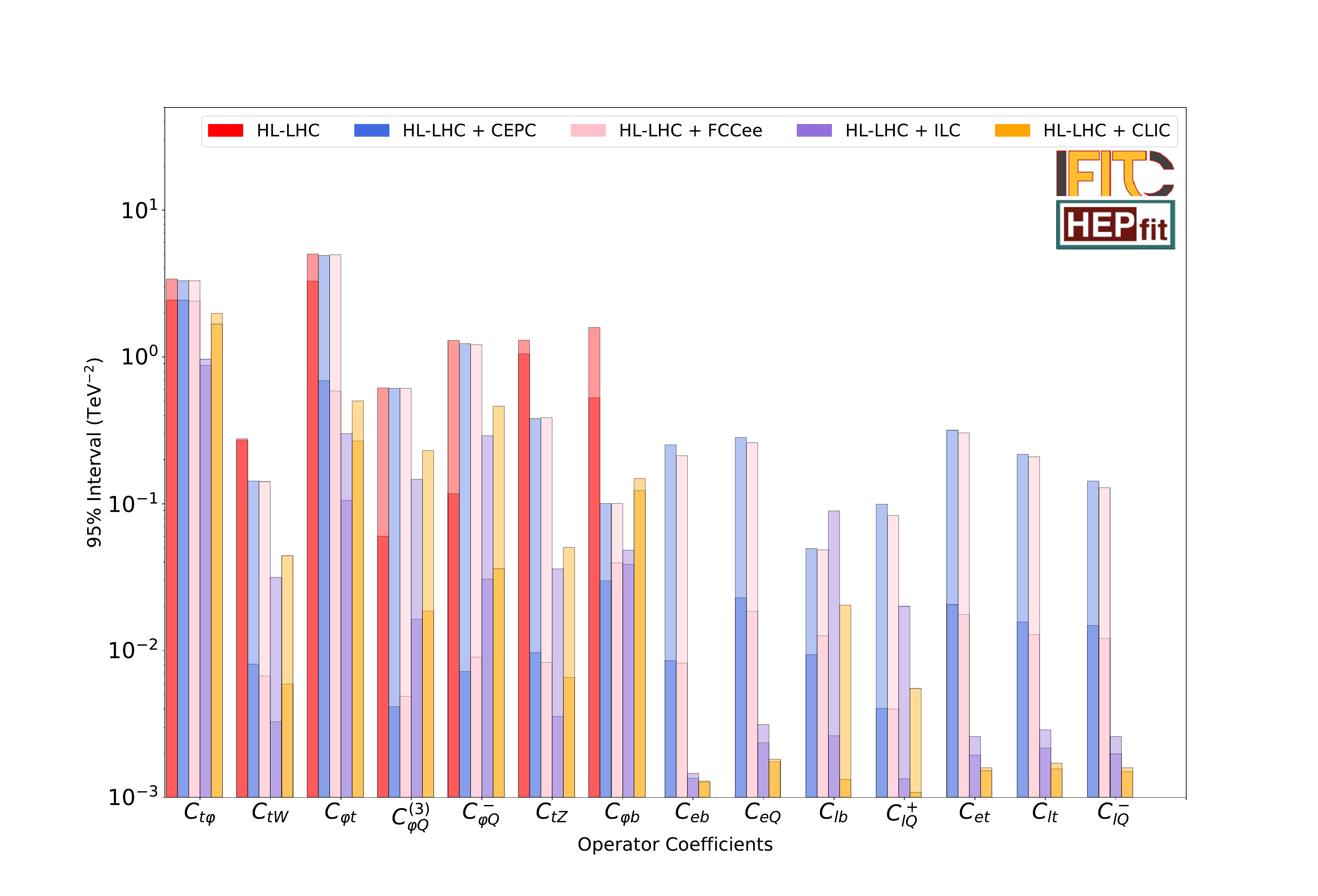}
\caption{\label{fig:CEPC_CC_ILC_CLIC_projection} Comparison of the constraints expected from a combination of HL-LHC and lepton collider data.
The limits on the $q\bar{q}t\bar{t}$ and $C_{tG}$ coefficients are not shown, since the $e^+e^-$ collider measurements considered are not sensitive to them, but all operators are included in the global fit.
The improvement expected from the HL-LHC on these coefficients is shown in Fig.~\ref{fig:hllhc_projection}. The solid bars provide the individual limits of the single-parameter fit and the shaded ones the marginalised limits of the global fit.
}
\end{figure}

In Fig.~\ref{fig:CEPC_CC_ILC_CLIC_projection}, we compare the bounds expected from the HL-LHC and from the final stages of the CEPC, FCC-$ee$, ILC and CLIC.
The centre-of-mass energies, integrated luminosities and beam polarisations envisaged for each of these projects are given in Table~\ref{tab:epem_setup}. 
The circular colliders (FCC-$ee$ and CECP) operated at and slightly above the $t\bar{t}$ threshold are expected to improve constraints on the bottom- and top-operators by factors 5 and 2 for some two-fermion operators.
Indeed, their ``TeraZ'' runs provide very competitive bounds (individual ones, in particular) on two-fermion bottom-operator coefficients.
Their constraining power on four-fermion operators is, however, limited by the energy reach.
Since, at these colliders, the two runs above the $t\bar{t}$-threshold are very close the two-fermion and four-fermion operators are harder to disentangle.
The global limits remain significantly above the individual bounds.

The linear colliders (ILC and CLIC), operated at two centre-of-mass energies above the $t\bar{t}$ threshold, can provide very tight bounds on all operators.
The bounds on four-fermion operators take advantage of the energy-growing sensitivity and become very competitive if $e^+e^-$ collision data at a centre-of-mass energy greater than 1 TeV is available.
The ILC1000 and CLIC3000 bounds of $\mathcal{O}(10^{-3})$ on the $e^+e^-t\bar{t}$ operators are by far the tightest top-sector SMEFT constraints that can be achieved at any future collider considered in this work.\footnote{A muon collider or advanced linear collider have the potential to improve these bounds further, but quantitative projections for integrated luminosity and experimental performance are currently not available.}

Furthermore, operation above the $e^+e^- \rightarrow t\bar{t}H$ production threshold provides a direct probe of the top-quark Yukawa.
The additional bar for $C_{t\phi}$, in Fig.~\ref{fig:hllhc_ILC_projection}, accounts for an ILC run at 550~GeV and shows the impact of the strongly enhanced cross section of the $e^+e^-\rightarrow t\bar{t}H$ process reaches the resonant peak boosts the sensitivity~\cite{Yonamine:2011jg} to the top-quark Yukawa coupling. Also the scenarios for 1~TeV and 1.5~TeV operation considered here yield competitive constraints on this process, that help to improve the bounds on $C_{t\phi}$ with respect to the HL-LHC, as shown in Fig.~\ref{fig:CEPC_CC_ILC_CLIC_projection}. The limits obtained for $C_{t\phi}$ have been expressed in terms of the top-quark yukawa coupling ($\delta y_t$) in Tab.~\ref{tab:top_yuk} using the relation $\delta y_t=-\frac{v^2}{\Lambda^2} C_{t\phi}$.

\begin{table}[]
    \centering
    \begin{tabular}{c|c|c|c|c|c|c|c|}
    \multicolumn{2}{c|}{ Values in \% units }   & LHC  & HL-LHC & ILC500 & ILC550 & ILC1000 & CLIC  \\\hline
    \multirow{2}{*}{$\delta y_t$} & Global fit  & 12.2 & 5.06   & 3.14   & 2.60   & 1.48   & 2.96  \\
                                  & Indiv. fit  & 10.2 & 3.70   & 2.82   & 2.34   & 1.41   & 2.52
    \end{tabular}
    \caption{
    Uncertainties for the top-quark yukawa coupling at 68\% probability for different scenarios, in percentage. The ILC500, ILC550 and CLIC scenarios also include the HL-LHC. The ILC1000 scenario includes also ILC500 and HL-LHC.
    }
    \label{tab:top_yuk}
\end{table}

%\section{Conclusions}
\section{Conclusion and Outlook}
In this work we performed a few global SMEFT fits for the Higgs and Electroweak sector, 4-fermion interactions, top-quark sector and pure bosonic CP-odd operators, each with a well defined subset of dimension-6 operators in the Warsaw basis. The focus was on the future lepton colliders with various running scenarios, that are being discussed in the process of Snowmass 2021. We conclude that future lepton colliders can advance significantly our understanding of the properties of various SM particles, by offering precise and coherent probes to new physics effects in a way that is independent of underlining UV models.

LHC will keep pushing the boundaries of precision measurements, capable of delivering 2-5\% precision for many Higgs effective couplings at the end of HL-LHC. Future $e^+e^-$ will not only be able to improve the precision by a factor of 2-10, but also provide a qualitatively new determination of the Higgs total width by treating it as a free parameter. The capabilities of all future $e^+e^-$ colliders considered in this work are shown to be similar for Higgs coupling determinations. Muon colliders can offer comparable precisions in the cases where, either the Higgs total width is constrained (not allowing any untagged Higgs exotic decay), or the 125 GeV run is combined. There are synergies, which play important roles in the global fits, on Higgs rare decays ($H\to\gamma\gamma,\gamma Z,\mu\mu$) as well as top-Yukawa coupling between HL-LHC and future lepton colliders.

Electroweak effective couplings for $W$ and $Z$ can be improved by a few orders of magnitude at future $e^+e^-$ colliders over what we know of today. Circular $e^+e^-$ can offer better precisions with the dedicated high luminosity run at $Z$-pole and $WW$ threshold. Linear $e^+e^-$ can offer competitive measurements on left-right asymmetries by either a dedicated $Z$-pole run or radiative return events at the same $ZH$ run using polarized beams. There are important synergies between EWPOs and direct Higgs observables. The $Z$-pole and $WW$ run at circular $e^+e^-$ can help improve Higgs coupling precisions by a factor of around two. While at linear $e^+e^-$ this improvement factor, which is much lower, already saturates after using the EWPOs by radiative events. 

The 4-fermion interactions can be probed at future $e^+e^-$ up to a scale of O(100) TeV when the underlining models are strongly coupled. The reaches are significantly better at linear $e^+e^-$ than circular $e^+e^-$ not only because of higher collision energies but also polarized beams which help lift degeneracies. There are important synergies with low-energy measurements without which certain degeneracies can not be lift. 

The measurements of top-quark mass and EW couplings will be improved significantly at future $e^+e^-$ when the top-pair threshold and open production runs are included. The degeneracies in $eett$ contact interactions can not be lift without running at two different energies well above $tt$ threshold. Many top-quark measurements at (HL-)LHC are helpful in the global fit for improving the precision of top-quark EW couplings. 

The advance in the SM theory predictions will be indispensable in order to match the precision that will become reachable at future lepton colliders. In general beyond NNLO electroweak corrections will be needed. The requirement is in particular strong for EWPOs by the $Z$-pole programs at cicular $e^+e^-$.

\Acknowledgements
We would like to thank A. Belloni and A. Freitas for continuous support
in coordinating the efforts for the needed inputs and facilitating discussions of this work with EF04 group members. We also would like to thank S. Dawson, A. Gritsan, J. M. Hernandez, A. Irels, Z. Liu, J. List, P. Meade, I. Ojalvo, R. Schwienhorst, C. Vernieri and D. Wacheroth for helpful discussions to this work. 

The work of J.B. has been supported by the FEDER/Junta de Andaluc\'ia project grant P18-FRJ-3735.
YD is supported by National Key Research and Development Program of China under Grant No. 2020YFC2201501, the National Science Foundation of China (NSFC) under Grants No. 12022514, No. 11875003 and No. 12047503, and CAS Project for Young Scientists in Basic Research YSBR-006, and the Key Research Program of the CAS Grant No. XDPB15.
JG is supported by National Natural Science Foundation of China (NSFC) under grant No.~12035008.
CG is supported  by the Helmholtz Association through the recruitment initiative program and by the Deutsche Forschungsgemeinschaft under Germany’s Excellence Strategy EXC 2121 “Quantum Universe” - 390833306.
JT was supported by the Japan Society for the Promotion of Science (JSPS) under Grants-in-Aid for Science Research 15H02083. The work of VM has been supported by the Italian Ministry of Research (MUR) under
the grant PRIN20172LNEEZ.

%%%%%%%%%%%%%%%%%%%%%%%%%%%%%%%%%%%%

%\newpage

%\begin{thebibliography}{99}
%\end{thebibliography}

\bibliographystyle{JHEP}
\bibliography{SMEFT21}

\end{document}